\documentclass[
reprint,
superscriptaddress,
nofootinbib,
 amsmath,amssymb,
 aps,
prx]{revtex4-2}

\usepackage{graphicx}
\usepackage{dcolumn}
\usepackage{bm}
\usepackage{amsmath}
\usepackage{comment}
\usepackage{layout}
\usepackage[dvipsnames]{xcolor}
\usepackage{xspace}
\usepackage{tabularx}
\usepackage{enumitem}

\newcommand{\nk}[1]{\textcolor{red}{(NK: #1)}}

\newcommand{\ow}[1]{\textcolor{teal}{(OW: #1)}}

\newcommand{\sqd}[1]{\textcolor{magenta}{(SQD: #1)}}

\newcommand{\Geant}{\textsc{Geant}4\xspace}

\begin{document}

\preprint{APS/123-QED}

\title{Position- and temperature-dependent responsivity in a KID-based phonon-mediated detector}

\author{Z.J.~Smith}
\affiliation{SLAC National Accelerator Laboratory, Menlo Park, California, USA}
\affiliation{Kavli Institute for Particle Astrophysics and Cosmology, Stanford University, Stanford, CA, USA}
\email{smithzj@stanford.edu}

\author{O.~Wen}
\affiliation{SLAC National Accelerator Laboratory, Menlo Park, California, USA}
\affiliation{Kavli Institute for Particle Astrophysics and Cosmology, Stanford University, Stanford, CA, USA}
\affiliation{California Institute of Technology, Pasadena, CA, USA}
\email{osmondw@stanford.edu}

\author{S.Q.~Dang}
\affiliation{SLAC National Accelerator Laboratory, Menlo Park, California, USA}
\affiliation{Kavli Institute for Particle Astrophysics and Cosmology, Stanford University, Stanford, CA, USA}
\affiliation{Fermi National Accelerator Laboratory, Batavia, IL, USA}

\author{E. Panner}
\affiliation{SLAC National Accelerator Laboratory, Menlo Park, California, USA}
\affiliation{Tufts University, Medford, Massachusetts, USA}

\author{J.~Anczarski}
\affiliation{SLAC National Accelerator Laboratory, Menlo Park, California, USA}
\affiliation{Kavli Institute for Particle Astrophysics and Cosmology, Stanford University, Stanford, CA, USA}

\author{T.~Aralis}
\affiliation{SLAC National Accelerator Laboratory, Menlo Park, California, USA}
\affiliation{Kavli Institute for Particle Astrophysics and Cosmology, Stanford University, Stanford, CA, USA}

\author{T. Dyson}
\affiliation{SLAC National Accelerator Laboratory, Menlo Park, California, USA}
\affiliation{Kavli Institute for Particle Astrophysics and Cosmology, Stanford University, Stanford, CA, USA}

\author{S.W.~Henderson}
\affiliation{SLAC National Accelerator Laboratory, Menlo Park, California, USA}
\affiliation{Kavli Institute for Particle Astrophysics and Cosmology, Stanford University, Stanford, CA, USA}

\author{H.~Magoon}
\affiliation{SLAC National Accelerator Laboratory, Menlo Park, California, USA}
\affiliation{Kavli Institute for Particle Astrophysics and Cosmology, Stanford University, Stanford, CA, USA}

\author{A.~Simchony}
\affiliation{SLAC National Accelerator Laboratory, Menlo Park, California, USA}
\affiliation{Kavli Institute for Particle Astrophysics and Cosmology, Stanford University, Stanford, CA, USA}

\author{C. Stone-Whitehead}
\affiliation{Colorado School of Mines, Golden, Colorado, USA}
\affiliation{Kavli Institute for Particle Astrophysics and Cosmology, Stanford University, Stanford, CA, USA}

\author{N. Tabassum}
\affiliation{SLAC National Accelerator Laboratory, Menlo Park, California, USA}
\affiliation{Kavli Institute for Particle Astrophysics and Cosmology, Stanford University, Stanford, CA, USA}

\author{B.A. Young}
\affiliation{Santa Clara University, Santa Clara, CA, USA}


\author{M. Munaf}
\affiliation{Fermi National Accelerator Laboratory, Batavia, IL, USA}
\affiliation{University of Chicago, Chicago, IL, USA}

\author{D. J. Temples}
\affiliation{Fermi National Accelerator Laboratory, Batavia, IL, USA}

\author{C. Cap}
\affiliation{California Institute of Technology, Pasadena, CA, USA}

\author{C.L. Chang}
\affiliation{Argonne National Laboratory, Lemont, IL, USA.}

\author{Y.-Y. Chang}
\affiliation{Lawrence Berkeley National Laboratory, Berkeley, CA, USA}

\author{M. Garcia-Sciveres}
\affiliation{Lawrence Berkeley National Laboratory, Berkeley, CA, USA}

\author{W. Ho}
\affiliation{Washington University in St. Louis, St. Louis, MO, USA}

\author{K. Li}
\affiliation{Lawrence Berkeley National Laboratory, Berkeley, CA, USA}

\author{X. Li}
\affiliation{Lawrence Berkeley National Laboratory, Berkeley, CA, USA}

\author{R. Linehan}
\affiliation{Fermi National Accelerator Laboratory, Batavia, IL, USA}

\author{M. Lisovenko}
\affiliation{Argonne National Laboratory, Lemont, IL, USA.}

\author{K. Okubo}
\affiliation{University of Wisconsin, Madison, WI, USA}

\author{Manabputra}
\affiliation{University of Wisconsin, Madison, WI, USA}

\author{K. Ramanathan}
\affiliation{Washington University in St. Louis, St. Louis, MO, USA}

\author{S. Ray}
\affiliation{California Institute of Technology, Pasadena, CA, USA}

\author{B. Sandoval}
\affiliation{California Institute of Technology, Pasadena, CA, USA}

\author{A. Suzuki}
\affiliation{Lawrence Berkeley National Laboratory, Berkeley, CA, USA}

\author{G. Wang}
\affiliation{Argonne National Laboratory, Lemont, IL, USA.}

\author{M. Williams}
\affiliation{Lawrence Berkeley National Laboratory, Berkeley, CA, USA}

\author{J.R. Xiong}
\affiliation{California Institute of Technology, Pasadena, CA, USA}

\author{V. Yefremenko}
\affiliation{Argonne National Laboratory, Lemont, IL, USA.}

\author{B. L. T. Plourde}
\affiliation{University of Wisconsin, Madison, WI, USA}

\author{S. R. Golwala}
\affiliation{California Institute of Technology, Pasadena, CA, USA}

\author{K. Stifter}
\affiliation{SLAC National Accelerator Laboratory, Menlo Park, California, USA}
\affiliation{Kavli Institute for Particle Astrophysics and Cosmology, Stanford University, Stanford, CA, USA}

\author{N.A. Kurinsky}
\affiliation{SLAC National Accelerator Laboratory, Menlo Park, California, USA}
\affiliation{Kavli Institute for Particle Astrophysics and Cosmology, Stanford University, Stanford, CA, USA}


\date{\today}

\begin{abstract}

We report on the position- and temperature-dependent response of a kinetic inductance phonon-mediated (KIPM) detector. To study position-dependence, we used a controllable mirror attached to a micro-electromechanical system (MEMS) to steer optical light with a 100 $\mathrm{\mu}$m spot size over the backside of the cryogenic detector. With both pulsed and continuous operation of the LED, we observe features of the detector housing and find that the signal is strongest when the LED is pointed directly at the KID. The ``off-KID'' response is measured to be 34\% of the ``on-KID" response and is largely independent of direction. We model phonon propagation and collection in the KIPM detector with G4CMP and assert that the isotropic angular dependence is unlikely to be due to discrete phonon absorption ``sinks,"  such as the microwave feedline or the device mounting points, but can be explained instead by an isotropic energy loss in the crystal. We furthermore utilize the temperature flexibility of the MEMS and KIPM detector system to investigate temperature-dependent responsivity, both at the sensor level (i.e., frequency shifts in the KID) and at the detector level (i.e., position dependence). We fit the KID's response to a two-component heat capacity and present the model as an alternative way to understand the KID's response to absorbed energy. We show on- and off-KID data at a range of temperatures and show that the phonon dynamics are not temperature dependent. We comment on the resulting implication for KID dynamics and design improvements.

\end{abstract}

\maketitle


\section{\label{sec:intro}Introduction}


Athermal phonon sensors are a crucial detector technology for low-mass dark matter searches \cite{essig2023snowmass} and low-energy neutrino science \cite{nucleus22,Bowen_2020,Billard_2018}. They may also serve as useful co-sensors for studying radiation effects on superconducting circuits \cite{yang2026}. Whereas electronic excitations are generally in the eV range or higher, athermal phonons with detector-length mean free paths are around 4\,meV ($\sim$ 1\,THz) and lower. Low-threshold particle detection relies on high-efficiency collection of this athermal phonon population, generally with meV-bandgap superconducting sensors. Although minimum threshold energies for superconductors are limited by the Cooper pair binding energy ($\sim$meV), actual sensor resolution may be limited by quasiparticle shot noise, thermal Johnson noise, or noise induced by two-level systems. Sub-eV energy resolution has been achieved with transition-edge sensors (TESs) coupled to phonon-absorbing fins that allow for near 100\% phonon collection efficiency \cite{TESSERACT_2025,newHVeV}. 


Kinetic inductance detectors (KIDs) are an alternative phonon-sensing architecture that may provide long-term gains in sensitivity as well as multiplexability. Monolithic KID designs that combine the inductive sensor element with the phonon-absorbing element have been explored for their design and fabrication simplicity. The core competing forces in designing a monolithic KID-based phonon-mediated (KIPM) detector are the following: (1) the sensor intrinsic resolution improves with smaller volumes, and (2) the phonon collection efficiency improves with greater sensor volume. Recently studied KIPM detectors have demonstrated few-eV intrinsic resolution but have been impaired by poor collection efficiency, ranging from 1\% to 10\% (see e.g. Refs~\cite{Moore:2012gya,temples2025}). The main hypothesis for the poor collection efficiency has been that the non-signal-contributing components over the surface of the detector (e.g., the CPW transmission line) act as phonon sinks. See Refs~\cite{Kurinsky,Temples_2024,temples2025} and references therein for further discussion about the effects of parasitic ``dead metal" on device performance.


In this paper, we demonstrate that the current monolithic KIPM detector design is limited by unrecoverable phonon losses in the crystal, not dead metal losses. We reach this conclusion by using an optical scanner based on a micro-electromechanical system (MEMS) mirror that allows us to distinguish between energy depositions that are directly behind the sensor (``on-KID'') and depositions that are far from the sensor (``off-KID''). We find that the phonon collection efficiency is isotropic and strongly position dependent, which points to a spatially homogeneous loss mechanism. 





Using the ratio of the on- and off-KID response as a proxy for the phonon loss, we also study how this loss depends on temperature. We interpret the temperature dependence of the KIPM response using both an excess-quasiparticle model and a heat capacity model.
The observation of rising pulse amplitudes with temperature, first presented in \cite{temples2025}, was a core motivator for this study.

%

We hereby present the design and analytical methods of our experiment, composed of a MEMS calibration system and a KID-based phonon-mediated detector in a temperature-controllable environment. In Section II, we describe the experimental setup, focusing on the design of the KIPM detector and the operation of the MEMS scanner. In Section III, we focus on the position-dependent response of the KIPM detector, both to continuous streams and pulsed depositions of optical light. In Section IV, we turn to the temperature-dependent on- and off-KID data, investigate the rising pulse amplitudes with temperature. In section V, we build a heat-capacity-based model for the response of a KID to absorbed energy and discuss phonon efficiency over the full temperature range. In Section VI, we present a forward-looking discussion for the implications of our results on the design of future KIPM detectors.




\section{Experimental Setup}

This section introduces components of the experimental setup. First, we describe the KIPM designed and fabricated at Caltech/JPL for phonon efficiency measurements. Second, we describe the SLAC-built optical calibration unit (referred to as the MEMS scanner), which uses an AWG-driven LED driver for pulse tuning and a steering controller for spatial positioning. Third, we describe the KID warm-readout and optical light source configurations and the datasets used in this analysis.

\subsection{KIPM Design}\label{sec:kipm-design}
The KIPM detector under investigation is shown in Fig.~\ref{fig:VNAscan_MEMS}(i) and consists of an aluminum lumped-element resonator patterned from a $30$\,nm-thick film on a $1$\,mm thick double-sided polished high-resistivity silicon wafer, later diced into a $2.2$\,cm by $2.2$\,cm die. The resonator features an $80\,\mu\text{m}$-wide by $5$\,mm-long inductive meander trace and nine interdigitated capacitive fingers, following previously reported KIPM design principles \cite{Chang_2018, Chang_2023,Wen2025}.

The resonator was designed to operate at $5.4$\,GHz in order to  align with the commonly found 4 to 8 GHz parametric amplifier range. This was a higher resonance frequency relative to past designs, and was chosen so the KID could be integrated with a qubit array already operating in that band. This frequency was reached by shortening the inductive meander in the vertical dimension, which halved the sensing volume relative to the earlier KIPM study by \citet{Temples_2024}. The final sensing volume ($V_s$) of the detector in this study was $11,900 \text{ }\mu\text{m}^3$. 

The KIPM detector was inductively coupled to a straight coplanar waveguide patterned from a $70$\,nm-thick niobium film. Reducing the niobium thickness to under $70$\,nm  could have resulted in feed-line impedance mismatches due to elevated kinetic inductance. The waveguide dimensions and path were chosen to minimize the non-sensing ``dead metal'' presence on the chip, which we hypothesized to be the limiting cause of the percent-level phonon collection efficiency found in Ref\;\cite {Temples_2024}. The volume of niobium was decreased from 1.2$\times 10^6$\,$\mathrm{\mu}$m$^3$ to 1.8$\times 10^5$\,$\mathrm{\mu}$m$^3$. The detector was mounted into a gold-plated copper housing using the same packaging as outlined in the previous study. The overall wiring diagram for the readout is shown in Appendix\;\ref{appendix:setup}. 

\subsection{MEMS Scanner}
\begin{figure*}[t]
    \centering
    \includegraphics[width=0.4\linewidth]{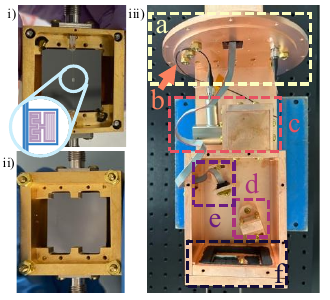}
    \includegraphics[width=0.55\linewidth]{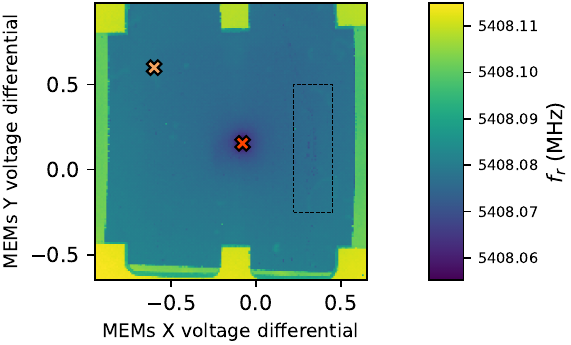}
    \caption{\textit{Left: }The front side of the KIPM detector (i) features an aluminum lumped element LC resonator and niobium coplanar waveguide with a minimal ground-plane (Top Left insert i). The back side of the KIPM detector (ii) is placed facing the beam from the MEMS calibration system (iii). This setup is mounted from the Cu lid of a cryoperm can (a), and transmits the light from a fiber optic cable (b), through a focusing unit (c) to a stationary mirror (d) and onto a scanning MEMS mirror (e). This MEMS mirror directs the focused beam across the backside of the mounted detector, after passing through a 2~mm thick KG3 IR filter(f). \textit{Right: }The resonant frequency response of the detector to MEMS operation. The X's mark the locations for which the on-\ and off-KID analysis of Section\;\ref{sec:calibration} was performed. The black box outlines a streak of residue visible in this scan that is also visible in the backside image of the detector.}
    
    \label{fig:VNAscan_MEMS}
\end{figure*}

To measure the position-dependent response of the KIPM to both continuous beams and pulses of optical light, we used a MEMS-based optical scanner described in detail in Refs~\cite{tabassum2025broadbandopticalmodulationcontrol,Stifter_2025}. The scanning setup and its components are shown in Fig.\;\ref{fig:VNAscan_MEMS}(iii). This assembly operates at 10\,mK within a light-tight cryoperm can, and allows for fiber-coupled light sources to be steered across a cryogenic payload.

As demonstrated in Ref~\cite{tabassum2025broadbandopticalmodulationcontrol}, the broad-band focuser achieves $\sim100\,\mathrm{\mu}m$ optical spot size across a maximum field of view of a 4\,cm $\times$ 4\,cm square and is compatible with a room temperature optical source with selectable wavelengths between 365 and 970\,nm.  For the measurements reported here, a 625\,nm LED mounted at room temperature was used, as this wavelength is optimally transmitted by our optical fiber and filter stack. 
The back-side of the KIPM detector is shown in Fig.\,\ref{fig:VNAscan_MEMS}(ii) and was mounted facing the MEMS mirror field of view. The boundaries of the KIPM housing and chip edges in MEMS voltage space were first measured on the bench using a CCD camera, to establish the operating range of the optical calibration unit. 


\subsection{Warm Readout and Optical Source}
The position-dependent response of the KIPM detector to LED pulses was measured by recording complex IQ transmission at the KID's unloaded resonance frequency as a function of time: $S_{21}(t;f_r)$. The measurements utilized the SLAC Microresonator Radio Frequency (SMuRF) Electronics system, operated in a mode that directly down converted the  SMuRF ADC output to return orthogonal IQ transmission \cite{Yu2023}. 
Although this operating mode was intended for debugging, it enabled IQ data acquisition at sample rates up to 2.4\,MHz for a fixed number of samples.

LED pulses were generated using a Thorlabs LED driver DC2200 externally modulated by a Moku:Pro waveform generator. 
Pulse amplitude, phase, and repetition frequency were controlled using the SplenDAQ software package \cite{Watkins2024}. Full details of the warm readout and optical configuration are provided in Appendix\;\ref{appendix:setup}.

\subsection{Datasets}
The data presented in this paper are from two separate cooldowns of an Oxford ProteoxMX in the SLAC Millikelvin Facility. They will be referred to as Run\;1 and Run\;2. The data from Run\;1 were taken from January 31, 2025 to April 9, 2025 and is primarily featured in Section \ref{sec:position_dependence}. The data from Run\;2 were taken from May 16, 2025 to June 28, 2025 and are presented in Sections \ref{sec:position_dependence} and \ref{sec:calibration}. A 197\,kHz shift in the resonant frequency and a 20,000 shift in $Q_i$ was observed between Run\;1 and Run\;2, but all position- and temperature-dependence we report were consistent across both runs.

\begin{figure*}
    \centering
    \includegraphics[width=\linewidth]{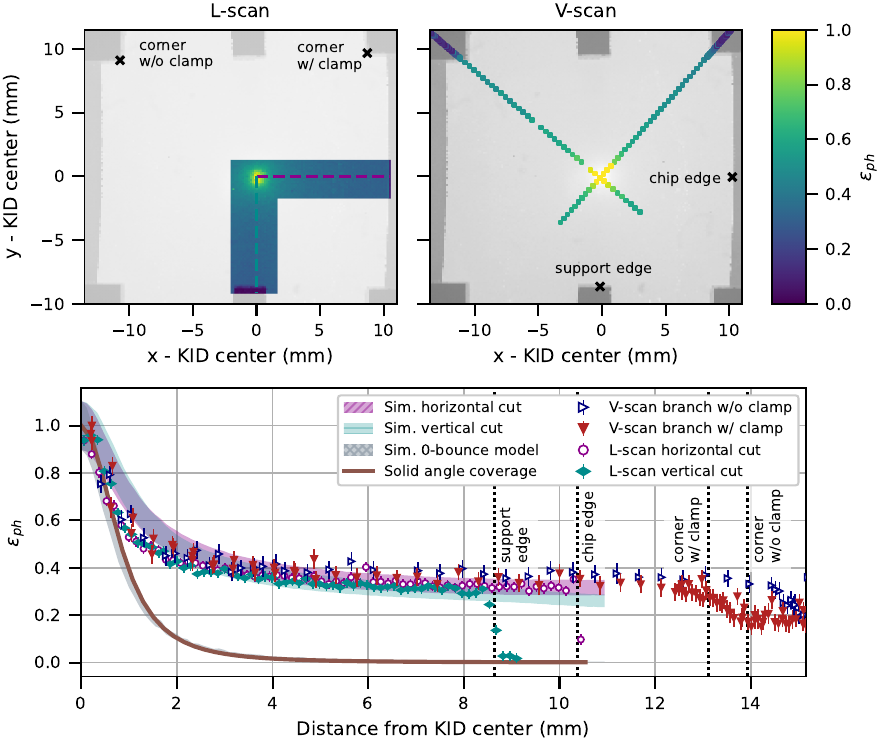} 
    \caption{\textit{Top Left: }Measured peak-normalized pulse response from the L-scan. The continuous scan is underlaid and shown in grayscale. The cyan and magenta lines indicate the contours along which the analysis for the bottom panel is performed. Relevant corner locations for the V-scan are annotated in black. \textit{Top Right: } Measured peak-normalized pulse response for the V-scan. Relevant features for the L-scan are annotated in black.  \textit{Bottom:} Peak-normalized pulse heights and simulated relative phonon collection efficiency $\epsilon_\textbf{ph}$ across the horizontal and vertical branches of the L-scan. Uncertainties in data are propagated from the standard deviation of resonance transmission. Relative energy collected from phonons that do not undergo reflections before being absorbed in the sensor is plotted against an analytical solid angle calculation. The dashed lines show the distance from KID center of each of the annotated features above.  The ``off-KID" location indicated in Fig.~\ref{fig:VNAscan_MEMS}(right) is approximately the same distance from the KID center as the annotated chip edge.  }
    \label{fig:pulsedMeasurements}
\end{figure*}



\section{Position-dependent detector response}
\label{sec:position_dependence}

\subsection{Constant Illumination Study}\label{res:continuous}
To understand the KIPM response to the steered beam position, we performed a position sweep over several days.  The fridge temperature was stable at $16 .8$ mK $\pm 0.3$ mK throughout the entirety of data collection.

At the start of the position sweep, the beam was steered to randomized positions over a grid with $\sim150\,\mu m$ spacing in both x and y dimensions (0.1 spacing in MEMs voltage space, see Appendix~\ref{appendix:setup}). A total of 26,404 positions were sampled. Randomizing the scan limited any local heating of the chip caused by consecutive measurements in the same location.
At each position, we waited 1 second before measuring the complex transmission through the resonator while sweeping the readout frequency.

The data from this VNA-style measurement at each position  was fit to the standard transmission model for a resonator\cite{Khalil2012}
\begin{equation}\label{eqn:S21}
S_{21}(f)
=
a e^{-2\pi i \tau (f-f_r)}
\left(
1
-
\frac{Q_r}{\hat{Q}_c}
\frac{1}{1 + 2iQ_r\frac{f-f_r}{f_r}}
\right),
\end{equation} 
to extract the resonance parameters. Here, $\hat{Q}_c$ is the complex coupling quality factor, $Q_r$ is the total  KID  quality factor, $f_r$ is the KID resonant frequency, and $f$ is the generator frequency. $Q_c$ can be extracted as $\operatorname{Re}(1/\hat{Q}_c)$ The frequency dependent cable delay time ($\tau$) is unrelated to KID operation, and $a$ is the off-resonance transmission. A map of the resonant frequency vs. incident LED position is shown in Fig.~\ref{fig:VNAscan_MEMS} Right. 

This constant illumination study has multiple advantages for initial characterization compared to the pulsed readout scheme described later in this section. In particular, the measurement is possible with a standard VNA and commercially available Mirrorcle\textsuperscript{TM} electronics, and thus represents a fairly device-agnostic method of MEMS-based detector calibration. Careful study of Fig.~\ref{fig:VNAscan_MEMS} shows a number of device features visually, including surface residue to the back-side of the device, housing reflections and edge imperfections near the outside of the device, and full reflection loss when the laser is incident on the device holder. This full scan can be also be used as a way of converting relative MEMS voltage units into physical position space, which is described in more detail in Appendix~\ref{appendix:setup}.



\subsection{Pulsed Illumination Study}\label{res:pulsed}

The position dependent phonon collection efficiency of the  KIPM detector was studied by directing a pulsed beam of 625 nm photons onto precise locations on the back surface of the detector. We performed two scans, which we call the ``L-scan" and the ``V-scan" based on their geometry. These measurements were all taken with the cryostat set to the lowest stable base temperature, between 10 and 11\,mK.

\begin{figure*}
    \centering
    \includegraphics[width=\linewidth]{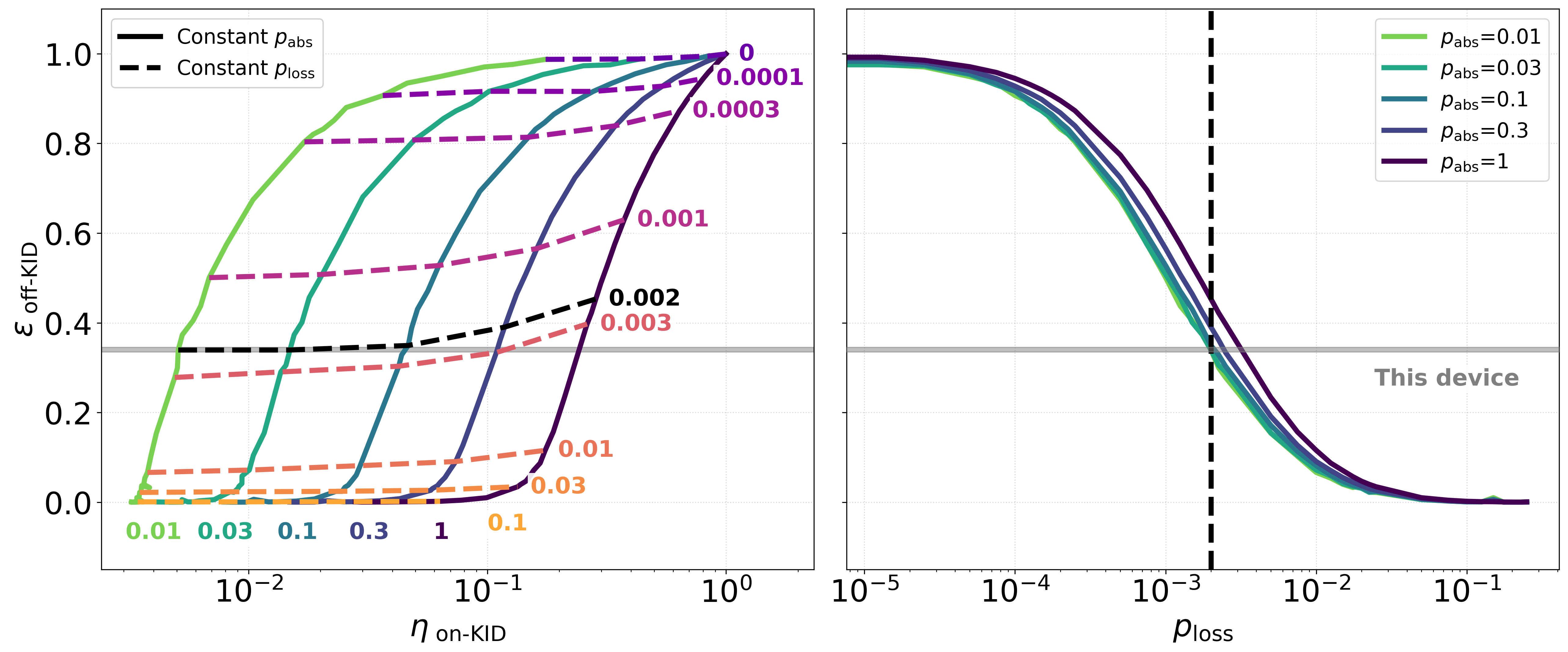}
    \caption{\textit{Left}: Simulations of absolute phonon collection efficiency with $e^-/h^+$ pairs generated below the inductor (``on-KID'') and phonon collection efficiency with $e^-/h^+$ pairs generated away from the inductor (``off-KID'') relative to the on-KID absolute efficiency. Solid contours show the phonon absorption probability at the substrate-sensor interface, $p_{\text{abs}}$ (values 0.01--1). Dashed contours show the probability of phonon loss at the top and bottom polished surfaces of the substrate, $p_\mathrm{loss}$ (values 0--0.1).
  \textit{Right}: Relative phonon collection efficiency for LED induced $e^-/h^+$ pairs generated off-KID as a function of $p_\mathrm{loss}$ for different $p_{\text{abs}}$. An estimate for $\epsilon_\mathrm{off-KID}$ in our device is calculated from the mean peak height data between 5 mm and 8 mm from the KID center in the horizontal branch of the L-scan presented in Fig.\ref{fig:pulsedMeasurements}.
    }
    \label{fig:positionDependenceVsCollectionEfficiency}
\end{figure*}


As shown in Fig.~\ref{fig:pulsedMeasurements}, the L-scan consisted of two rectangular regions: one extended horizontally from the resonator position across bare silicon to the edge of the chip,  the other extended vertically down, following  the niobium coplanar waveguide to the edge of the copper support. The V-scan started at the center of the resonator and extended diagonally towards the corner supports. Known attributes in the detector mounting scheme enabled direct identification of special features observed in the scans. For example, as seen in Fig.~\ref{fig:VNAscan_MEMS}, one corner was compressed by a clamp from the front side, while the other was left un-clamped. This asymmetry was used to better understand origins of phonon energy loss in our detector. 


Both scans followed the same measurement procedure. At each beam position, a VNA-style measurement was recorded with the LED off to determine the resonator's central frequency $f_r$. Three single-tone measurements  were then taken slightly de-tuned  from the resonant frequency to correct for any phase drift. Our generator frequency  was then tuned to be on resonance $(f=f_r)$,  and the LED  was set to pulse at maximum voltage amplitude of 2.5 V for 12\,$\mu\text{s}$ pulse widths at a 20\,Hz repetition rate. In this configuration, we recorded 21 second time streams at a 200 kHz sample rate. Finally, the LED was turned off before moving to the next position through a randomized scan order to minimize local heating from consecutive measurements at nearby locations. The time stream data was converted to individual pulses, from which the average pulse heights were extracted as described in Appendix~\ref{appendix:setup}.

The analysis was performed in units of $\delta f_r /f_r$, which can be extracted from a small $S_{21}$ pulse trajectory in the IQ plane with the following approximation:
\begin{equation}
\delta S_{21} = \frac{Q_r^2}{Q_c}\left(\delta\frac{1}{Q_i}-2i\frac{\delta f_r}{f_r}\right),\label{eq:dS21}
\end{equation}
where $Q_i$ is the internal quality factor of the KID. Eq.\;\ref{eq:dS21} can be derived by linearizing Eq.\;\ref{eqn:S21} about the resonance transmission and by assuming the data are converted to an idealized $S_{21}$ where the cable delay $\tau$, off-resonance transmission $a$, and the complex nature of $\hat{Q}_c$ are removed. A core assumption in this section is that $\delta f_r/f_r$ is proportional to the quasiparticle deviation $\delta n_\mathrm{qp}$, which is proportional to the absorbed energy $\delta E$. We break down this assumption in detail in Section\;\ref{sec:calibration}.

In Fig.~\ref{fig:pulsedMeasurements} we present the peak-normalized, position dependent component of the phonon collection efficiency 
\begin{equation}\label{eq:eps_ph}
    \epsilon_\mathrm{ph}(R) = \frac{\frac{\delta f}{f}(R)}{\frac{\delta f}{f} (R=0)}
\end{equation}
computed from both the L and V scans, with $R$ the radial distance from the KID. This is the relative phonon collection efficiency of the KID with the peak phonon collection efficiency, $\eta_\text{on-KID}$, removed to show only the part of efficiency determined by distance from the KID. We discuss this variable in the context of simplified phonon modeling in more detail in Appendix~\ref{appendix:analytic-response-modeling}.

The most notable feature in Fig.\,\ref{fig:pulsedMeasurements} is the hot-spot below the KID, which was also present in the constant illumination study. We had expected to see evidence of the feedline in the L scan (which runs along the cyan line in Fig.~\ref{fig:pulsedMeasurements} (upper left)) and of excess loss to the copper clamps (corners of the V scan), but instead see a remarkably isotropic response that decays only with distance from the center of the detector. We estimate $\epsilon_\mathrm{off-KID} \approx \epsilon_\mathrm{ph}(5~$mm$~<R<8~$mm$)$ for our device to be $0.340\pm0.005$ from taking the mean of relative peak height data between 5\,mm and 8\,mm from the KID in the horizontal branch of the L-scan presented in Fig.~\ref{fig:pulsedMeasurements}. The precise value of this number can be attributed to a loss mechanism that has not been modeled in past work on phonon sensing (see e.g. Refs\;\cite{newHVeV,TESSERACT_2025}), as discussed in the next section. 


\subsection{Comparison to Simulation}\label{sec:comparisonToSimluation}

To investigate the effect of potential loss mechanisms on our detector's position-dependent phonon response, we constructed a Monte-Carlo model of phonon collection using the \Geant Condensed Matter Physics (G4CMP) package \cite{KELSEY2023168473}. We used the G4CMP framework to simulate (1) the production and transport of charge carriers and phonons in the substrate bulk and (2) the absorption and reflection of phonons at substrate boundaries. We note that the default assumptions in G4CMP are that reflection at interfaces is lossless, and thus our results represent a significant break from this assumption which we justify further here and discuss in more detail in Appendix~\ref{appendix:simulation-details}.

We compare the relative peak height data from the pulsed measurements to simulated relative phonon collection efficiency across vertical and horizontal spatial extents in Fig.~\ref{fig:pulsedMeasurements}. A qualitative convergence between data and simulation at $R\gtrsim2$~mm from the KID center occurs when we introduce a position-independent source of phonon loss to the model. The weak divergence in the experimental relative peak height data between the vertical branch along the feedline and the horizontal branch away from the feedline suggests that loss to the feedline's dead metal is subdominant but non-zero. In addition, if we only calculate $\epsilon_\mathrm{ph}(r)$ from phonons which do not undergo reflection before being absorbed, we obtain a curve in good agreement with the relative solid angle subtended by the Al inductor of the KID resonator. 

As shown in Fig.\;\ref{fig:pulsedMeasurements}, our `best-guess' model for $\epsilon_\text{ph}$ includes a 10.3\% phonon absorption probability in the Al film, a 0.2\% phonon loss probability at the top and bottom polished faces of the crystal substrate and a 10\% phonon loss probability at the feedline, with no phonon-reemission from the feedline considered. The qualitative agreement between the simulated and experimental values of $\epsilon_\mathrm{ph}(r)$ is not unique to this set of loss mechanisms or probabilities. Calculations for the phonon absorption probability are presented in Appendix \ref{appendix:analytic-response-modeling}, and additional studies of phonon loss can be found in Appendix \ref{appendix:simulation-details}. Our simulations show that, in any scenario, phonon energy losses at the mounting clamps and patterned films are not sufficient to describe the spatially uniform loss seen in the data.

Finally, we can extend our model to explore the parameter space of position-dependent phonon collection efficiency for a low-coverage single-sensor KIPM detector, assuming a position-independent loss mechanism. In Fig.~\ref{fig:positionDependenceVsCollectionEfficiency}, we present the peak phonon collection efficiency ($\eta_\mathrm{on-KID}$) and the spatially-offset relative phonon collection efficiency ($\epsilon_\mathrm{off-KID}$) for a range of simulation models, parametrized by the phonon absorption probability at the sensor, $p_{abs}$, and the spatially uniform phonon loss probability, $p_\mathrm{loss}$. Here $\epsilon_\mathrm{off-KID}$ is a proxy for position dependence, with a highly position-dependent phonon response corresponding to a low $\epsilon_\mathrm{off-KID}$. 



 We note that at a given $p_\mathrm{loss}$, raising $p_\mathrm{abs}$ primarily affects the absolute phonon collection efficiency and not the position dependence. In Fig.~\ref{fig:positionDependenceVsCollectionEfficiency} (right), we extract the profile of  $\epsilon_\mathrm{off-KID}$ as a function of $p_\mathrm{loss}$ and find that the position dependence is an effective indicator of loss strength across a wide range of absorption probabilities. Given $\epsilon_\mathrm{off-KID} = 0.340\pm0.005$ from the previous section, we infer a loss probability of approximately 0.2\%, indicated by the black, dashed line in Fig.~\ref{fig:pulsedMeasurements}.  With the assumption that $p_\mathrm{loss}$ is a substrate property,  projections with varying $p_\mathrm{abs}$ provides an intuitive bound on the effects of raising the sensor film thickness.

\begin{figure*}[t]
  \centering

  \begin{minipage}[c]{0.66\textwidth}
    \centering
    \includegraphics[width=\linewidth]{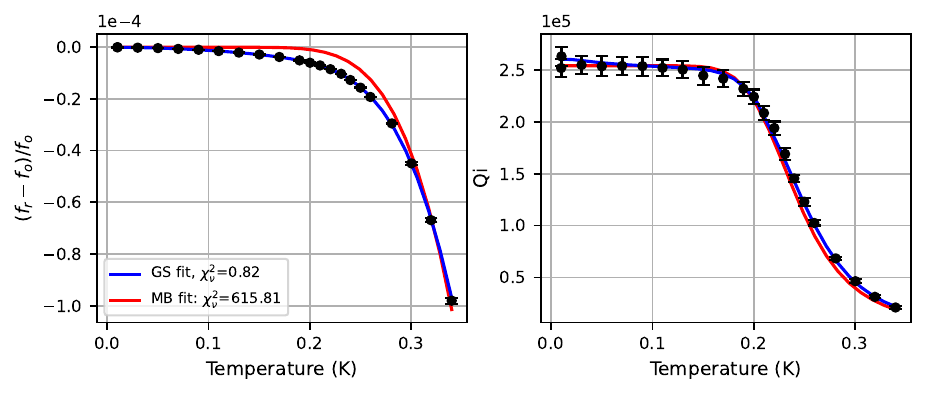}

  \end{minipage}\hfill
  \begin{minipage}[c]{0.32\textwidth}
    \centering
    \scriptsize
    \setlength{\tabcolsep}{2pt}
    \resizebox{\linewidth}{!}{%
      \begin{tabular}{|c|c|c|}
        \hline
        \textbf{Parameter} & \textbf{GS} & \textbf{MB} \\
        \hline
        $f_{0}$ (GHz) & 5.4079 & 5.4079 \\
        \hline
        $Q_{i,0}/10^5$ & $2.61 \pm 0.02$ & $2.541 \pm 0.001$ \\
        \hline
        $\alpha$ & $0.051 \pm 0.003$ & $0.053 \pm 0.001$ \\
        \hline
        $\Delta_0$ ($\mathrm{\mu}$eV) & $196.1 \pm 0.7$ & $188.5 \pm 0.5$ \\
        \hline
        $\Gamma$ ($\mathrm{\mu}$eV) & $2.17 \pm 0.06$ & --- \\
        \hline
        $\chi_\nu^2$ & $0.82$ & $615.1$ \\
        \hline
      \end{tabular}%
    }

  \end{minipage}

  \caption{\textit{Left and Middle}: Resonator frequency and quality factor as a function
  of temperature, measured in Run 2. Fits to an idealized
  Mattis-Bardeen (MB) model and a gap-smeared (GS) model are shown.
  Frequencies in the left plot panel are relative to the
  zero-temperature best-fit resonant frequency extracted from the
  models. The $\chi_\nu^2$ values of the joint $f_r$ and $Q_i$ fit
  are given in the legend. \textit{Right}: Best-fit model parameters.}
  \label{fig:VNAscan}
\end{figure*}

\section{Temperature-dependent detector response}\label{sec:calibration}
\subsection{Mattis-Bardeen fits with gap-smearing} \label{sec:MB}



Standard characterization of the frequency response of a KID includes resonator fits across a wide range of bath temperatures. Fits to the Mattis-Bardeen (MB) equations are then used to extract film parameters such as superconducting gap $\Delta_0$, kinetic inductance fraction $\alpha$, and the zero-temperature values of the resonance frequency $f_0$ and internal quality factor $Q_{i,0}$  \cite{GaoThesis,Gao2008}. During Run 2, we collected $S_{21}(f)$ data at temperatures ranging from 10\,mK to 340\,mK and fit the data to Eq \ref{eqn:S21} to extract resonator parameters.  The cryostat was allowed to equilibrate for 40 minutes before beginning each $S_{21}(f)$ measurement. 

The temperature dependence of the fractional resonant frequency shift $(f_r-f_0)/f_0$ and internal quality factor is shown in Fig.~\ref{fig:VNAscan}. The error bars were determined by using typical $S_{21}$ scatter measured by the VNA and by computing the range of parameters that would cause $\Delta \chi^2 =1$. We fit the frequency and dissipation data from Fig.~\ref{fig:VNAscan} to the integrated MB model as well as a numerical evaluation of the model with gap-smearing \cite{Nam1967}, with best fit parameters listed in Fig.~\ref{fig:VNAscan} Right. Both models are described in Appendix\;\ref{app:gs}, as well as a comparison to the numerical evaluation of the MB model without gap smearing. We find that the MB fits are unable to account for the frequency shifts above 100\,mK and so rely on the gap-smeared model for the rest of the analysis. 



We attribute the need for gap smearing in our analysis to disorder in our thin Al films and imperfect magnetic field shielding. At $30$\,nm, the films are too thin for pure BCS theory and we postulate there exists an appreciable density-of-states population below the nominal gap energy $2\Delta$. The need for gap-smearing in an Al film is unusual and may reflect anomalous film quality, as in Refs ~\cite{li2025, Zobrist2019Hafnium}. It is also noteworthy that there is an uptick at low temperatures in the gap-smeared $Q_i$ model as a result of the suppressed thermal occupation of the sub-gap states: $T\lesssim (\Gamma/\Delta) T_c$, where $\Gamma$ is the Dynes parameter. In the following sub-section, we use this gap-smearing model to convert between $f_r$ and $T$ and thus investigate the thermal response of the KID. 






\subsection{Temperature-dependent pulse amplitudes}\label{res:tempsweeps}

Temperature-dependent pulse-shape variation in KIPM detectors was first reported in Refs~\cite{Temples_2024,temples2025}. The data from Run 2 are the same data reported in Ref~\cite{temples2025}, with an update to the analysis pipeline. Here we present the data in more detail and also provide a model that may explain the variation in pulse amplitude.

At each temperature, we collected pulse data in a manner similar to what was described in Section \ref{res:pulsed}. We first collected data with the beam pointed directly behind the resonator (position indicated by  dark-orange ``X" in Fig.~\ref{fig:VNAscan_MEMS} Right), then steered the beam to an off-KID location (position indicated by the light-orange ``X" in Fig.~\ref{fig:VNAscan_MEMS} Right) and repeated the measurement. LED pulses were generated from a $2.5$\, V square pulse. The data was taken during one cool down, but in two separate sessions separated by two weeks, causing potential differences in thermalization. The data taken at 200\,mK, 220\,mK, etc. are visibly larger in peak height than the data at 190\,mK, 210\,mK, etc., but we treat all data equally in this analysis. 
\begin{figure*}
    \centering
    \includegraphics{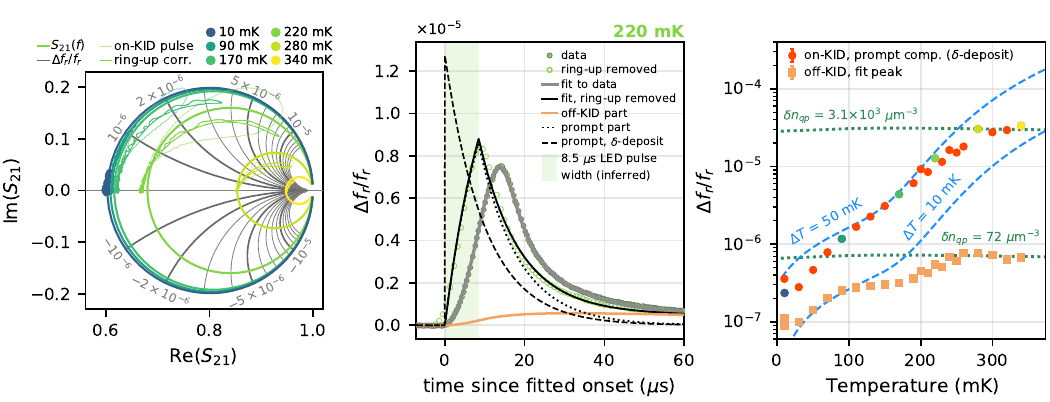}
\caption{\textit{Left}: On-KID pulse shapes at five different temperatures. The faded curves are the raw data rotated into this idealized $S_{21}$ plane, and the solid color curves are ring-up corrected. The thicker colored circles are $S_{21}(f)$ curves from Eq~\ref{eqn:S21} with the pre-factor divided out, and the complex coupling corrected for using the $Q_c$, $Q_r$, and $f_r$ values obtained at each temperature. The gray lines indicate contours of constant $\Delta f_r/f_r$ over a range of temperatures. The colored circles and the grey lines form a Smith chart. \textit{Middle}: Conversion process of pulse trajectory into a peak amplitude. $\Delta f_r/f_r$ has been ring-up corrected and fit to a model given by prompt + off-KID pulse templates, as described in Appendix~\ref{app:pulse_fitting}. The amplitude of the prompt component to a $\delta$-function is then derived from the inferred LED pulse width. \textit{Right}: Peak $\Delta f_r/f_r$ of observed pulses at temperatures spanning 10\,mK to 340\,mK, both on-KID (circles) and off-KID (squares). The temperatures shown in the left panel are highlighted in their respective colors. The blue curves are computed from the $\Delta f_r /f_r$ that results from  a $\Delta T$ of 50\,mK and 10\,mK at a particular bath temperature as given by the gap-smeared MB fit.  
}

    \label{fig:pulseShapeModeling}
\end{figure*}

We observed a wide variation in $S_{21}$ trajectories over a temperature range from 10\,mK to 340\,mK; see the faded curves in Fig.~\ref{fig:pulseShapeModeling} Left. The trajectories are beyond the regime where $\delta f_r/f_r$ can be extracted from linear perturbations in $\delta S_{21}$. In the steady state, an $S_{21}$ trajectory can be used to extract $\Delta f_r/f_r$ in the following way\footnote{We shift from $\delta$ to $\Delta$ to indicate that pulse shapes in $S_{21}$ are no longer small and linear with energy. That said, it is still true that $\Delta f_r\ll f_r, f_0$.} : 
\begin{equation}
    S_{21}(t; f_g) =1-\frac{Q_r}{Q_c}\frac{1}{1+2iQ_r\frac{\Delta f_r}{f_r}},\label{eq:nonlineardS21}
\end{equation}
where $\Delta f_r/f_r\equiv (f_g-f_r(t))/f_r(t)$ and both $f_r$ and $Q_r$ are decreasing with time over the pulse. $f_g$ is the generator frequency, which is set at the resonant frequency at a particular temperature. We plot lines of constant $\Delta f_r/f_r$ in Fig.~\ref{fig:pulseShapeModeling}~Left with the gray curves. These gray curves and the circles of constant $Q_i$ form a Smith chart, where $1/Q_i$ and $\Delta f_r/f_r$ correspond to the real and imaginary components of the resonator's impedance.  $\Delta f_r/f_r$ as well as $\Delta(1/Q_i)$ can be extracted directly from the measured $S_{21}(t)$ via an inversion of Eq. \ref{eq:nonlineardS21}. The derivation of the inversion is given in Ref~\cite{Wen2025} and can also be derived from Ref~\cite{Zobrist2021}. 

The raw pulses are shown as faded lines and move clockwise. We note that the large initial excursions outside the $S_{21}(f)$ curves have to do with the resonator response time. The ring-up time can be corrected for from the steady-state formula for $\Delta f_r/f_r$ to extract the instantaneous $\Delta f_r/f_r$ at any given point in time. Corrections for ring-up time in $\Delta f_r/f_r$ and $Q$ are described in detail in Appendix\;\ref{app:pulse_fitting}. We apply this correction and observe in Fig. \ref{fig:pulseShapeModeling} Left that the trajectory is unified along a particular arc for both the rise and fall. In the timestream shown in Fig. \ref{fig:pulseShapeModeling} Middle, we see that the time dynamics of the pulse are changed as well. We note that this correction only has a meaningful impact on pulse height modeling in the high-temperature data, and thus does not change the conclusions of the base-temperature data discussed in the previous sections.

After compensating for these measurement effects, we fit pulses to a two-component model. The off-KID data is first fit to a single rise time, and anywhere from one to three fall times. The on-KID data is then fit to the sum of a ``prompt'' component and the ``off-KID'' component. Versions of this fitting procedure have been studied in Refs~\cite{Florang2026,Moore2012,Temples_2024}. The off-KID pulse heights are estimated from the peaks of the fits to the pulses. The on-KID amplitude is derived from the prompt component that is corrected to incorporate the effect of the finite LED pulse width and the energy that leaks out of the sensor during that time. A global fit over all temperatures is performed to the width of the LED pulse and produces an inferred width of 8.5\,$\mu$s. This width is smaller than the $12\,\mu$s that was set at the room temperature electronics. We attribute this to $\mu$s-level rise times in the LED driver. Detailed discussion of these fits and the temperature trends for both amplitude and time constants can be found in Appendix~\ref{app:pulse_fitting}.

Following this procedure, we thus produce an estimate for the response of the KID to a sudden phonon impulse. We claim that this quantity describes the responsivity of the KID to the initial burst of phonons that have undergone minimal downconversion and scattering off surfaces, and is thus an optimal estimator for the `zero-loss' phonon amplitude. In contrast, the off-KID pulse amplitude is the mean response of the sensor to phonons which have undergone an average number of down-conversion steps. This two-component model allows us to measure the relative efficiency of ballistic vs diffuse phonon collection as a function of temperature as shown in Fig.\;\ref{fig:pulseShapeModeling} Right.

\begin{figure*}
\includegraphics
{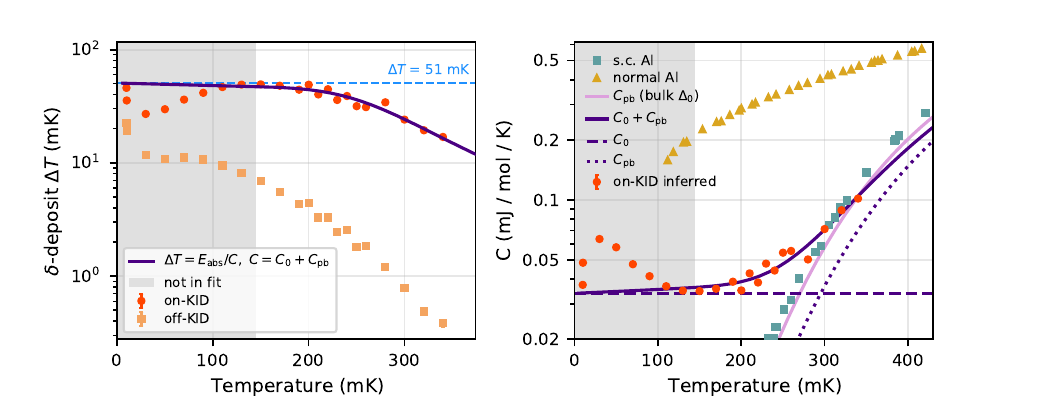}
\caption{ \textit{Left}: $\Delta T$ as inferred from $\Delta f_r /f_r$ for both on-KID and off-KID, see Appendix \ref{app:dT} for this conversion. A model for the heat capacity derived from a MB-like integral is discussed in the text. \text{Right}: Comparison of the heat capacity as measured in this work with the values in \cite{Al_C}. The full fit to the heat capacity model described in \ref{sec:heat_capacity}, as well as its components are shown as solid, dashed, and dotted dark purple lines. This fit was calculated assuming the values in Fig. \ref{fig:VNAscan} Right, while the pink line was calculated assuming a superconducting gap of $176.7$\,$\mathrm{\mu}$eV as measured in \cite{Al_C}. }
\label{fig:C_comparison}
\end{figure*}

Both on- and off-KID data in Fig.~\ref{fig:pulseShapeModeling}~Right show a precipitous rise in peak height at high temperatures: a factor of 100 in the on-KID data, and a factor of 10 in the off-KID data. With the green dotted lines, we plot the expected $\Delta f_r/f_r$ response versus temperature for $\delta n_\text{qp}$ values derived from the average of the four highest temperature data points for both on- and off-KID. The discrepancy between the data and the constant-$\delta n_\text{qp}$ curves at low temperatures is evidence that the sensor is behaving in an unexpected way at low temperatures, where the peak $\Delta f_r/f_r$ is not proportional to the absorbed energy. This behavior was previously noted in Ref~\cite{temples2025}, and is contrary to expected KID behavior.

We explore an alternative explanation for the variation in pulse amplitude by re-casting $\Delta f_r/f_r$ into an equivalent temperature change $\Delta T$. We can use the MB fits described in section~\ref{sec:MB} to convert $\Delta f_r/f_r$ to $\Delta T$ by treating this relation as a look-up table-- this process is described in more detail in Appendix~\ref{app:dT}. When we over-plot lines of constant $\Delta T$, we find that both the on- and off-KID data follow these curves at low temperature. While the agreement with data is qualitative, this trend is remarkably close to observed behavior, in contrast to the expected trend in $\delta n_
{qp}$. The latter line is the expected response to a fixed change in quasiparticle density according to the standard treatment found in the literature~\cite{Gao2008,Wen2025}.



Following this lead, we can instead plot inferred temperature change as a function of base temperature, seen in Fig.~\ref{fig:C_comparison}. The two most notable features of our detector response are thus the following: (1)\;at high temperatures, the on-KID $\Delta f_r/f_r$ peaks follow lines of constant $\delta n_\text{qp}$ (Figure~\ref{fig:pulseShapeModeling}~Right) and (2)\;at low temperatures, the on-KID peaks follow lines of constant $\Delta T$. These two features motivate the modeling we describe in the next section.

\section{Implications for Detector Modeling}

 In the previous sections, we have discussed the main results from the two datasets taken with this experimental setup. We have shown two main results: first, that the KID response is position dependent, but that the position dependence is consistent with a finite and isotropic loss term. Second, that the KID response at low temperature can be best described as a constant temperature change across a large range of temperature. 
 
 This latter fact suggests that the temperature scan data should be thought of as tracking the heat capacity of the KID until that response would exceed the expected quasiparticle result, and thus that the response is still in proportion to the energy absorbed; in the thermal case, $\delta E = C\delta T$, and in the quasiparticle case, $\delta E = 2\Delta \delta n_{qp}V$. In this section, we first validate the former assumption, and compare an inferred heat capacity measurement to expected behavior. We then stitch these regimes together to show that our results can be used to separate a temperature-independent phonon response from a temperature-dependent sensor response.

\subsection{Heat capacity modeling of $\Delta T$ pulse heights }\label{sec:heat_capacity}

As mentioned in Section \ref{res:tempsweeps}, the peak $\Delta f_r/f_r$ data are transformed into peak $\Delta T$. Here, we assume that the energy absorbed into the KIPM detector from events on-KID is temperature independent. The simulations presented in Fig.~\ref{fig:pulsedMeasurements} and Ref.~\cite{temples2025} indicate that direct-hit and few-bounce phonons dominate on-KID events, suggesting the on-KID events are affected minimally by phonon loss in the silicon, as discussed in the previous section. Second, we assume that $\Delta T$ as converted from the $\delta$-deposit-corrected prompt component of the fitted $\Delta f_r/f_r$ timestream can be used to estimate the energy absorbed by detector as discussed in Appendix~\ref{app:pulse_fitting}. 

The temperature dependence of $\Delta T$ above $150$\,mK, which remains constant between $150$\,mK and $280$\,mK then decreases with increasing temperature, suggests the following model: 
\begin{equation}
\Delta T(T) = \frac{E_\text{abs}}{C(T)},\quad C(T) = C_0 +  C_\mathrm{pb}(\Delta_0,\Gamma,T).
\label{eq:C_model}
\end{equation} 
Here, $C_\mathrm{pb}$ is the heat capacity of a superconductor due to pair-breaking and $C_0$ is a temperature-independent heat capacity. When the absorbed energy is predominantly used to break Cooper pairs, the change in quasiparticle density is proportional to the absorbed energy, and the pair-breaking heat capacity is $C_\mathrm{pb}=\partial U_\mathrm{qp}/\partial T$, where $U_\mathrm{qp}$ is the internal energy of the quasiparticles in the film. We evaluate $\partial U_\mathrm{qp}/\partial T$ numerically, using a BCS-style integral with a gap-smeared density of states. See  Appendix\;\ref{app:heat_capacity} for more discussion.


We fit $\Delta T$ to Equation \ref{eq:C_model}, with free parameters $ E_\text{abs}$ and $C_0$, and fix $C_\mathrm{pb}$ using $\Delta_0$ and $\Gamma$ from Fig.~\ref{fig:VNAscan}~Right. We exclude temperatures below 150\,mK due to the excursion of the $\Delta T$ points at low temperatures. We find $C_0 = 34\,\mu\text{J}/\text{mol}/\text{K}$, equivalently $7.8\,\text{eV}/\text{ng}/\text{mK}$ or $21\,\text{meV}/\mu\text{m}^3/\text{mK}$, with an uncertainty of 12\%. Multiplying by the constant temperature rise of $\Delta T = 51\,$mK fitted at intermediate temperatures ($T \ge 145\,$mK) gives an absorbed energy $E_\text{abs} = 12.8 \pm 1.2\,\text{keV}$.

In Fig.~\ref{fig:C_comparison}, we plot our $C_\mathrm{pb}(T)$ model, derived from the steady-state resonator fits, and the bulk Al heat capacity from~\cite{Al_C} to demonstrate that the behaviors are similar. In the regime where $C_\mathrm{pb}(T)$ dominates over $C_0$, the $C_\mathrm{pb}(T)$ model also agrees with the standard MB prediction, shown as the green dotted line in Fig. \ref{fig:pulseShapeModeling}.  This simultaneous agreement in this temperature regime with both the bulk heat-capacity and MB approaches reflects the approximate equality of the complex conductivity response to thermally generated quasiparticles and to quasiparticles created by pair-breaking energy input demonstrated in Ref~\cite{Gao2008}. 



Moreover, the pair-breaking heat-capacity model is explicitly present in that equality. Using the thermal picture, the fractional frequency response is~\cite{GaoThesis}
\begin{equation}
\frac{\Delta f_r}{f_r} \approx -\frac{\partial \sigma_2 /\partial T}{2\,\sigma_2}\, \frac{E_{abs}}{C(T)}
\label{eq:dff_from_C_pb}
\end{equation}
where $\sigma_2$ is the imaginary part of the complex conductivity. In the regime in which $C_\mathrm{pb}(T)$ dominates over $C_0$ in the denominator,  both the $\partial \sigma_2 /\partial T$ and $C(T)$ terms are proportional to $\exp(-\Delta/k_BT)$, and thus the fractional frequency response is  relatively temperature independent. Between 150~mK and 280~mK, the $C_0$ term dominates the denominator, causing the fractional frequency response to fall exponentially with decreasing temperature, as seen in Fig.~\ref{fig:pulseShapeModeling}~Right.  

One candidate explanation for $C_0$ is an elevated quasiparticle temperature, which is a widely observed phenomenon in KIDs~\cite{Zmuidzinas2012} and could be due to readout power, blackbody radiation, and/or ionizing radiation.  In this case, however, Equation~\ref{eq:dff_from_C_pb} would continue to hold with $T$ saturating at some minimum value $T_{qp}$ as the bath temperature decreases.  The $\Delta f_r/f_r$ response would continue to remain on the green dotted line in Fig.~\ref{fig:pulseShapeModeling} Right, though with $T_{qp}$ larger than that indicated on the horizontal axis, and the $\Delta f_r/f_r$ response would not drop as it does.  

An alternative explanation more consistent with the data is a population of TLS.  As elaborated in Appendix~\ref{app:heat_capacity_tls}, a population of TLS with a log uniform density of states in the TLS energy $E_{TLS}$ results in a heat capacity constant with temperature (over some temperature regime), as needed here.  Moreover, Appendix~\ref{app:heat_capacity_tls} also shows that an additional population of TLS with a narrow energy distribution yields a Schottky anomalous heat capacity and can explain the deviations from Equation~\ref{eq:C_model} below 150~mK.  Further work is needed to identify a viable physical manifestation of such TLS and to test the model conclusively. 
 
\subsection{Temperature-dependent position sensitivity}\label{sec:fiveB}

In light of our improved understanding of the temperature dependence of the on-KID response, we can use the pulse shape modeling from Appendix~\ref{app:pulse_fitting} to draw some conclusions about phonon propagation in the substrate. Core to this modeling is determining which amplitudes estimate the integrated phonon energy received, which estimate phonon power, and whether any are influenced by additional degrees of freedom beyond thermal and phonon collection modeling. Here, to match the terminology in Appendix~\ref{app:pulse_fitting},  we refer to the on-KID prompt amplitude as $B_1$, the on-KID slow amplitude as $B_2$, and the off-KID amplitude as $A_1$. To simplify this treatment, we neglect the long fall times and define the on- and off-KID fall times as the shortest fall time of each component. See Figs.~\ref{fig:riseFits},\ref{fig:fallFits} for more details on this behavior.

First, we try to use all of the modeling information to back out a quantity comparable to that described in previous sections. If we assume that the prompt fall time component of the on-KID pulse is the true sensor response $\tau_r$, then by construction, our prompt amplitude is an energy estimator. In contrast, the slow component of the off-KID response $\tau_{f}$, which lasts substantially longer than the sensor response time at high temperature, is proportional to instantaneous phonon power. 

\begin{figure}
\includegraphics[width=\columnwidth]{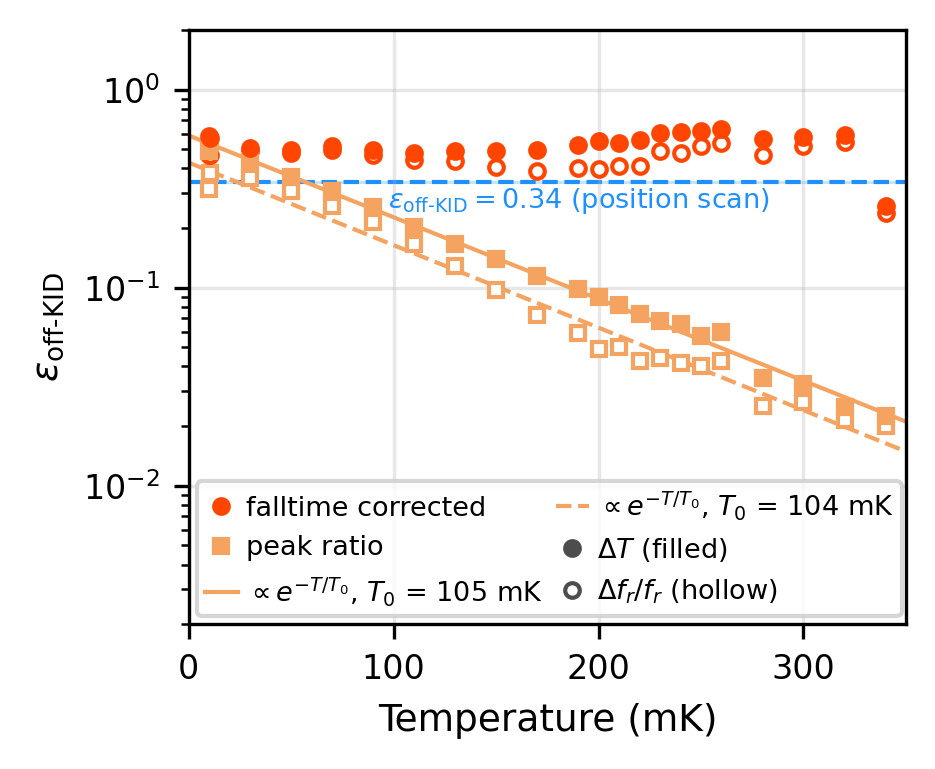}
\caption{Comparison of the temperature dependence of the relative phonon energy collection efficiency ($\epsilon_\mathrm{off-KID}$) calculated using two different energy estimators: peak ratios (squares) and $\epsilon_\mathrm{off-KID}$ as calculated from Eq. \ref{eq:off-KID-ps} (circles). Amplitudes are calculated from responses converted to $\Delta T$ (solid) and $\Delta f_r/f_r$ (hollow). We also show exponential trend lines fit to the peak ratio to illustrate the clear exponential trend in that estimator.}
\label{fig:eph_vs_temp}
\end{figure}

Thus, following the model of Appendix~\ref{app:pulse_fitting}, we  can use the fit parameters to construct the estimate
\begin{equation}\label{eq:off-KID-ps}
    \epsilon_\mathrm{off-KID} = \frac{E_\mathrm{abs,\,off-KID}}{E_\mathrm{abs,\,on-KID}} = \frac{A_1(1-r)}{B_1 r + B_2(1-r)}
\end{equation}
where $r=\tau_r/\tau_f$ is the ratio of sensor response time to phonon collection time. Fig.~\ref{fig:eph_vs_temp} shows $\epsilon_\mathrm{off-KID}$ for our data, which exhibit a flat temperature dependence. This behavior implies that the loss observed in the low temperature data does not change with temperature, and that the main effect we see as temperature increases is a change in sensor responsivity, which is highly pronounced when the sensor is faster than the diffuse phonon collection.



We include both this estimator and the amplitude ratio quantities to show that, in the low temperature limit, they are roughly equivalent as the sensor response is entirely determined by the KID response below 100~mK. As discussed in the previous section, the low-temperature regime has not been particularly well understood prior to this result, and in Appendix~\ref{app:pulse_fitting}, this is where many of the pulses begin to exhibit extremely long fall times. At the lowest temperatures, the on- and off-KID pulse shapes are roughly comparable, highly determined by long timescale decays, and their amplitudes alone are good estimators of total energy collection.


At higher temperatures, in particular above 100~mK, this is not the case, and a cursory inspection of amplitudes without considering operating regime would lead one to believe that phonon mean free path is changing -- this is not the nominal expectation\cite{Tamura,Herring1954}. Instead, the dynamics of the system conspire to produce a smoothly varying, continuous trend in fall time ratio between the sensor and substrate response. If we make the assumption that phonon collection timescales are to first order the constant off-KID fall time and are temperature-independent, we can then conclude that the on-KID response time, and thus sensor response, is an exponential function of temperature. This behavior is otherwise obscured by the pole mixing discussed in detail in Appendix~\ref{app:pulse_fitting}.  This amplitude ratio does, on the other hand, emphasize the stark separation between prompt and diffuse phonon responses, which would not be visible in the limit of high efficiency phonon collection.

\section{Discussion}\label{section:discussion}


In this paper, we have shown that the current KIPM detector design is inefficient in two ways: (1) the low surface coverage causes the detector to be particularly susceptible to loss mechanisms in the substrate and surface, and (2) an additional heat capacity becomes dominant at low temperature, resulting in an exponential decrease in the $\delta f_r/f_r$ response at low temperature. This investigation helps clarify open questions about low phonon efficiency and unexpected pulse shape trends in KIPM detectors, discussed first in a previous work from our consortium\cite{temples2025}. With explanations in hand for the unexpected detector response, we can explore their implications for future developments with this particular KIPM detector architecture.


First, we observe a loss that is independent of position in the x-y plane, indicating that the dead metal and the detector mounting clamps are not likely to be the dominant phonon sinks. According to Fig.~\ref{fig:positionDependenceVsCollectionEfficiency}, the homogeneous loss needed to account for $\epsilon_\mathrm{off-KID}=34\%$ is only 0.2\% per crystal transit, and the phonon absorption probability $p_\mathrm{abs}$ limits phonon collection efficiency. Fig.~\ref{fig:eph_vs_temp} implies that this loss is not temperature dependent, but that there is a large difference in the prompt vs diffuse phonon response that is enhanced as the sensor becomes substantially faster than the diffuse phonon collection time.


The data thus provide a clear path for improving the KIPM detector phonon collection efficiency. As discussed in Appendix~\ref{appendix:analytic-response-modeling}, the escape probability for athermal phonons in Al is relatively high for films thinner than 100~nm. Film thickness for the primary absorber needs to be improved such that phonon absorption is maximized when the phonon encounters a sensor; in this respect 30\,nm is much less than the 600\,nm thickness that is typically used in detectors that rely on quasiparticle trapping (see e.g. Refs~\cite{Kurinsky,newHVeV}).  Efforts are being made to design a phonon-absorber-assisted KIPM detector, wherein the phonon absorber is a separate entity from the sensing element, allowing for more beneficial volume scalings both in collection efficiency and intrinsic energy resolution.

Regarding the temperature-dependent response of the KID itself, we find that low temperature operation suppresses the detector response. This suppression can be explained by additional heat capacities: a constant heat capacity that dominates over the superconductor heat capacity below 280~mK and explains the data well from 280~mK to 150~mK; and, a Schottky anomalous heat capacity that dominates below 150~mK and fits all the data in that regime.  Such additional heat capacities could be provided by populations of TLS with, respectively, a broad, log-uniform density of states and a narrow, peaked density of states.  A more extensive investigation across multiple detectors and even multiple materials is thus warranted. 

We note that further work currently in preparation suggests that the current designs are highly susceptible to flux trapping, as evidenced by strong variation in Q with shielding environment. Increasing film thickness will both improve phonon collection and reduce flux trapping in these films.

The monolithic KIPM detector design may yet be a suitable path for sub-eV energy sensitivity. Intrinsic sensor energy resolution has been demonstrated at the few eV level \cite{Temples_2024}. In order for overall detector resolution to reach that level, greater collection efficiency is needed. Although thicker films may lead to lower kinetic inductance fractions, the countervailing improvements in readout power handling and film quality serve as rough justification for further investigation into monolithic KID designs for phonon-mediated detection. 


\begin{acknowledgments}
The authors thank Mike Kelsey for discussions regarding existing loss modeling in G4CMP. This work was supported by ARO grant W911NF-22-1-0257. NK was supported in part by the US Department of Energy Early Career Research Program (ECRP) under FWP 100872. ZS was supported in part by a DOE HEPCAT fellowship. SD and computational resources for this work were supported by ARO grant W911NF-23-S-0001. This work was supported in part by KIPAC funding. The MEMs system for this work was produced with support from a SLAC LDRD led by KS. 
\end{acknowledgments}

\bibliographystyle{aipnum4-1}
\bibliography{refs}

\begin{thebibliography}{43}%
\makeatletter
\providecommand \@ifxundefined [1]{%
 \@ifx{#1\undefined}
}%
\providecommand \@ifnum [1]{%
 \ifnum #1\expandafter \@firstoftwo
 \else \expandafter \@secondoftwo
 \fi
}%
\providecommand \@ifx [1]{%
 \ifx #1\expandafter \@firstoftwo
 \else \expandafter \@secondoftwo
 \fi
}%
\providecommand \natexlab [1]{#1}%
\providecommand \enquote  [1]{``#1''}%
\providecommand \bibnamefont  [1]{#1}%
\providecommand \bibfnamefont [1]{#1}%
\providecommand \citenamefont [1]{#1}%
\providecommand \href@noop [0]{\@secondoftwo}%
\providecommand \href [0]{\begingroup \@sanitize@url \@href}%
\providecommand \@href[1]{\@@startlink{#1}\@@href}%
\providecommand \@@href[1]{\endgroup#1\@@endlink}%
\providecommand \@sanitize@url [0]{\catcode `\\12\catcode `\$12\catcode `\&12\catcode `\#12\catcode `\^12\catcode `\_12\catcode `\%12\relax}%
\providecommand \@@startlink[1]{}%
\providecommand \@@endlink[0]{}%
\providecommand \url  [0]{\begingroup\@sanitize@url \@url }%
\providecommand \@url [1]{\endgroup\@href {#1}{\urlprefix }}%
\providecommand \urlprefix  [0]{URL }%
\providecommand \Eprint [0]{\href }%
\providecommand \doibase [0]{http://dx.doi.org/}%
\providecommand \selectlanguage [0]{\@gobble}%
\providecommand \bibinfo  [0]{\@secondoftwo}%
\providecommand \bibfield  [0]{\@secondoftwo}%
\providecommand \translation [1]{[#1]}%
\providecommand \BibitemOpen [0]{}%
\providecommand \bibitemStop [0]{}%
\providecommand \bibitemNoStop [0]{.\EOS\space}%
\providecommand \EOS [0]{\spacefactor3000\relax}%
\providecommand \BibitemShut  [1]{\csname bibitem#1\endcsname}%
\let\auto@bib@innerbib\@empty
\bibitem [{\citenamefont {Essig}\ \emph {et~al.}(2023)\citenamefont {Essig}, \citenamefont {Giovanetti}, \citenamefont {Kurinsky}, \citenamefont {McKinsey}, \citenamefont {Ramanathan}, \citenamefont {Stifter}, \citenamefont {Yu}, \citenamefont {Aboubrahim}, \citenamefont {Adams}, \citenamefont {Alves}, \citenamefont {Aralis}, \citenamefont {Araújo}, \citenamefont {Baxter}, \citenamefont {Berghaus}, \citenamefont {Berlin}, \citenamefont {Blanco}, \citenamefont {Bloch}, \citenamefont {Bonivento}, \citenamefont {Bunker}, \citenamefont {Burdin}, \citenamefont {Caminata}, \citenamefont {Carmona-Benitez}, \citenamefont {Chaplinsky}, \citenamefont {Chen}, \citenamefont {Derenzo}, \citenamefont {de~Viveiros}, \citenamefont {Dick}, \citenamefont {Marco}, \citenamefont {Du}, \citenamefont {Dutta}, \citenamefont {Ebadi}, \citenamefont {Emken}, \citenamefont {Esposito}, \citenamefont {Etzion}, \citenamefont {Feng}, \citenamefont {Fernandez}, \citenamefont {Ge}, \citenamefont {Ghosh}, \citenamefont {Giroux},
  \citenamefont {Hamaide}, \citenamefont {Hertel}, \citenamefont {Herrera}, \citenamefont {Hochberg}, \citenamefont {Kahn}, \citenamefont {Kavanagh}, \citenamefont {Khan}, \citenamefont {Kluck}, \citenamefont {Kravitz}, \citenamefont {von Krosigk}, \citenamefont {Kumar}, \citenamefont {Lawson}, \citenamefont {Lehmann}, \citenamefont {Lin}, \citenamefont {Liao}, \citenamefont {Lyon}, \citenamefont {Majewski}, \citenamefont {Manzari}, \citenamefont {Monroe}, \citenamefont {Monzani}, \citenamefont {Morrissey}, \citenamefont {Norcini}, \citenamefont {Orly}, \citenamefont {Parikh}, \citenamefont {Park}, \citenamefont {Patel}, \citenamefont {Paul}, \citenamefont {Pèrez-Ríos}, \citenamefont {Phipps}, \citenamefont {Pocar}, \citenamefont {Ritz}, \citenamefont {Sarkis}, \citenamefont {Schuster}, \citenamefont {Schwetz}, \citenamefont {Shaw}, \citenamefont {Shin}, \citenamefont {Singal}, \citenamefont {Singh}, \citenamefont {Slone}, \citenamefont {Sorensen}, \citenamefont {Sun}, \citenamefont {Szydagis},
  \citenamefont {Temples}, \citenamefont {Testera}, \citenamefont {Thieme}, \citenamefont {Toro}, \citenamefont {Trickle}, \citenamefont {Uemura}, \citenamefont {Velan}, \citenamefont {Vitagliano}, \citenamefont {Wagner}, \citenamefont {Wang}, \citenamefont {Westerdale},\ and\ \citenamefont {Zurek}}]{essig2023snowmass}%
  \BibitemOpen
  \bibfield  {author} {\bibinfo {author} {\bibfnamefont {R.}~\bibnamefont {Essig}}, \bibinfo {author} {\bibfnamefont {G.~K.}\ \bibnamefont {Giovanetti}}, \bibinfo {author} {\bibfnamefont {N.}~\bibnamefont {Kurinsky}}, \bibinfo {author} {\bibfnamefont {D.}~\bibnamefont {McKinsey}}, \bibinfo {author} {\bibfnamefont {K.}~\bibnamefont {Ramanathan}}, \bibinfo {author} {\bibfnamefont {K.}~\bibnamefont {Stifter}}, \bibinfo {author} {\bibfnamefont {T.-T.}\ \bibnamefont {Yu}}, \bibinfo {author} {\bibfnamefont {A.}~\bibnamefont {Aboubrahim}}, \bibinfo {author} {\bibfnamefont {D.}~\bibnamefont {Adams}}, \bibinfo {author} {\bibfnamefont {D.~S.~M.}\ \bibnamefont {Alves}}, \bibinfo {author} {\bibfnamefont {T.}~\bibnamefont {Aralis}}, \bibinfo {author} {\bibfnamefont {H.~M.}\ \bibnamefont {Araújo}}, \bibinfo {author} {\bibfnamefont {D.}~\bibnamefont {Baxter}}, \bibinfo {author} {\bibfnamefont {K.~V.}\ \bibnamefont {Berghaus}}, \bibinfo {author} {\bibfnamefont {A.}~\bibnamefont {Berlin}}, \bibinfo {author} {\bibfnamefont
  {C.}~\bibnamefont {Blanco}}, \bibinfo {author} {\bibfnamefont {I.~M.}\ \bibnamefont {Bloch}}, \bibinfo {author} {\bibfnamefont {W.~M.}\ \bibnamefont {Bonivento}}, \bibinfo {author} {\bibfnamefont {R.}~\bibnamefont {Bunker}}, \bibinfo {author} {\bibfnamefont {S.}~\bibnamefont {Burdin}}, \bibinfo {author} {\bibfnamefont {A.}~\bibnamefont {Caminata}}, \bibinfo {author} {\bibfnamefont {M.~C.}\ \bibnamefont {Carmona-Benitez}}, \bibinfo {author} {\bibfnamefont {L.}~\bibnamefont {Chaplinsky}}, \bibinfo {author} {\bibfnamefont {T.~Y.}\ \bibnamefont {Chen}}, \bibinfo {author} {\bibfnamefont {S.~E.}\ \bibnamefont {Derenzo}}, \bibinfo {author} {\bibfnamefont {L.}~\bibnamefont {de~Viveiros}}, \bibinfo {author} {\bibfnamefont {R.}~\bibnamefont {Dick}}, \bibinfo {author} {\bibfnamefont {N.~D.}\ \bibnamefont {Marco}}, \bibinfo {author} {\bibfnamefont {P.}~\bibnamefont {Du}}, \bibinfo {author} {\bibfnamefont {B.}~\bibnamefont {Dutta}}, \bibinfo {author} {\bibfnamefont {R.}~\bibnamefont {Ebadi}}, \bibinfo {author}
  {\bibfnamefont {T.}~\bibnamefont {Emken}}, \bibinfo {author} {\bibfnamefont {A.}~\bibnamefont {Esposito}}, \bibinfo {author} {\bibfnamefont {E.}~\bibnamefont {Etzion}}, \bibinfo {author} {\bibfnamefont {J.~L.}\ \bibnamefont {Feng}}, \bibinfo {author} {\bibfnamefont {N.}~\bibnamefont {Fernandez}}, \bibinfo {author} {\bibfnamefont {S.~F.}\ \bibnamefont {Ge}}, \bibinfo {author} {\bibfnamefont {S.}~\bibnamefont {Ghosh}}, \bibinfo {author} {\bibfnamefont {G.}~\bibnamefont {Giroux}}, \bibinfo {author} {\bibfnamefont {L.}~\bibnamefont {Hamaide}}, \bibinfo {author} {\bibfnamefont {S.~A.}\ \bibnamefont {Hertel}}, \bibinfo {author} {\bibfnamefont {G.}~\bibnamefont {Herrera}}, \bibinfo {author} {\bibfnamefont {Y.}~\bibnamefont {Hochberg}}, \bibinfo {author} {\bibfnamefont {Y.}~\bibnamefont {Kahn}}, \bibinfo {author} {\bibfnamefont {B.~J.}\ \bibnamefont {Kavanagh}}, \bibinfo {author} {\bibfnamefont {A.~N.}\ \bibnamefont {Khan}}, \bibinfo {author} {\bibfnamefont {H.}~\bibnamefont {Kluck}}, \bibinfo {author}
  {\bibfnamefont {S.}~\bibnamefont {Kravitz}}, \bibinfo {author} {\bibfnamefont {B.}~\bibnamefont {von Krosigk}}, \bibinfo {author} {\bibfnamefont {J.}~\bibnamefont {Kumar}}, \bibinfo {author} {\bibfnamefont {I.~T.}\ \bibnamefont {Lawson}}, \bibinfo {author} {\bibfnamefont {B.~V.}\ \bibnamefont {Lehmann}}, \bibinfo {author} {\bibfnamefont {T.}~\bibnamefont {Lin}}, \bibinfo {author} {\bibfnamefont {J.}~\bibnamefont {Liao}}, \bibinfo {author} {\bibfnamefont {S.~A.}\ \bibnamefont {Lyon}}, \bibinfo {author} {\bibfnamefont {P.~M.}\ \bibnamefont {Majewski}}, \bibinfo {author} {\bibfnamefont {C.~A.}\ \bibnamefont {Manzari}}, \bibinfo {author} {\bibfnamefont {J.}~\bibnamefont {Monroe}}, \bibinfo {author} {\bibfnamefont {M.~E.}\ \bibnamefont {Monzani}}, \bibinfo {author} {\bibfnamefont {D.~E.}\ \bibnamefont {Morrissey}}, \bibinfo {author} {\bibfnamefont {D.}~\bibnamefont {Norcini}}, \bibinfo {author} {\bibfnamefont {A.}~\bibnamefont {Orly}}, \bibinfo {author} {\bibfnamefont {A.}~\bibnamefont {Parikh}}, \bibinfo
  {author} {\bibfnamefont {J.~C.}\ \bibnamefont {Park}}, \bibinfo {author} {\bibfnamefont {P.~K.}\ \bibnamefont {Patel}}, \bibinfo {author} {\bibfnamefont {S.}~\bibnamefont {Paul}}, \bibinfo {author} {\bibfnamefont {J.}~\bibnamefont {Pèrez-Ríos}}, \bibinfo {author} {\bibfnamefont {A.}~\bibnamefont {Phipps}}, \bibinfo {author} {\bibfnamefont {A.}~\bibnamefont {Pocar}}, \bibinfo {author} {\bibfnamefont {A.}~\bibnamefont {Ritz}}, \bibinfo {author} {\bibfnamefont {Y.}~\bibnamefont {Sarkis}}, \bibinfo {author} {\bibfnamefont {P.}~\bibnamefont {Schuster}}, \bibinfo {author} {\bibfnamefont {T.}~\bibnamefont {Schwetz}}, \bibinfo {author} {\bibfnamefont {S.}~\bibnamefont {Shaw}}, \bibinfo {author} {\bibfnamefont {S.}~\bibnamefont {Shin}}, \bibinfo {author} {\bibfnamefont {A.}~\bibnamefont {Singal}}, \bibinfo {author} {\bibfnamefont {R.}~\bibnamefont {Singh}}, \bibinfo {author} {\bibfnamefont {O.}~\bibnamefont {Slone}}, \bibinfo {author} {\bibfnamefont {P.}~\bibnamefont {Sorensen}}, \bibinfo {author} {\bibfnamefont
  {C.}~\bibnamefont {Sun}}, \bibinfo {author} {\bibfnamefont {M.}~\bibnamefont {Szydagis}}, \bibinfo {author} {\bibfnamefont {D.~J.}\ \bibnamefont {Temples}}, \bibinfo {author} {\bibfnamefont {G.}~\bibnamefont {Testera}}, \bibinfo {author} {\bibfnamefont {K.}~\bibnamefont {Thieme}}, \bibinfo {author} {\bibfnamefont {N.}~\bibnamefont {Toro}}, \bibinfo {author} {\bibfnamefont {T.}~\bibnamefont {Trickle}}, \bibinfo {author} {\bibfnamefont {S.}~\bibnamefont {Uemura}}, \bibinfo {author} {\bibfnamefont {V.}~\bibnamefont {Velan}}, \bibinfo {author} {\bibfnamefont {E.}~\bibnamefont {Vitagliano}}, \bibinfo {author} {\bibfnamefont {F.}~\bibnamefont {Wagner}}, \bibinfo {author} {\bibfnamefont {G.}~\bibnamefont {Wang}}, \bibinfo {author} {\bibfnamefont {S.}~\bibnamefont {Westerdale}}, \ and\ \bibinfo {author} {\bibfnamefont {K.~M.}\ \bibnamefont {Zurek}},\ }\href {https://arxiv.org/abs/2203.08297} {\enquote {\bibinfo {title} {Snowmass2021 cosmic frontier: The landscape of low-threshold dark matter direct detection in the
  next decade},}\ } (\bibinfo {year} {2023}),\ \Eprint {http://arxiv.org/abs/2203.08297} {arXiv:2203.08297 [hep-ph]} \BibitemShut {NoStop}%
\bibitem [{\citenamefont {collaboration}\ \emph {et~al.}(2022)\citenamefont {collaboration}, \citenamefont {Goupy}, \citenamefont {Abele}, \citenamefont {Angloher}, \citenamefont {Bento}, \citenamefont {Canonica}, \citenamefont {Cappella}, \citenamefont {Cardani}, \citenamefont {Casali}, \citenamefont {Cerulli}, \citenamefont {Colantoni}, \citenamefont {Cruciani}, \citenamefont {Castello}, \citenamefont {del Gallo~Roccagiovine}, \citenamefont {Doblhammer}, \citenamefont {Dorer}, \citenamefont {Erhart}, \citenamefont {Friendl}, \citenamefont {Garai}, \citenamefont {Ghete}, \citenamefont {Hauff}, \citenamefont {Jeanneau}, \citenamefont {Jericha}, \citenamefont {Kaznacheeva}, \citenamefont {Kinast}, \citenamefont {Kluck}, \citenamefont {Langenkämper}, \citenamefont {Lasserre}, \citenamefont {Lhuillier}, \citenamefont {Mancuso}, \citenamefont {Mauri}, \citenamefont {Mazzolari}, \citenamefont {Mazzucato}, \citenamefont {Neyrial}, \citenamefont {Nones}, \citenamefont {Oberauer}, \citenamefont {Ortmann},
  \citenamefont {Pattavina}, \citenamefont {Peters}, \citenamefont {Petricca}, \citenamefont {Potzel}, \citenamefont {Pröbst}, \citenamefont {Pucci}, \citenamefont {Riendl}, \citenamefont {Rogly}, \citenamefont {Romagnoni}, \citenamefont {Rothe}, \citenamefont {Schermer}, \citenamefont {Schieck}, \citenamefont {Schönert}, \citenamefont {Schwertner}, \citenamefont {Scola}, \citenamefont {Sidikov}, \citenamefont {Stodolsky}, \citenamefont {Strauss}, \citenamefont {Tamisari}, \citenamefont {Tomei}, \citenamefont {Vignati}, \citenamefont {Vivier}, \citenamefont {Wagner}, \citenamefont {Wex},\ and\ \citenamefont {.}}]{nucleus22}%
  \BibitemOpen
  \bibfield  {author} {\bibinfo {author} {\bibfnamefont {T.~N.}\ \bibnamefont {collaboration}}, \bibinfo {author} {\bibfnamefont {C.}~\bibnamefont {Goupy}}, \bibinfo {author} {\bibfnamefont {H.}~\bibnamefont {Abele}}, \bibinfo {author} {\bibfnamefont {G.}~\bibnamefont {Angloher}}, \bibinfo {author} {\bibfnamefont {A.}~\bibnamefont {Bento}}, \bibinfo {author} {\bibfnamefont {L.}~\bibnamefont {Canonica}}, \bibinfo {author} {\bibfnamefont {F.}~\bibnamefont {Cappella}}, \bibinfo {author} {\bibfnamefont {L.}~\bibnamefont {Cardani}}, \bibinfo {author} {\bibfnamefont {N.}~\bibnamefont {Casali}}, \bibinfo {author} {\bibfnamefont {R.}~\bibnamefont {Cerulli}}, \bibinfo {author} {\bibfnamefont {I.}~\bibnamefont {Colantoni}}, \bibinfo {author} {\bibfnamefont {A.}~\bibnamefont {Cruciani}}, \bibinfo {author} {\bibfnamefont {G.~D.}\ \bibnamefont {Castello}}, \bibinfo {author} {\bibfnamefont {M.}~\bibnamefont {del Gallo~Roccagiovine}}, \bibinfo {author} {\bibfnamefont {A.}~\bibnamefont {Doblhammer}}, \bibinfo {author}
  {\bibfnamefont {S.}~\bibnamefont {Dorer}}, \bibinfo {author} {\bibfnamefont {A.}~\bibnamefont {Erhart}}, \bibinfo {author} {\bibfnamefont {M.}~\bibnamefont {Friendl}}, \bibinfo {author} {\bibfnamefont {A.}~\bibnamefont {Garai}}, \bibinfo {author} {\bibfnamefont {V.~M.}\ \bibnamefont {Ghete}}, \bibinfo {author} {\bibfnamefont {D.}~\bibnamefont {Hauff}}, \bibinfo {author} {\bibfnamefont {F.}~\bibnamefont {Jeanneau}}, \bibinfo {author} {\bibfnamefont {E.}~\bibnamefont {Jericha}}, \bibinfo {author} {\bibfnamefont {M.}~\bibnamefont {Kaznacheeva}}, \bibinfo {author} {\bibfnamefont {A.}~\bibnamefont {Kinast}}, \bibinfo {author} {\bibfnamefont {H.}~\bibnamefont {Kluck}}, \bibinfo {author} {\bibfnamefont {A.}~\bibnamefont {Langenkämper}}, \bibinfo {author} {\bibfnamefont {T.}~\bibnamefont {Lasserre}}, \bibinfo {author} {\bibfnamefont {D.}~\bibnamefont {Lhuillier}}, \bibinfo {author} {\bibfnamefont {M.}~\bibnamefont {Mancuso}}, \bibinfo {author} {\bibfnamefont {B.}~\bibnamefont {Mauri}}, \bibinfo {author}
  {\bibfnamefont {A.}~\bibnamefont {Mazzolari}}, \bibinfo {author} {\bibfnamefont {E.}~\bibnamefont {Mazzucato}}, \bibinfo {author} {\bibfnamefont {H.}~\bibnamefont {Neyrial}}, \bibinfo {author} {\bibfnamefont {C.}~\bibnamefont {Nones}}, \bibinfo {author} {\bibfnamefont {L.}~\bibnamefont {Oberauer}}, \bibinfo {author} {\bibfnamefont {T.}~\bibnamefont {Ortmann}}, \bibinfo {author} {\bibfnamefont {L.}~\bibnamefont {Pattavina}}, \bibinfo {author} {\bibfnamefont {L.}~\bibnamefont {Peters}}, \bibinfo {author} {\bibfnamefont {F.}~\bibnamefont {Petricca}}, \bibinfo {author} {\bibfnamefont {W.}~\bibnamefont {Potzel}}, \bibinfo {author} {\bibfnamefont {F.}~\bibnamefont {Pröbst}}, \bibinfo {author} {\bibfnamefont {F.}~\bibnamefont {Pucci}}, \bibinfo {author} {\bibfnamefont {F.}~\bibnamefont {Riendl}}, \bibinfo {author} {\bibfnamefont {R.}~\bibnamefont {Rogly}}, \bibinfo {author} {\bibfnamefont {M.}~\bibnamefont {Romagnoni}}, \bibinfo {author} {\bibfnamefont {J.}~\bibnamefont {Rothe}}, \bibinfo {author} {\bibfnamefont
  {N.}~\bibnamefont {Schermer}}, \bibinfo {author} {\bibfnamefont {J.}~\bibnamefont {Schieck}}, \bibinfo {author} {\bibfnamefont {S.}~\bibnamefont {Schönert}}, \bibinfo {author} {\bibfnamefont {C.}~\bibnamefont {Schwertner}}, \bibinfo {author} {\bibfnamefont {L.}~\bibnamefont {Scola}}, \bibinfo {author} {\bibfnamefont {G.~S.}\ \bibnamefont {Sidikov}}, \bibinfo {author} {\bibfnamefont {L.}~\bibnamefont {Stodolsky}}, \bibinfo {author} {\bibfnamefont {R.}~\bibnamefont {Strauss}}, \bibinfo {author} {\bibfnamefont {M.}~\bibnamefont {Tamisari}}, \bibinfo {author} {\bibfnamefont {C.}~\bibnamefont {Tomei}}, \bibinfo {author} {\bibfnamefont {M.}~\bibnamefont {Vignati}}, \bibinfo {author} {\bibfnamefont {M.}~\bibnamefont {Vivier}}, \bibinfo {author} {\bibfnamefont {V.}~\bibnamefont {Wagner}}, \bibinfo {author} {\bibfnamefont {A.}~\bibnamefont {Wex}}, \ and\ \bibinfo {author} {\bibnamefont {.}},\ }\href {https://arxiv.org/abs/2211.04189} {\enquote {\bibinfo {title} {Exploring coherent elastic neutrino-nucleus
  scattering of reactor neutrinos with the nucleus experiment},}\ } (\bibinfo {year} {2022}),\ \Eprint {http://arxiv.org/abs/2211.04189} {arXiv:2211.04189 [physics.ins-det]} \BibitemShut {NoStop}%
\bibitem [{\citenamefont {Bowen}\ and\ \citenamefont {Huber}(2020)}]{Bowen_2020}%
  \BibitemOpen
  \bibfield  {author} {\bibinfo {author} {\bibfnamefont {M.}~\bibnamefont {Bowen}}\ and\ \bibinfo {author} {\bibfnamefont {P.}~\bibnamefont {Huber}},\ }\href {\doibase 10.1103/physrevd.102.053008} {\bibfield  {journal} {\bibinfo  {journal} {Physical Review D}\ }\textbf {\bibinfo {volume} {102}} (\bibinfo {year} {2020}),\ 10.1103/physrevd.102.053008}\BibitemShut {NoStop}%
\bibitem [{\citenamefont {Billard}, \citenamefont {Johnston},\ and\ \citenamefont {Kavanagh}(2018)}]{Billard_2018}%
  \BibitemOpen
  \bibfield  {author} {\bibinfo {author} {\bibfnamefont {J.}~\bibnamefont {Billard}}, \bibinfo {author} {\bibfnamefont {J.}~\bibnamefont {Johnston}}, \ and\ \bibinfo {author} {\bibfnamefont {B.~J.}\ \bibnamefont {Kavanagh}},\ }\href {\doibase 10.1088/1475-7516/2018/11/016} {\bibfield  {journal} {\bibinfo  {journal} {Journal of Cosmology and Astroparticle Physics}\ }\textbf {\bibinfo {volume} {2018}},\ \bibinfo {pages} {016–016} (\bibinfo {year} {2018})}\BibitemShut {NoStop}%
\bibitem [{\citenamefont {Yang}\ \emph {et~al.}(2026)\citenamefont {Yang}, \citenamefont {Carroll}, \citenamefont {Mason}, \citenamefont {Schwartz}, \citenamefont {O'Hara}, \citenamefont {Lund}, \citenamefont {Gottschalk}, \citenamefont {Stephenson}, \citenamefont {Friedman}, \citenamefont {Yumiceva}, \citenamefont {Hackley}, \citenamefont {Graninger}, \citenamefont {Rotella}, \citenamefont {Warner}, \citenamefont {Cochran}, \citenamefont {Bruce}, \citenamefont {Wagner}, \citenamefont {Wenner}, \citenamefont {Steers}, \citenamefont {Moore}, \citenamefont {Marakov},\ and\ \citenamefont {Christensen}}]{yang2026}%
  \BibitemOpen
  \bibfield  {author} {\bibinfo {author} {\bibfnamefont {J.}~\bibnamefont {Yang}}, \bibinfo {author} {\bibfnamefont {T.~J.}\ \bibnamefont {Carroll}}, \bibinfo {author} {\bibfnamefont {P.}~\bibnamefont {Mason}}, \bibinfo {author} {\bibfnamefont {R.}~\bibnamefont {Schwartz}}, \bibinfo {author} {\bibfnamefont {K.~M.}\ \bibnamefont {O'Hara}}, \bibinfo {author} {\bibfnamefont {J.}~\bibnamefont {Lund}}, \bibinfo {author} {\bibfnamefont {M.}~\bibnamefont {Gottschalk}}, \bibinfo {author} {\bibfnamefont {T.}~\bibnamefont {Stephenson}}, \bibinfo {author} {\bibfnamefont {L.~H.}\ \bibnamefont {Friedman}}, \bibinfo {author} {\bibfnamefont {F.}~\bibnamefont {Yumiceva}}, \bibinfo {author} {\bibfnamefont {J.}~\bibnamefont {Hackley}}, \bibinfo {author} {\bibfnamefont {A.~L.}\ \bibnamefont {Graninger}}, \bibinfo {author} {\bibfnamefont {C.}~\bibnamefont {Rotella}}, \bibinfo {author} {\bibfnamefont {P.}~\bibnamefont {Warner}}, \bibinfo {author} {\bibfnamefont {J.~M.}\ \bibnamefont {Cochran}}, \bibinfo {author} {\bibfnamefont
  {A.~V.}\ \bibnamefont {Bruce}}, \bibinfo {author} {\bibfnamefont {M.}~\bibnamefont {Wagner}}, \bibinfo {author} {\bibfnamefont {J.}~\bibnamefont {Wenner}}, \bibinfo {author} {\bibfnamefont {S.}~\bibnamefont {Steers}}, \bibinfo {author} {\bibfnamefont {C.}~\bibnamefont {Moore}}, \bibinfo {author} {\bibfnamefont {A.}~\bibnamefont {Marakov}}, \ and\ \bibinfo {author} {\bibfnamefont {B.~G.}\ \bibnamefont {Christensen}},\ }\href {https://arxiv.org/abs/2602.23544} {\enquote {\bibinfo {title} {High-temporal-resolution measurements of the impacts of ionizing radiation on superconducting qubits},}\ } (\bibinfo {year} {2026}),\ \Eprint {http://arxiv.org/abs/2602.23544} {arXiv:2602.23544 [quant-ph]} \BibitemShut {NoStop}%
\bibitem [{\citenamefont {Anthony-Petersen}\ \emph {et~al.}(2025)\citenamefont {Anthony-Petersen}, \citenamefont {Chang}, \citenamefont {Chang}, \citenamefont {Chaplinsky}, \citenamefont {Fink}, \citenamefont {Garcia-Sciveres}, \citenamefont {Guo}, \citenamefont {Hertel}, \citenamefont {Li}, \citenamefont {Lin}, \citenamefont {Lisovenko}, \citenamefont {Mahapatra}, \citenamefont {Matava}, \citenamefont {McKinsey}, \citenamefont {Osterman}, \citenamefont {Patel}, \citenamefont {Penning}, \citenamefont {Platt}, \citenamefont {Pyle}, \citenamefont {Qi}, \citenamefont {Reed}, \citenamefont {Rydstrom}, \citenamefont {Romani}, \citenamefont {Sadoulet}, \citenamefont {Serfass}, \citenamefont {Sorensen}, \citenamefont {Suerfu}, \citenamefont {Velan}, \citenamefont {Wang}, \citenamefont {Wang}, \citenamefont {Watkins},\ and\ \citenamefont {Williams}}]{TESSERACT_2025}%
  \BibitemOpen
  \bibfield  {author} {\bibinfo {author} {\bibfnamefont {R.}~\bibnamefont {Anthony-Petersen}}, \bibinfo {author} {\bibfnamefont {C.~L.}\ \bibnamefont {Chang}}, \bibinfo {author} {\bibfnamefont {Y.-Y.}\ \bibnamefont {Chang}}, \bibinfo {author} {\bibfnamefont {L.}~\bibnamefont {Chaplinsky}}, \bibinfo {author} {\bibfnamefont {C.~W.}\ \bibnamefont {Fink}}, \bibinfo {author} {\bibfnamefont {M.}~\bibnamefont {Garcia-Sciveres}}, \bibinfo {author} {\bibfnamefont {W.}~\bibnamefont {Guo}}, \bibinfo {author} {\bibfnamefont {S.~A.}\ \bibnamefont {Hertel}}, \bibinfo {author} {\bibfnamefont {X.}~\bibnamefont {Li}}, \bibinfo {author} {\bibfnamefont {J.}~\bibnamefont {Lin}}, \bibinfo {author} {\bibfnamefont {M.}~\bibnamefont {Lisovenko}}, \bibinfo {author} {\bibfnamefont {R.}~\bibnamefont {Mahapatra}}, \bibinfo {author} {\bibfnamefont {W.}~\bibnamefont {Matava}}, \bibinfo {author} {\bibfnamefont {D.~N.}\ \bibnamefont {McKinsey}}, \bibinfo {author} {\bibfnamefont {D.~Z.}\ \bibnamefont {Osterman}}, \bibinfo {author}
  {\bibfnamefont {P.~K.}\ \bibnamefont {Patel}}, \bibinfo {author} {\bibfnamefont {B.}~\bibnamefont {Penning}}, \bibinfo {author} {\bibfnamefont {M.}~\bibnamefont {Platt}}, \bibinfo {author} {\bibfnamefont {M.}~\bibnamefont {Pyle}}, \bibinfo {author} {\bibfnamefont {Y.}~\bibnamefont {Qi}}, \bibinfo {author} {\bibfnamefont {M.}~\bibnamefont {Reed}}, \bibinfo {author} {\bibfnamefont {I.}~\bibnamefont {Rydstrom}}, \bibinfo {author} {\bibfnamefont {R.~K.}\ \bibnamefont {Romani}}, \bibinfo {author} {\bibfnamefont {B.}~\bibnamefont {Sadoulet}}, \bibinfo {author} {\bibfnamefont {B.}~\bibnamefont {Serfass}}, \bibinfo {author} {\bibfnamefont {P.}~\bibnamefont {Sorensen}}, \bibinfo {author} {\bibfnamefont {B.}~\bibnamefont {Suerfu}}, \bibinfo {author} {\bibfnamefont {V.}~\bibnamefont {Velan}}, \bibinfo {author} {\bibfnamefont {G.}~\bibnamefont {Wang}}, \bibinfo {author} {\bibfnamefont {Y.}~\bibnamefont {Wang}}, \bibinfo {author} {\bibfnamefont {S.~L.}\ \bibnamefont {Watkins}}, \ and\ \bibinfo {author} {\bibfnamefont
  {M.~R.}\ \bibnamefont {Williams}},\ }\href {\doibase 10.1063/5.0247343} {\bibfield  {journal} {\bibinfo  {journal} {Applied Physics Letters}\ }\textbf {\bibinfo {volume} {126}},\ \bibinfo {pages} {102601} (\bibinfo {year} {2025})}\BibitemShut {NoStop}%
\bibitem [{\citenamefont {Kennard}\ \emph {et~al.}(2026)\citenamefont {Kennard}, \citenamefont {Pradeep}, \citenamefont {Buchanan}, \citenamefont {Fu}, \citenamefont {Simchony}, \citenamefont {Wang}, \citenamefont {Michielin}, \citenamefont {Aralis}, \citenamefont {Cudmore}, \citenamefont {Cushman}, \citenamefont {Diamond}, \citenamefont {Figueroa-Feliciano}, \citenamefont {Fink}, \citenamefont {Harms}, \citenamefont {Hines}, \citenamefont {Hong}, \citenamefont {Huber}, \citenamefont {Kubik}, \citenamefont {Kurinsky}, \citenamefont {Mahapatra}, \citenamefont {Novati}, \citenamefont {Pandey}, \citenamefont {Patel}, \citenamefont {Peng}, \citenamefont {Platt}, \citenamefont {Pressman-Cyna}, \citenamefont {Rau}, \citenamefont {Ren}, \citenamefont {Reynolds}, \citenamefont {Ryan}, \citenamefont {Saab}, \citenamefont {Sadek}, \citenamefont {Schmidt}, \citenamefont {Smith}, \citenamefont {Stevens}, \citenamefont {Stifter}, \citenamefont {Stukel}, \citenamefont {Viol}, \citenamefont {Wang}, \citenamefont {Wilson},
  \citenamefont {Young}, \citenamefont {Zatschler}, \citenamefont {Zenger},\ and\ \citenamefont {Zuñiga-Reyes}}]{newHVeV}%
  \BibitemOpen
  \bibfield  {author} {\bibinfo {author} {\bibfnamefont {K.}~\bibnamefont {Kennard}}, \bibinfo {author} {\bibfnamefont {A.}~\bibnamefont {Pradeep}}, \bibinfo {author} {\bibfnamefont {M.}~\bibnamefont {Buchanan}}, \bibinfo {author} {\bibfnamefont {H.}~\bibnamefont {Fu}}, \bibinfo {author} {\bibfnamefont {A.}~\bibnamefont {Simchony}}, \bibinfo {author} {\bibfnamefont {Q.}~\bibnamefont {Wang}}, \bibinfo {author} {\bibfnamefont {E.}~\bibnamefont {Michielin}}, \bibinfo {author} {\bibfnamefont {T.}~\bibnamefont {Aralis}}, \bibinfo {author} {\bibfnamefont {E.}~\bibnamefont {Cudmore}}, \bibinfo {author} {\bibfnamefont {P.}~\bibnamefont {Cushman}}, \bibinfo {author} {\bibfnamefont {M.}~\bibnamefont {Diamond}}, \bibinfo {author} {\bibfnamefont {E.}~\bibnamefont {Figueroa-Feliciano}}, \bibinfo {author} {\bibfnamefont {C.}~\bibnamefont {Fink}}, \bibinfo {author} {\bibfnamefont {S.}~\bibnamefont {Harms}}, \bibinfo {author} {\bibfnamefont {B.~A.}\ \bibnamefont {Hines}}, \bibinfo {author} {\bibfnamefont {Z.}~\bibnamefont
  {Hong}}, \bibinfo {author} {\bibfnamefont {M.~E.}\ \bibnamefont {Huber}}, \bibinfo {author} {\bibfnamefont {A.}~\bibnamefont {Kubik}}, \bibinfo {author} {\bibfnamefont {N.}~\bibnamefont {Kurinsky}}, \bibinfo {author} {\bibfnamefont {R.}~\bibnamefont {Mahapatra}}, \bibinfo {author} {\bibfnamefont {V.}~\bibnamefont {Novati}}, \bibinfo {author} {\bibfnamefont {L.}~\bibnamefont {Pandey}}, \bibinfo {author} {\bibfnamefont {P.~K.}\ \bibnamefont {Patel}}, \bibinfo {author} {\bibfnamefont {W.}~\bibnamefont {Peng}}, \bibinfo {author} {\bibfnamefont {M.}~\bibnamefont {Platt}}, \bibinfo {author} {\bibfnamefont {R.}~\bibnamefont {Pressman-Cyna}}, \bibinfo {author} {\bibfnamefont {W.}~\bibnamefont {Rau}}, \bibinfo {author} {\bibfnamefont {R.}~\bibnamefont {Ren}}, \bibinfo {author} {\bibfnamefont {T.}~\bibnamefont {Reynolds}}, \bibinfo {author} {\bibfnamefont {J.}~\bibnamefont {Ryan}}, \bibinfo {author} {\bibfnamefont {T.}~\bibnamefont {Saab}}, \bibinfo {author} {\bibfnamefont {D.}~\bibnamefont {Sadek}}, \bibinfo
  {author} {\bibfnamefont {B.}~\bibnamefont {Schmidt}}, \bibinfo {author} {\bibfnamefont {Z.}~\bibnamefont {Smith}}, \bibinfo {author} {\bibfnamefont {S.}~\bibnamefont {Stevens}}, \bibinfo {author} {\bibfnamefont {K.}~\bibnamefont {Stifter}}, \bibinfo {author} {\bibfnamefont {M.}~\bibnamefont {Stukel}}, \bibinfo {author} {\bibfnamefont {J.}~\bibnamefont {Viol}}, \bibinfo {author} {\bibfnamefont {Y.}~\bibnamefont {Wang}}, \bibinfo {author} {\bibfnamefont {M.~J.}\ \bibnamefont {Wilson}}, \bibinfo {author} {\bibfnamefont {B.}~\bibnamefont {Young}}, \bibinfo {author} {\bibfnamefont {S.}~\bibnamefont {Zatschler}}, \bibinfo {author} {\bibfnamefont {H.}~\bibnamefont {Zenger}}, \ and\ \bibinfo {author} {\bibfnamefont {A.}~\bibnamefont {Zuñiga-Reyes}},\ }\href {https://arxiv.org/abs/2601.16307} {\enquote {\bibinfo {title} {Performance of a supercdms hvev detector with sub-ev energy resolution and single charge-sensitivity},}\ } (\bibinfo {year} {2026}),\ \Eprint {http://arxiv.org/abs/2601.16307} {arXiv:2601.16307
  [physics.ins-det]} \BibitemShut {NoStop}%
\bibitem [{\citenamefont {Moore}(2012)}]{Moore:2012gya}%
  \BibitemOpen
  \bibfield  {author} {\bibinfo {author} {\bibfnamefont {D.~C.}\ \bibnamefont {Moore}},\ }\emph {\bibinfo {title} {{A Search for Low-Mass Dark Matter with the Cryogenic Dark Matter Search and the Development of Highly Multiplexed Phonon-Mediated Particle Detectors}}},\ \href {\doibase 10.2172/1247705} {Ph.D. thesis},\ \bibinfo  {school} {Caltech} (\bibinfo {year} {2012})\BibitemShut {NoStop}%
\bibitem [{\citenamefont {Temples}\ \emph {et~al.}(2025)\citenamefont {Temples}, \citenamefont {Smith}, \citenamefont {Dang}, \citenamefont {Aralis}, \citenamefont {Cap}, \citenamefont {Chang}, \citenamefont {Chang}, \citenamefont {Garcia-Sciveres}, \citenamefont {Golwala}, \citenamefont {Ho}, \citenamefont {Kurinsky}, \citenamefont {Li}, \citenamefont {Li}, \citenamefont {Lisovenko}, \citenamefont {Panner}, \citenamefont {Ramanathan}, \citenamefont {Ray}, \citenamefont {Sandoval}, \citenamefont {Suzuki}, \citenamefont {Wang}, \citenamefont {Wen}, \citenamefont {Williams}, \citenamefont {Xiong},\ and\ \citenamefont {Yefremenko}}]{temples2025}%
  \BibitemOpen
  \bibfield  {author} {\bibinfo {author} {\bibfnamefont {D.~J.}\ \bibnamefont {Temples}}, \bibinfo {author} {\bibfnamefont {Z.~J.}\ \bibnamefont {Smith}}, \bibinfo {author} {\bibfnamefont {S.~Q.}\ \bibnamefont {Dang}}, \bibinfo {author} {\bibfnamefont {T.}~\bibnamefont {Aralis}}, \bibinfo {author} {\bibfnamefont {C.}~\bibnamefont {Cap}}, \bibinfo {author} {\bibfnamefont {C.}~\bibnamefont {Chang}}, \bibinfo {author} {\bibfnamefont {Y.-Y.}\ \bibnamefont {Chang}}, \bibinfo {author} {\bibfnamefont {M.}~\bibnamefont {Garcia-Sciveres}}, \bibinfo {author} {\bibfnamefont {S.}~\bibnamefont {Golwala}}, \bibinfo {author} {\bibfnamefont {W.}~\bibnamefont {Ho}}, \bibinfo {author} {\bibfnamefont {N.}~\bibnamefont {Kurinsky}}, \bibinfo {author} {\bibfnamefont {K.}~\bibnamefont {Li}}, \bibinfo {author} {\bibfnamefont {X.}~\bibnamefont {Li}}, \bibinfo {author} {\bibfnamefont {M.}~\bibnamefont {Lisovenko}}, \bibinfo {author} {\bibfnamefont {E.}~\bibnamefont {Panner}}, \bibinfo {author} {\bibfnamefont {K.}~\bibnamefont
  {Ramanathan}}, \bibinfo {author} {\bibfnamefont {S.}~\bibnamefont {Ray}}, \bibinfo {author} {\bibfnamefont {B.}~\bibnamefont {Sandoval}}, \bibinfo {author} {\bibfnamefont {A.}~\bibnamefont {Suzuki}}, \bibinfo {author} {\bibfnamefont {G.}~\bibnamefont {Wang}}, \bibinfo {author} {\bibfnamefont {O.}~\bibnamefont {Wen}}, \bibinfo {author} {\bibfnamefont {M.}~\bibnamefont {Williams}}, \bibinfo {author} {\bibfnamefont {J.~R.}\ \bibnamefont {Xiong}}, \ and\ \bibinfo {author} {\bibfnamefont {V.}~\bibnamefont {Yefremenko}},\ }\href {https://arxiv.org/abs/2509.25544} {\enquote {\bibinfo {title} {Development status of the kipm detector consortium},}\ } (\bibinfo {year} {2025}),\ \Eprint {http://arxiv.org/abs/2509.25544} {arXiv:2509.25544 [physics.ins-det]} \BibitemShut {NoStop}%
\bibitem [{\citenamefont {Kurinsky}(2018)}]{Kurinsky}%
  \BibitemOpen
  \bibfield  {author} {\bibinfo {author} {\bibfnamefont {N.~A.}\ \bibnamefont {Kurinsky}},\ }\emph {\bibinfo {title} {The low-mass limit : dark matter detectors with eV-scale energy resolution}},\ \href@noop {} {Ph.D. thesis},\ \bibinfo  {school} {Stanford University} (\bibinfo {year} {2018})\BibitemShut {NoStop}%
\bibitem [{\citenamefont {Temples}\ \emph {et~al.}(2024)\citenamefont {Temples}, \citenamefont {Wen}, \citenamefont {Ramanathan}, \citenamefont {Aralis}, \citenamefont {Chang}, \citenamefont {Golwala}, \citenamefont {Hsu}, \citenamefont {Bathurst}, \citenamefont {Baxter}, \citenamefont {Bowring}, \citenamefont {Chen}, \citenamefont {Figueroa-Feliciano}, \citenamefont {Hollister}, \citenamefont {James}, \citenamefont {Kennard}, \citenamefont {Kurinsky}, \citenamefont {Lewis}, \citenamefont {Lukens}, \citenamefont {Novati}, \citenamefont {Ren},\ and\ \citenamefont {Schmidt}}]{Temples_2024}%
  \BibitemOpen
  \bibfield  {author} {\bibinfo {author} {\bibfnamefont {D.~J.}\ \bibnamefont {Temples}}, \bibinfo {author} {\bibfnamefont {O.}~\bibnamefont {Wen}}, \bibinfo {author} {\bibfnamefont {K.}~\bibnamefont {Ramanathan}}, \bibinfo {author} {\bibfnamefont {T.}~\bibnamefont {Aralis}}, \bibinfo {author} {\bibfnamefont {Y.-Y.}\ \bibnamefont {Chang}}, \bibinfo {author} {\bibfnamefont {S.}~\bibnamefont {Golwala}}, \bibinfo {author} {\bibfnamefont {L.}~\bibnamefont {Hsu}}, \bibinfo {author} {\bibfnamefont {C.}~\bibnamefont {Bathurst}}, \bibinfo {author} {\bibfnamefont {D.}~\bibnamefont {Baxter}}, \bibinfo {author} {\bibfnamefont {D.}~\bibnamefont {Bowring}}, \bibinfo {author} {\bibfnamefont {R.}~\bibnamefont {Chen}}, \bibinfo {author} {\bibfnamefont {E.}~\bibnamefont {Figueroa-Feliciano}}, \bibinfo {author} {\bibfnamefont {M.}~\bibnamefont {Hollister}}, \bibinfo {author} {\bibfnamefont {C.}~\bibnamefont {James}}, \bibinfo {author} {\bibfnamefont {K.}~\bibnamefont {Kennard}}, \bibinfo {author} {\bibfnamefont
  {N.}~\bibnamefont {Kurinsky}}, \bibinfo {author} {\bibfnamefont {S.}~\bibnamefont {Lewis}}, \bibinfo {author} {\bibfnamefont {P.}~\bibnamefont {Lukens}}, \bibinfo {author} {\bibfnamefont {V.}~\bibnamefont {Novati}}, \bibinfo {author} {\bibfnamefont {R.}~\bibnamefont {Ren}}, \ and\ \bibinfo {author} {\bibfnamefont {B.}~\bibnamefont {Schmidt}},\ }\href {\doibase 10.1103/physrevapplied.22.044045} {\bibfield  {journal} {\bibinfo  {journal} {Physical Review Applied}\ }\textbf {\bibinfo {volume} {22}} (\bibinfo {year} {2024}),\ 10.1103/physrevapplied.22.044045}\BibitemShut {NoStop}%
\bibitem [{\citenamefont {Chang}\ \emph {et~al.}(2018)\citenamefont {Chang}, \citenamefont {Cornell}, \citenamefont {Aralis}, \citenamefont {Bumble},\ and\ \citenamefont {Golwala}}]{Chang_2018}%
  \BibitemOpen
  \bibfield  {author} {\bibinfo {author} {\bibfnamefont {Y.}~\bibnamefont {Chang}}, \bibinfo {author} {\bibfnamefont {B.}~\bibnamefont {Cornell}}, \bibinfo {author} {\bibfnamefont {T.}~\bibnamefont {Aralis}}, \bibinfo {author} {\bibfnamefont {B.}~\bibnamefont {Bumble}}, \ and\ \bibinfo {author} {\bibfnamefont {S.}~\bibnamefont {Golwala}},\ }\href {\doibase 10.1007/s10909-018-1900-9} {\bibfield  {journal} {\bibinfo  {journal} {Journal of Low Temperature Physics}\ }\textbf {\bibinfo {volume} {193}} (\bibinfo {year} {2018}),\ 10.1007/s10909-018-1900-9}\BibitemShut {NoStop}%
\bibitem [{\citenamefont {Chang}(2023)}]{Chang_2023}%
  \BibitemOpen
  \bibfield  {author} {\bibinfo {author} {\bibfnamefont {Y.-Y.}\ \bibnamefont {Chang}},\ }\emph {\bibinfo {title} {{SuperCDMS} {HVeV} Run 2 low-mass dark matter search, highly multiplexed phonon-mediated particle detector with kinetic inductance detector, and the blackbody radiation in cryogenic experiments}},\ \href@noop {} {Ph.D. thesis},\ \bibinfo  {school} {California Institute of Technology} (\bibinfo {year} {2023})\BibitemShut {NoStop}%
\bibitem [{\citenamefont {Wen}(2025)}]{Wen2025}%
  \BibitemOpen
  \bibfield  {author} {\bibinfo {author} {\bibfnamefont {O.}~\bibnamefont {Wen}},\ }\emph {\bibinfo {title} {{Strategic Planning and Sensitivity-Enhancing Tactics for Detecting Low-Mass Particle Dark Matter with Phonon-Mediated Detectors}}},\ \href {\doibase 10.7907/g23k-c067} {Ph.D. thesis},\ \bibinfo  {school} {Caltech, Pasadena (main)} (\bibinfo {year} {2025})\BibitemShut {NoStop}%
\bibitem [{\citenamefont {Tabassum}\ \emph {et~al.}(2025)\citenamefont {Tabassum}, \citenamefont {Aralis}, \citenamefont {Anczarski}, \citenamefont {Baxter}, \citenamefont {Cabrera}, \citenamefont {Chapla}, \citenamefont {Entin}, \citenamefont {Hsu}, \citenamefont {Magoon}, \citenamefont {Nunez}, \citenamefont {Ryan}, \citenamefont {Salatino}, \citenamefont {Simchony}, \citenamefont {Smith}, \citenamefont {Stevens}, \citenamefont {Perez}, \citenamefont {Stueber}, \citenamefont {Young}, \citenamefont {Kurinsky},\ and\ \citenamefont {Stifter}}]{tabassum2025broadbandopticalmodulationcontrol}%
  \BibitemOpen
  \bibfield  {author} {\bibinfo {author} {\bibfnamefont {N.}~\bibnamefont {Tabassum}}, \bibinfo {author} {\bibfnamefont {T.}~\bibnamefont {Aralis}}, \bibinfo {author} {\bibfnamefont {J.}~\bibnamefont {Anczarski}}, \bibinfo {author} {\bibfnamefont {D.}~\bibnamefont {Baxter}}, \bibinfo {author} {\bibfnamefont {B.}~\bibnamefont {Cabrera}}, \bibinfo {author} {\bibfnamefont {R.}~\bibnamefont {Chapla}}, \bibinfo {author} {\bibfnamefont {N.}~\bibnamefont {Entin}}, \bibinfo {author} {\bibfnamefont {L.}~\bibnamefont {Hsu}}, \bibinfo {author} {\bibfnamefont {H.~W.}\ \bibnamefont {Magoon}}, \bibinfo {author} {\bibfnamefont {A.}~\bibnamefont {Nunez}}, \bibinfo {author} {\bibfnamefont {J.~L.}\ \bibnamefont {Ryan}}, \bibinfo {author} {\bibfnamefont {M.}~\bibnamefont {Salatino}}, \bibinfo {author} {\bibfnamefont {A.}~\bibnamefont {Simchony}}, \bibinfo {author} {\bibfnamefont {Z.~J.}\ \bibnamefont {Smith}}, \bibinfo {author} {\bibfnamefont {S.}~\bibnamefont {Stevens}}, \bibinfo {author} {\bibfnamefont {G.}~\bibnamefont
  {Perez}}, \bibinfo {author} {\bibfnamefont {H.}~\bibnamefont {Stueber}}, \bibinfo {author} {\bibfnamefont {B.~A.}\ \bibnamefont {Young}}, \bibinfo {author} {\bibfnamefont {N.~A.}\ \bibnamefont {Kurinsky}}, \ and\ \bibinfo {author} {\bibfnamefont {K.}~\bibnamefont {Stifter}},\ }\href {https://arxiv.org/abs/2504.06995} {\enquote {\bibinfo {title} {Broadband optical modulation and control at millikelvin temperatures},}\ } (\bibinfo {year} {2025}),\ \Eprint {http://arxiv.org/abs/2504.06995} {arXiv:2504.06995 [physics.ins-det]} \BibitemShut {NoStop}%
\bibitem [{\citenamefont {Stifter}\ \emph {et~al.}(2025)\citenamefont {Stifter}, \citenamefont {Magoon}, \citenamefont {Anderson}, \citenamefont {Temples}, \citenamefont {Kurinsky}, \citenamefont {Stoughton}, \citenamefont {Hernandez}, \citenamefont {Nuñez}, \citenamefont {Anyang}, \citenamefont {Linehan}, \citenamefont {Young}, \citenamefont {Barry}, \citenamefont {Baxter}, \citenamefont {Bowring}, \citenamefont {Cancelo}, \citenamefont {Chou}, \citenamefont {Dibert}, \citenamefont {Figueroa-Feliciano}, \citenamefont {Hsu}, \citenamefont {Khatiwada}, \citenamefont {Mork}, \citenamefont {Stefanazzi}, \citenamefont {Tabassum}, \citenamefont {Uemura},\ and\ \citenamefont {Young}}]{Stifter_2025}%
  \BibitemOpen
  \bibfield  {author} {\bibinfo {author} {\bibfnamefont {K.}~\bibnamefont {Stifter}}, \bibinfo {author} {\bibfnamefont {H.}~\bibnamefont {Magoon}}, \bibinfo {author} {\bibfnamefont {A.~J.}\ \bibnamefont {Anderson}}, \bibinfo {author} {\bibfnamefont {D.~J.}\ \bibnamefont {Temples}}, \bibinfo {author} {\bibfnamefont {N.~A.}\ \bibnamefont {Kurinsky}}, \bibinfo {author} {\bibfnamefont {C.}~\bibnamefont {Stoughton}}, \bibinfo {author} {\bibfnamefont {I.}~\bibnamefont {Hernandez}}, \bibinfo {author} {\bibfnamefont {A.}~\bibnamefont {Nuñez}}, \bibinfo {author} {\bibfnamefont {K.}~\bibnamefont {Anyang}}, \bibinfo {author} {\bibfnamefont {R.}~\bibnamefont {Linehan}}, \bibinfo {author} {\bibfnamefont {M.~R.}\ \bibnamefont {Young}}, \bibinfo {author} {\bibfnamefont {P.}~\bibnamefont {Barry}}, \bibinfo {author} {\bibfnamefont {D.}~\bibnamefont {Baxter}}, \bibinfo {author} {\bibfnamefont {D.}~\bibnamefont {Bowring}}, \bibinfo {author} {\bibfnamefont {G.}~\bibnamefont {Cancelo}}, \bibinfo {author} {\bibfnamefont
  {A.}~\bibnamefont {Chou}}, \bibinfo {author} {\bibfnamefont {K.~R.}\ \bibnamefont {Dibert}}, \bibinfo {author} {\bibfnamefont {E.}~\bibnamefont {Figueroa-Feliciano}}, \bibinfo {author} {\bibfnamefont {L.}~\bibnamefont {Hsu}}, \bibinfo {author} {\bibfnamefont {R.}~\bibnamefont {Khatiwada}}, \bibinfo {author} {\bibfnamefont {S.~D.}\ \bibnamefont {Mork}}, \bibinfo {author} {\bibfnamefont {L.}~\bibnamefont {Stefanazzi}}, \bibinfo {author} {\bibfnamefont {N.}~\bibnamefont {Tabassum}}, \bibinfo {author} {\bibfnamefont {S.}~\bibnamefont {Uemura}}, \ and\ \bibinfo {author} {\bibfnamefont {B.~A.}\ \bibnamefont {Young}},\ }\href {\doibase 10.1117/1.jom.5.2.024503} {\bibfield  {journal} {\bibinfo  {journal} {Journal of Optical Microsystems}\ }\textbf {\bibinfo {volume} {5}} (\bibinfo {year} {2025}),\ 10.1117/1.jom.5.2.024503}\BibitemShut {NoStop}%
\bibitem [{\citenamefont {Yu}\ \emph {et~al.}(2023)\citenamefont {Yu}, \citenamefont {Ahmed}, \citenamefont {Frisch}, \citenamefont {Henderson}, \citenamefont {Silva-Feaver}, \citenamefont {Arnold}, \citenamefont {Brown}, \citenamefont {Connors}, \citenamefont {Cukierman}, \citenamefont {D’Ewart}, \citenamefont {Dober}, \citenamefont {Dusatko}, \citenamefont {Haller}, \citenamefont {Herbst}, \citenamefont {Hilton}, \citenamefont {Hubmayr}, \citenamefont {Irwin}, \citenamefont {Kuo}, \citenamefont {Mates}, \citenamefont {Ruckman}, \citenamefont {Ullom}, \citenamefont {Vale}, \citenamefont {Van~Winkle}, \citenamefont {Vasquez},\ and\ \citenamefont {Young}}]{Yu2023}%
  \BibitemOpen
  \bibfield  {author} {\bibinfo {author} {\bibfnamefont {C.}~\bibnamefont {Yu}}, \bibinfo {author} {\bibfnamefont {Z.}~\bibnamefont {Ahmed}}, \bibinfo {author} {\bibfnamefont {J.~C.}\ \bibnamefont {Frisch}}, \bibinfo {author} {\bibfnamefont {S.~W.}\ \bibnamefont {Henderson}}, \bibinfo {author} {\bibfnamefont {M.}~\bibnamefont {Silva-Feaver}}, \bibinfo {author} {\bibfnamefont {K.}~\bibnamefont {Arnold}}, \bibinfo {author} {\bibfnamefont {D.}~\bibnamefont {Brown}}, \bibinfo {author} {\bibfnamefont {J.}~\bibnamefont {Connors}}, \bibinfo {author} {\bibfnamefont {A.~J.}\ \bibnamefont {Cukierman}}, \bibinfo {author} {\bibfnamefont {J.~M.}\ \bibnamefont {D’Ewart}}, \bibinfo {author} {\bibfnamefont {B.~J.}\ \bibnamefont {Dober}}, \bibinfo {author} {\bibfnamefont {J.~E.}\ \bibnamefont {Dusatko}}, \bibinfo {author} {\bibfnamefont {G.}~\bibnamefont {Haller}}, \bibinfo {author} {\bibfnamefont {R.}~\bibnamefont {Herbst}}, \bibinfo {author} {\bibfnamefont {G.~C.}\ \bibnamefont {Hilton}}, \bibinfo {author} {\bibfnamefont
  {J.}~\bibnamefont {Hubmayr}}, \bibinfo {author} {\bibfnamefont {K.~D.}\ \bibnamefont {Irwin}}, \bibinfo {author} {\bibfnamefont {C.-L.}\ \bibnamefont {Kuo}}, \bibinfo {author} {\bibfnamefont {J.~A.~B.}\ \bibnamefont {Mates}}, \bibinfo {author} {\bibfnamefont {L.}~\bibnamefont {Ruckman}}, \bibinfo {author} {\bibfnamefont {J.}~\bibnamefont {Ullom}}, \bibinfo {author} {\bibfnamefont {L.}~\bibnamefont {Vale}}, \bibinfo {author} {\bibfnamefont {D.~D.}\ \bibnamefont {Van~Winkle}}, \bibinfo {author} {\bibfnamefont {J.}~\bibnamefont {Vasquez}}, \ and\ \bibinfo {author} {\bibfnamefont {E.}~\bibnamefont {Young}},\ }\href {\doibase 10.1063/5.0125084} {\bibfield  {journal} {\bibinfo  {journal} {Review of Scientific Instruments}\ }\textbf {\bibinfo {volume} {94}},\ \bibinfo {pages} {014712} (\bibinfo {year} {2023})}\BibitemShut {NoStop}%
\bibitem [{\citenamefont {Watkins}(2024)}]{Watkins2024}%
  \BibitemOpen
  \bibfield  {author} {\bibinfo {author} {\bibfnamefont {S.~L.}\ \bibnamefont {Watkins}},\ }\href {\doibase 10.1007/s10909-023-03021-w} {\bibfield  {journal} {\bibinfo  {journal} {J. Low Temp. Phys.}\ }\textbf {\bibinfo {volume} {214}},\ \bibinfo {pages} {133} (\bibinfo {year} {2024})},\ \Eprint {http://arxiv.org/abs/2310.01279} {arXiv:2310.01279 [physics.ins-det]} \BibitemShut {NoStop}%
\bibitem [{\citenamefont {Khalil}\ \emph {et~al.}(2012)\citenamefont {Khalil}, \citenamefont {Stoutimore}, \citenamefont {Wellstood},\ and\ \citenamefont {Osborn}}]{Khalil2012}%
  \BibitemOpen
  \bibfield  {author} {\bibinfo {author} {\bibfnamefont {M.~S.}\ \bibnamefont {Khalil}}, \bibinfo {author} {\bibfnamefont {M.~J.~A.}\ \bibnamefont {Stoutimore}}, \bibinfo {author} {\bibfnamefont {F.~C.}\ \bibnamefont {Wellstood}}, \ and\ \bibinfo {author} {\bibfnamefont {K.~D.}\ \bibnamefont {Osborn}},\ }\href {\doibase 10.1063/1.3692073} {\bibfield  {journal} {\bibinfo  {journal} {Journal of Applied Physics}\ }\textbf {\bibinfo {volume} {111}},\ \bibinfo {pages} {054510} (\bibinfo {year} {2012})}\BibitemShut {NoStop}%
\bibitem [{\citenamefont {Kelsey}\ \emph {et~al.}(2023)\citenamefont {Kelsey}, \citenamefont {Agnese}, \citenamefont {Alam}, \citenamefont {Langroudy}, \citenamefont {Azadbakht}, \citenamefont {Brandt}, \citenamefont {Bunker}, \citenamefont {Cabrera}, \citenamefont {Chang}, \citenamefont {Coombes}, \citenamefont {Cormier}, \citenamefont {Diamond}, \citenamefont {Edwards}, \citenamefont {Figueroa-Feliciano}, \citenamefont {Gao}, \citenamefont {Harrington}, \citenamefont {Hong}, \citenamefont {Hui}, \citenamefont {Kurinsky}, \citenamefont {Lawrence}, \citenamefont {Loer}, \citenamefont {Masten}, \citenamefont {Michaud}, \citenamefont {Michielin}, \citenamefont {Miller}, \citenamefont {Novati}, \citenamefont {Oblath}, \citenamefont {Orrell}, \citenamefont {Perry}, \citenamefont {Redl}, \citenamefont {Reynolds}, \citenamefont {Saab}, \citenamefont {Sadoulet}, \citenamefont {Serniak}, \citenamefont {Singh}, \citenamefont {Speaks}, \citenamefont {Stanford}, \citenamefont {Stevens}, \citenamefont {Strube},
  \citenamefont {Toback}, \citenamefont {Ullom}, \citenamefont {VanDevender}, \citenamefont {Vissers}, \citenamefont {Wilson}, \citenamefont {Wilson}, \citenamefont {Zatschler},\ and\ \citenamefont {Zatschler}}]{KELSEY2023168473}%
  \BibitemOpen
  \bibfield  {author} {\bibinfo {author} {\bibfnamefont {M.}~\bibnamefont {Kelsey}}, \bibinfo {author} {\bibfnamefont {R.}~\bibnamefont {Agnese}}, \bibinfo {author} {\bibfnamefont {Y.}~\bibnamefont {Alam}}, \bibinfo {author} {\bibfnamefont {I.~A.}\ \bibnamefont {Langroudy}}, \bibinfo {author} {\bibfnamefont {E.}~\bibnamefont {Azadbakht}}, \bibinfo {author} {\bibfnamefont {D.}~\bibnamefont {Brandt}}, \bibinfo {author} {\bibfnamefont {R.}~\bibnamefont {Bunker}}, \bibinfo {author} {\bibfnamefont {B.}~\bibnamefont {Cabrera}}, \bibinfo {author} {\bibfnamefont {Y.-Y.}\ \bibnamefont {Chang}}, \bibinfo {author} {\bibfnamefont {H.}~\bibnamefont {Coombes}}, \bibinfo {author} {\bibfnamefont {R.}~\bibnamefont {Cormier}}, \bibinfo {author} {\bibfnamefont {M.}~\bibnamefont {Diamond}}, \bibinfo {author} {\bibfnamefont {E.}~\bibnamefont {Edwards}}, \bibinfo {author} {\bibfnamefont {E.}~\bibnamefont {Figueroa-Feliciano}}, \bibinfo {author} {\bibfnamefont {J.}~\bibnamefont {Gao}}, \bibinfo {author} {\bibfnamefont
  {P.}~\bibnamefont {Harrington}}, \bibinfo {author} {\bibfnamefont {Z.}~\bibnamefont {Hong}}, \bibinfo {author} {\bibfnamefont {M.}~\bibnamefont {Hui}}, \bibinfo {author} {\bibfnamefont {N.}~\bibnamefont {Kurinsky}}, \bibinfo {author} {\bibfnamefont {R.}~\bibnamefont {Lawrence}}, \bibinfo {author} {\bibfnamefont {B.}~\bibnamefont {Loer}}, \bibinfo {author} {\bibfnamefont {M.}~\bibnamefont {Masten}}, \bibinfo {author} {\bibfnamefont {E.}~\bibnamefont {Michaud}}, \bibinfo {author} {\bibfnamefont {E.}~\bibnamefont {Michielin}}, \bibinfo {author} {\bibfnamefont {J.}~\bibnamefont {Miller}}, \bibinfo {author} {\bibfnamefont {V.}~\bibnamefont {Novati}}, \bibinfo {author} {\bibfnamefont {N.}~\bibnamefont {Oblath}}, \bibinfo {author} {\bibfnamefont {J.}~\bibnamefont {Orrell}}, \bibinfo {author} {\bibfnamefont {W.}~\bibnamefont {Perry}}, \bibinfo {author} {\bibfnamefont {P.}~\bibnamefont {Redl}}, \bibinfo {author} {\bibfnamefont {T.}~\bibnamefont {Reynolds}}, \bibinfo {author} {\bibfnamefont {T.}~\bibnamefont {Saab}},
  \bibinfo {author} {\bibfnamefont {B.}~\bibnamefont {Sadoulet}}, \bibinfo {author} {\bibfnamefont {K.}~\bibnamefont {Serniak}}, \bibinfo {author} {\bibfnamefont {J.}~\bibnamefont {Singh}}, \bibinfo {author} {\bibfnamefont {Z.}~\bibnamefont {Speaks}}, \bibinfo {author} {\bibfnamefont {C.}~\bibnamefont {Stanford}}, \bibinfo {author} {\bibfnamefont {J.}~\bibnamefont {Stevens}}, \bibinfo {author} {\bibfnamefont {J.}~\bibnamefont {Strube}}, \bibinfo {author} {\bibfnamefont {D.}~\bibnamefont {Toback}}, \bibinfo {author} {\bibfnamefont {J.}~\bibnamefont {Ullom}}, \bibinfo {author} {\bibfnamefont {B.}~\bibnamefont {VanDevender}}, \bibinfo {author} {\bibfnamefont {M.}~\bibnamefont {Vissers}}, \bibinfo {author} {\bibfnamefont {M.}~\bibnamefont {Wilson}}, \bibinfo {author} {\bibfnamefont {J.}~\bibnamefont {Wilson}}, \bibinfo {author} {\bibfnamefont {B.}~\bibnamefont {Zatschler}}, \ and\ \bibinfo {author} {\bibfnamefont {S.}~\bibnamefont {Zatschler}},\ }\href {\doibase https://doi.org/10.1016/j.nima.2023.168473}
  {\bibfield  {journal} {\bibinfo  {journal} {Nuclear Instruments and Methods in Physics Research Section A: Accelerators, Spectrometers, Detectors and Associated Equipment}\ }\textbf {\bibinfo {volume} {1055}},\ \bibinfo {pages} {168473} (\bibinfo {year} {2023})}\BibitemShut {NoStop}%
\bibitem [{\citenamefont {{Gao}}(2008)}]{GaoThesis}%
  \BibitemOpen
  \bibfield  {author} {\bibinfo {author} {\bibfnamefont {J.}~\bibnamefont {{Gao}}},\ }\emph {\bibinfo {title} {{The Physics of Superconducting Microwave Resonators}}},\ \href@noop {} {Ph.D. thesis},\ \bibinfo  {school} {California Institute of Technology} (\bibinfo {year} {2008})\BibitemShut {NoStop}%
\bibitem [{\citenamefont {Gao}\ \emph {et~al.}(2008)\citenamefont {Gao}, \citenamefont {Zmuidzinas}, \citenamefont {Vayonakis}, \citenamefont {Day}, \citenamefont {Mazin},\ and\ \citenamefont {Leduc}}]{Gao2008}%
  \BibitemOpen
  \bibfield  {author} {\bibinfo {author} {\bibfnamefont {J.}~\bibnamefont {Gao}}, \bibinfo {author} {\bibfnamefont {J.}~\bibnamefont {Zmuidzinas}}, \bibinfo {author} {\bibfnamefont {A.}~\bibnamefont {Vayonakis}}, \bibinfo {author} {\bibfnamefont {P.}~\bibnamefont {Day}}, \bibinfo {author} {\bibfnamefont {B.}~\bibnamefont {Mazin}}, \ and\ \bibinfo {author} {\bibfnamefont {H.}~\bibnamefont {Leduc}},\ }\href {\doibase 10.1007/s10909-007-9688-z} {\bibfield  {journal} {\bibinfo  {journal} {Journal of Low Temperature Physics}\ }\textbf {\bibinfo {volume} {151}},\ \bibinfo {pages} {557} (\bibinfo {year} {2008})}\BibitemShut {NoStop}%
\bibitem [{\citenamefont {Nam}(1967)}]{Nam1967}%
  \BibitemOpen
  \bibfield  {author} {\bibinfo {author} {\bibfnamefont {S.~B.}\ \bibnamefont {Nam}},\ }\href {\doibase 10.1103/PhysRev.156.470} {\bibfield  {journal} {\bibinfo  {journal} {Phys. Rev.}\ }\textbf {\bibinfo {volume} {156}},\ \bibinfo {pages} {470} (\bibinfo {year} {1967})}\BibitemShut {NoStop}%
\bibitem [{\citenamefont {Li}, \citenamefont {Suzuki},\ and\ \citenamefont {Garcia-Sciveres}(2025)}]{li2025}%
  \BibitemOpen
  \bibfield  {author} {\bibinfo {author} {\bibfnamefont {X.}~\bibnamefont {Li}}, \bibinfo {author} {\bibfnamefont {A.}~\bibnamefont {Suzuki}}, \ and\ \bibinfo {author} {\bibfnamefont {M.}~\bibnamefont {Garcia-Sciveres}},\ }\href {https://arxiv.org/abs/2502.19818} {\enquote {\bibinfo {title} {Low tc hafnium kinetic inductance device with high internal quality factor},}\ } (\bibinfo {year} {2025}),\ \Eprint {http://arxiv.org/abs/2502.19818} {arXiv:2502.19818 [hep-ex]} \BibitemShut {NoStop}%
\bibitem [{\citenamefont {Zobrist}\ \emph {et~al.}(2019)\citenamefont {Zobrist}, \citenamefont {Coiffard}, \citenamefont {Bumble}, \citenamefont {Swimmer}, \citenamefont {Steiger}, \citenamefont {Daal}, \citenamefont {Collura}, \citenamefont {Walter}, \citenamefont {Bockstiegel}, \citenamefont {Fruitwala}, \citenamefont {Lipartito},\ and\ \citenamefont {Mazin}}]{Zobrist2019Hafnium}%
  \BibitemOpen
  \bibfield  {author} {\bibinfo {author} {\bibfnamefont {N.}~\bibnamefont {Zobrist}}, \bibinfo {author} {\bibfnamefont {G.}~\bibnamefont {Coiffard}}, \bibinfo {author} {\bibfnamefont {B.}~\bibnamefont {Bumble}}, \bibinfo {author} {\bibfnamefont {N.}~\bibnamefont {Swimmer}}, \bibinfo {author} {\bibfnamefont {S.}~\bibnamefont {Steiger}}, \bibinfo {author} {\bibfnamefont {M.}~\bibnamefont {Daal}}, \bibinfo {author} {\bibfnamefont {G.}~\bibnamefont {Collura}}, \bibinfo {author} {\bibfnamefont {A.~B.}\ \bibnamefont {Walter}}, \bibinfo {author} {\bibfnamefont {C.}~\bibnamefont {Bockstiegel}}, \bibinfo {author} {\bibfnamefont {N.}~\bibnamefont {Fruitwala}}, \bibinfo {author} {\bibfnamefont {I.}~\bibnamefont {Lipartito}}, \ and\ \bibinfo {author} {\bibfnamefont {B.~A.}\ \bibnamefont {Mazin}},\ }\href {\doibase 10.1063/1.5127768} {\bibfield  {journal} {\bibinfo  {journal} {Applied Physics Letters}\ }\textbf {\bibinfo {volume} {115}},\ \bibinfo {pages} {213503} (\bibinfo {year} {2019})},\ \Eprint
  {http://arxiv.org/abs/1911.06434} {arXiv:1911.06434 [physics.ins-det]} \BibitemShut {NoStop}%
\bibitem [{\citenamefont {Zobrist}\ \emph {et~al.}(2021)\citenamefont {Zobrist}, \citenamefont {Klimovich}, \citenamefont {Eom}, \citenamefont {Coiffard}, \citenamefont {Daal}, \citenamefont {Swimmer}, \citenamefont {Steiger}, \citenamefont {Bumble}, \citenamefont {LeDuc}, \citenamefont {Day},\ and\ \citenamefont {Mazin}}]{Zobrist2021}%
  \BibitemOpen
  \bibfield  {author} {\bibinfo {author} {\bibfnamefont {N.}~\bibnamefont {Zobrist}}, \bibinfo {author} {\bibfnamefont {N.}~\bibnamefont {Klimovich}}, \bibinfo {author} {\bibfnamefont {B.}~\bibnamefont {Eom}}, \bibinfo {author} {\bibfnamefont {G.}~\bibnamefont {Coiffard}}, \bibinfo {author} {\bibfnamefont {M.}~\bibnamefont {Daal}}, \bibinfo {author} {\bibfnamefont {N.~J.}\ \bibnamefont {Swimmer}}, \bibinfo {author} {\bibfnamefont {S.}~\bibnamefont {Steiger}}, \bibinfo {author} {\bibfnamefont {B.}~\bibnamefont {Bumble}}, \bibinfo {author} {\bibfnamefont {H.~G.}\ \bibnamefont {LeDuc}}, \bibinfo {author} {\bibfnamefont {P.~K.}\ \bibnamefont {Day}}, \ and\ \bibinfo {author} {\bibfnamefont {B.~A.}\ \bibnamefont {Mazin}},\ }\href {\doibase 10.1117/1.JATIS.7.1.010501} {\bibfield  {journal} {\bibinfo  {journal} {Journal of Astronomical Telescopes, Instruments, and Systems}\ }\textbf {\bibinfo {volume} {7}},\ \bibinfo {pages} {010501} (\bibinfo {year} {2021})}\BibitemShut {NoStop}%
\bibitem [{\citenamefont {Florang}\ \emph {et~al.}(2026)\citenamefont {Florang}, \citenamefont {Dean}, \citenamefont {Fowler}, \citenamefont {Haugen}, \citenamefont {Hoogerheide}, \citenamefont {Jardin}, \citenamefont {Mumm}, \citenamefont {Nakamura}, \citenamefont {Natale}, \citenamefont {O'Neil}, \citenamefont {Ortiz}, \citenamefont {Paster}, \citenamefont {Rao}, \citenamefont {Roy}, \citenamefont {Swetz}, \citenamefont {Ullom}, \citenamefont {Vissers},\ and\ \citenamefont {Szypryt}}]{Florang2026}%
  \BibitemOpen
  \bibfield  {author} {\bibinfo {author} {\bibfnamefont {I.~F.}\ \bibnamefont {Florang}}, \bibinfo {author} {\bibfnamefont {J.~W.}\ \bibnamefont {Dean}}, \bibinfo {author} {\bibfnamefont {J.~W.}\ \bibnamefont {Fowler}}, \bibinfo {author} {\bibfnamefont {T.-E.}\ \bibnamefont {Haugen}}, \bibinfo {author} {\bibfnamefont {S.~F.}\ \bibnamefont {Hoogerheide}}, \bibinfo {author} {\bibfnamefont {D.}~\bibnamefont {Jardin}}, \bibinfo {author} {\bibfnamefont {H.~P.}\ \bibnamefont {Mumm}}, \bibinfo {author} {\bibfnamefont {N.}~\bibnamefont {Nakamura}}, \bibinfo {author} {\bibfnamefont {M.~R.}\ \bibnamefont {Natale}}, \bibinfo {author} {\bibfnamefont {G.~C.}\ \bibnamefont {O'Neil}}, \bibinfo {author} {\bibfnamefont {N.~J.}\ \bibnamefont {Ortiz}}, \bibinfo {author} {\bibfnamefont {J.}~\bibnamefont {Paster}}, \bibinfo {author} {\bibfnamefont {T.~M.}\ \bibnamefont {Rao}}, \bibinfo {author} {\bibfnamefont {A.}~\bibnamefont {Roy}}, \bibinfo {author} {\bibfnamefont {D.~S.}\ \bibnamefont {Swetz}}, \bibinfo {author}
  {\bibfnamefont {J.~N.}\ \bibnamefont {Ullom}}, \bibinfo {author} {\bibfnamefont {M.~R.}\ \bibnamefont {Vissers}}, \ and\ \bibinfo {author} {\bibfnamefont {P.}~\bibnamefont {Szypryt}},\ }\href {\doibase 10.1109/TASC.2026.3680384} {\bibfield  {journal} {\bibinfo  {journal} {IEEE Transactions on Applied Superconductivity}\ }\textbf {\bibinfo {volume} {36}},\ \bibinfo {pages} {1} (\bibinfo {year} {2026})}\BibitemShut {NoStop}%
\bibitem [{\citenamefont {Moore}\ \emph {et~al.}(2012)\citenamefont {Moore}, \citenamefont {Golwala}, \citenamefont {Bumble}, \citenamefont {Cornell}, \citenamefont {Day}, \citenamefont {LeDuc},\ and\ \citenamefont {Zmuidzinas}}]{Moore2012}%
  \BibitemOpen
  \bibfield  {author} {\bibinfo {author} {\bibfnamefont {D.~C.}\ \bibnamefont {Moore}}, \bibinfo {author} {\bibfnamefont {S.~R.}\ \bibnamefont {Golwala}}, \bibinfo {author} {\bibfnamefont {B.}~\bibnamefont {Bumble}}, \bibinfo {author} {\bibfnamefont {B.}~\bibnamefont {Cornell}}, \bibinfo {author} {\bibfnamefont {P.~K.}\ \bibnamefont {Day}}, \bibinfo {author} {\bibfnamefont {H.~G.}\ \bibnamefont {LeDuc}}, \ and\ \bibinfo {author} {\bibfnamefont {J.}~\bibnamefont {Zmuidzinas}},\ }\href {\doibase 10.1063/1.4726279} {\bibfield  {journal} {\bibinfo  {journal} {Applied Physics Letters}\ }\textbf {\bibinfo {volume} {100}},\ \bibinfo {pages} {232601} (\bibinfo {year} {2012})}\BibitemShut {NoStop}%
\bibitem [{\citenamefont {Phillips}(1959)}]{Al_C}%
  \BibitemOpen
  \bibfield  {author} {\bibinfo {author} {\bibfnamefont {N.~E.}\ \bibnamefont {Phillips}},\ }\href {\doibase 10.1103/PhysRev.114.676} {\bibfield  {journal} {\bibinfo  {journal} {Phys. Rev.}\ }\textbf {\bibinfo {volume} {114}},\ \bibinfo {pages} {676} (\bibinfo {year} {1959})}\BibitemShut {NoStop}%
\bibitem [{\citenamefont {{Zmuidzinas}}(2012)}]{Zmuidzinas2012}%
  \BibitemOpen
  \bibfield  {author} {\bibinfo {author} {\bibfnamefont {J.}~\bibnamefont {{Zmuidzinas}}},\ }\href {\doibase 10.1146/annurev-conmatphys-020911-125022} {\bibfield  {journal} {\bibinfo  {journal} {ARCMP}\ }\textbf {\bibinfo {volume} {3}},\ \bibinfo {pages} {169} (\bibinfo {year} {2012})}\BibitemShut {NoStop}%
\bibitem [{\citenamefont {Tamura}(1985)}]{Tamura}%
  \BibitemOpen
  \bibfield  {author} {\bibinfo {author} {\bibfnamefont {S.-i.}\ \bibnamefont {Tamura}},\ }\href {\doibase 10.1103/PhysRevB.31.2574} {\bibfield  {journal} {\bibinfo  {journal} {Phys. Rev. B}\ }\textbf {\bibinfo {volume} {31}},\ \bibinfo {pages} {2574(R)} (\bibinfo {year} {1985})}\BibitemShut {NoStop}%
\bibitem [{\citenamefont {Herring}(1954)}]{Herring1954}%
  \BibitemOpen
  \bibfield  {author} {\bibinfo {author} {\bibfnamefont {C.}~\bibnamefont {Herring}},\ }\href {\doibase 10.1103/PhysRev.95.954} {\bibfield  {journal} {\bibinfo  {journal} {Physical Review}\ }\textbf {\bibinfo {volume} {95}},\ \bibinfo {pages} {954} (\bibinfo {year} {1954})}\BibitemShut {NoStop}%
\bibitem [{\citenamefont {Ren}\ \emph {et~al.}(2021)\citenamefont {Ren}, \citenamefont {Bathurst}, \citenamefont {Chang}, \citenamefont {Chen}, \citenamefont {Fink}, \citenamefont {Hong}, \citenamefont {Kurinsky}, \citenamefont {Mast}, \citenamefont {Mishra}, \citenamefont {Novati}, \citenamefont {Spahn}, \citenamefont {Meyer~zu Theenhausen}, \citenamefont {Watkins}, \citenamefont {Williams}, \citenamefont {Wilson}, \citenamefont {Zaytsev}, \citenamefont {Bauer}, \citenamefont {Bunker}, \citenamefont {Figueroa-Feliciano}, \citenamefont {Hollister}, \citenamefont {Hsu}, \citenamefont {Lukens}, \citenamefont {Mahapatra}, \citenamefont {Mirabolfathi}, \citenamefont {Nebolsky}, \citenamefont {Platt}, \citenamefont {Ponce}, \citenamefont {Pyle}, \citenamefont {Reynolds},\ and\ \citenamefont {Saab}}]{Ren_2021}%
  \BibitemOpen
  \bibfield  {author} {\bibinfo {author} {\bibfnamefont {R.}~\bibnamefont {Ren}}, \bibinfo {author} {\bibfnamefont {C.}~\bibnamefont {Bathurst}}, \bibinfo {author} {\bibfnamefont {Y.}~\bibnamefont {Chang}}, \bibinfo {author} {\bibfnamefont {R.}~\bibnamefont {Chen}}, \bibinfo {author} {\bibfnamefont {C.}~\bibnamefont {Fink}}, \bibinfo {author} {\bibfnamefont {Z.}~\bibnamefont {Hong}}, \bibinfo {author} {\bibfnamefont {N.}~\bibnamefont {Kurinsky}}, \bibinfo {author} {\bibfnamefont {N.}~\bibnamefont {Mast}}, \bibinfo {author} {\bibfnamefont {N.}~\bibnamefont {Mishra}}, \bibinfo {author} {\bibfnamefont {V.}~\bibnamefont {Novati}}, \bibinfo {author} {\bibfnamefont {G.}~\bibnamefont {Spahn}}, \bibinfo {author} {\bibfnamefont {H.}~\bibnamefont {Meyer~zu Theenhausen}}, \bibinfo {author} {\bibfnamefont {S.}~\bibnamefont {Watkins}}, \bibinfo {author} {\bibfnamefont {Z.}~\bibnamefont {Williams}}, \bibinfo {author} {\bibfnamefont {M.}~\bibnamefont {Wilson}}, \bibinfo {author} {\bibfnamefont {A.}~\bibnamefont {Zaytsev}},
  \bibinfo {author} {\bibfnamefont {D.}~\bibnamefont {Bauer}}, \bibinfo {author} {\bibfnamefont {R.}~\bibnamefont {Bunker}}, \bibinfo {author} {\bibfnamefont {E.}~\bibnamefont {Figueroa-Feliciano}}, \bibinfo {author} {\bibfnamefont {M.}~\bibnamefont {Hollister}}, \bibinfo {author} {\bibfnamefont {L.}~\bibnamefont {Hsu}}, \bibinfo {author} {\bibfnamefont {P.}~\bibnamefont {Lukens}}, \bibinfo {author} {\bibfnamefont {R.}~\bibnamefont {Mahapatra}}, \bibinfo {author} {\bibfnamefont {N.}~\bibnamefont {Mirabolfathi}}, \bibinfo {author} {\bibfnamefont {B.}~\bibnamefont {Nebolsky}}, \bibinfo {author} {\bibfnamefont {M.}~\bibnamefont {Platt}}, \bibinfo {author} {\bibfnamefont {F.}~\bibnamefont {Ponce}}, \bibinfo {author} {\bibfnamefont {M.}~\bibnamefont {Pyle}}, \bibinfo {author} {\bibfnamefont {T.}~\bibnamefont {Reynolds}}, \ and\ \bibinfo {author} {\bibfnamefont {T.}~\bibnamefont {Saab}},\ }\href {\doibase 10.1103/physrevd.104.032010} {\bibfield  {journal} {\bibinfo  {journal} {Physical Review D}\ }\textbf {\bibinfo
  {volume} {104}} (\bibinfo {year} {2021}),\ 10.1103/physrevd.104.032010}\BibitemShut {NoStop}%
\bibitem [{\citenamefont {Yelton}\ \emph {et~al.}(2024)\citenamefont {Yelton}, \citenamefont {Larson}, \citenamefont {Iaia}, \citenamefont {Dodge}, \citenamefont {La~Magna}, \citenamefont {Baity}, \citenamefont {Pechenezhskiy}, \citenamefont {McDermott}, \citenamefont {Kurinsky}, \citenamefont {Catelani},\ and\ \citenamefont {Plourde}}]{Yelton2024}%
  \BibitemOpen
  \bibfield  {author} {\bibinfo {author} {\bibfnamefont {E.}~\bibnamefont {Yelton}}, \bibinfo {author} {\bibfnamefont {C.~P.}\ \bibnamefont {Larson}}, \bibinfo {author} {\bibfnamefont {V.}~\bibnamefont {Iaia}}, \bibinfo {author} {\bibfnamefont {K.}~\bibnamefont {Dodge}}, \bibinfo {author} {\bibfnamefont {G.}~\bibnamefont {La~Magna}}, \bibinfo {author} {\bibfnamefont {P.~G.}\ \bibnamefont {Baity}}, \bibinfo {author} {\bibfnamefont {I.~V.}\ \bibnamefont {Pechenezhskiy}}, \bibinfo {author} {\bibfnamefont {R.}~\bibnamefont {McDermott}}, \bibinfo {author} {\bibfnamefont {N.~A.}\ \bibnamefont {Kurinsky}}, \bibinfo {author} {\bibfnamefont {G.}~\bibnamefont {Catelani}}, \ and\ \bibinfo {author} {\bibfnamefont {B.~L.~T.}\ \bibnamefont {Plourde}},\ }\href {\doibase 10.1103/PhysRevB.110.024519} {\bibfield  {journal} {\bibinfo  {journal} {Phys. Rev. B}\ }\textbf {\bibinfo {volume} {110}},\ \bibinfo {pages} {024519} (\bibinfo {year} {2024})}\BibitemShut {NoStop}%
\bibitem [{\citenamefont {Ramanathan}\ and\ \citenamefont {Kurinsky}(2020)}]{Ramanathan_2020}%
  \BibitemOpen
  \bibfield  {author} {\bibinfo {author} {\bibfnamefont {K.}~\bibnamefont {Ramanathan}}\ and\ \bibinfo {author} {\bibfnamefont {N.}~\bibnamefont {Kurinsky}},\ }\href {\doibase 10.1103/PhysRevD.102.063026} {\bibfield  {journal} {\bibinfo  {journal} {Phys. Rev. D}\ }\textbf {\bibinfo {volume} {102}},\ \bibinfo {pages} {063026} (\bibinfo {year} {2020})}\BibitemShut {NoStop}%
\bibitem [{\citenamefont {Green}(2008)}]{Green2008}%
  \BibitemOpen
  \bibfield  {author} {\bibinfo {author} {\bibnamefont {Green}},\ }\href {\doibase 10.1016/j.solmat.2008.06.009} {\bibfield  {journal} {\bibinfo  {journal} {Solar Energy Materials and Solar Cells}\ }\textbf {\bibinfo {volume} {92}},\ \bibinfo {pages} {1305} (\bibinfo {year} {2008})}\BibitemShut {NoStop}%
\bibitem [{\citenamefont {Mattis}\ and\ \citenamefont {Bardeen}(1958)}]{MattisBardeen1958}%
  \BibitemOpen
  \bibfield  {author} {\bibinfo {author} {\bibfnamefont {D.~C.}\ \bibnamefont {Mattis}}\ and\ \bibinfo {author} {\bibfnamefont {J.}~\bibnamefont {Bardeen}},\ }\href {\doibase 10.1103/PhysRev.111.412} {\bibfield  {journal} {\bibinfo  {journal} {Phys. Rev.}\ }\textbf {\bibinfo {volume} {111}},\ \bibinfo {pages} {412} (\bibinfo {year} {1958})}\BibitemShut {NoStop}%
\bibitem [{\citenamefont {de~Visser}\ \emph {et~al.}(2014)\citenamefont {de~Visser}, \citenamefont {Goldie}, \citenamefont {Diener}, \citenamefont {Withington}, \citenamefont {Baselmans},\ and\ \citenamefont {Klapwijk}}]{deVisser2014}%
  \BibitemOpen
  \bibfield  {author} {\bibinfo {author} {\bibfnamefont {P.~J.}\ \bibnamefont {de~Visser}}, \bibinfo {author} {\bibfnamefont {D.~J.}\ \bibnamefont {Goldie}}, \bibinfo {author} {\bibfnamefont {P.}~\bibnamefont {Diener}}, \bibinfo {author} {\bibfnamefont {S.}~\bibnamefont {Withington}}, \bibinfo {author} {\bibfnamefont {J.~J.~A.}\ \bibnamefont {Baselmans}}, \ and\ \bibinfo {author} {\bibfnamefont {T.~M.}\ \bibnamefont {Klapwijk}},\ }\href {\doibase 10.1103/PhysRevLett.112.047004} {\bibfield  {journal} {\bibinfo  {journal} {Phys. Rev. Lett.}\ }\textbf {\bibinfo {volume} {112}},\ \bibinfo {pages} {047004} (\bibinfo {year} {2014})}\BibitemShut {NoStop}%
\bibitem [{\citenamefont {\ifmmode \check{Z}\else \v{Z}\fi{}emli\ifmmode~\check{c}\else \v{c}\fi{}ka}\ \emph {et~al.}(2015)\citenamefont {\ifmmode \check{Z}\else \v{Z}\fi{}emli\ifmmode~\check{c}\else \v{c}\fi{}ka}, \citenamefont {Neilinger}, \citenamefont {Trgala}, \citenamefont {Reh\'ak}, \citenamefont {Manca}, \citenamefont {Grajcar}, \citenamefont {Szab\'o}, \citenamefont {Samuely}, \citenamefont {Ga\ifmmode~\check{z}\else \v{z}\fi{}i}, \citenamefont {H\"ubner}, \citenamefont {Vinokur},\ and\ \citenamefont {Il'ichev}}]{Zemlicka2015}%
  \BibitemOpen
  \bibfield  {author} {\bibinfo {author} {\bibfnamefont {M.}~\bibnamefont {\ifmmode \check{Z}\else \v{Z}\fi{}emli\ifmmode~\check{c}\else \v{c}\fi{}ka}}, \bibinfo {author} {\bibfnamefont {P.}~\bibnamefont {Neilinger}}, \bibinfo {author} {\bibfnamefont {M.}~\bibnamefont {Trgala}}, \bibinfo {author} {\bibfnamefont {M.}~\bibnamefont {Reh\'ak}}, \bibinfo {author} {\bibfnamefont {D.}~\bibnamefont {Manca}}, \bibinfo {author} {\bibfnamefont {M.}~\bibnamefont {Grajcar}}, \bibinfo {author} {\bibfnamefont {P.}~\bibnamefont {Szab\'o}}, \bibinfo {author} {\bibfnamefont {P.}~\bibnamefont {Samuely}}, \bibinfo {author} {\bibfnamefont {i.~c.~v.}\ \bibnamefont {Ga\ifmmode~\check{z}\else \v{z}\fi{}i}}, \bibinfo {author} {\bibfnamefont {U.}~\bibnamefont {H\"ubner}}, \bibinfo {author} {\bibfnamefont {V.~M.}\ \bibnamefont {Vinokur}}, \ and\ \bibinfo {author} {\bibfnamefont {E.}~\bibnamefont {Il'ichev}},\ }\href {\doibase 10.1103/PhysRevB.92.224506} {\bibfield  {journal} {\bibinfo  {journal} {Phys. Rev. B}\ }\textbf {\bibinfo
  {volume} {92}},\ \bibinfo {pages} {224506} (\bibinfo {year} {2015})}\BibitemShut {NoStop}%
\bibitem [{\citenamefont {Gavidia}\ \emph {et~al.}(2026)\citenamefont {Gavidia}, \citenamefont {Golwala}, \citenamefont {Beyer}, \citenamefont {Cunnane}, \citenamefont {Day}, \citenamefont {Defrance}, \citenamefont {Frez}, \citenamefont {Huang}, \citenamefont {Kim}, \citenamefont {Martin}, \citenamefont {Sayers}, \citenamefont {Shu}, \citenamefont {Yu},\ and\ \citenamefont {Sadou}}]{Gavidia2026}%
  \BibitemOpen
  \bibfield  {author} {\bibinfo {author} {\bibfnamefont {A.}~\bibnamefont {Gavidia}}, \bibinfo {author} {\bibfnamefont {S.}~\bibnamefont {Golwala}}, \bibinfo {author} {\bibfnamefont {A.~D.}\ \bibnamefont {Beyer}}, \bibinfo {author} {\bibfnamefont {D.}~\bibnamefont {Cunnane}}, \bibinfo {author} {\bibfnamefont {P.~K.}\ \bibnamefont {Day}}, \bibinfo {author} {\bibfnamefont {F.}~\bibnamefont {Defrance}}, \bibinfo {author} {\bibfnamefont {C.~F.}\ \bibnamefont {Frez}}, \bibinfo {author} {\bibfnamefont {X.}~\bibnamefont {Huang}}, \bibinfo {author} {\bibfnamefont {J.}~\bibnamefont {Kim}}, \bibinfo {author} {\bibfnamefont {J.-M.}\ \bibnamefont {Martin}}, \bibinfo {author} {\bibfnamefont {J.}~\bibnamefont {Sayers}}, \bibinfo {author} {\bibfnamefont {S.}~\bibnamefont {Shu}}, \bibinfo {author} {\bibfnamefont {S.}~\bibnamefont {Yu}}, \ and\ \bibinfo {author} {\bibfnamefont {Y.}~\bibnamefont {Sadou}},\ }\href {\doibase 10.1109/TASC.2026.3692397} {\bibfield  {journal} {\bibinfo  {journal} {IEEE Transactions on Applied
  Superconductivity}\ }\textbf {\bibinfo {volume} {36}},\ \bibinfo {pages} {2402206} (\bibinfo {year} {2026})}\BibitemShut {NoStop}%
\bibitem [{\citenamefont {Dynes}, \citenamefont {Narayanamurti},\ and\ \citenamefont {Garno}(1978)}]{Dynes1978}%
  \BibitemOpen
  \bibfield  {author} {\bibinfo {author} {\bibfnamefont {R.~C.}\ \bibnamefont {Dynes}}, \bibinfo {author} {\bibfnamefont {V.}~\bibnamefont {Narayanamurti}}, \ and\ \bibinfo {author} {\bibfnamefont {J.~P.}\ \bibnamefont {Garno}},\ }\href {\doibase 10.1103/PhysRevLett.41.1509} {\bibfield  {journal} {\bibinfo  {journal} {Phys. Rev. Lett.}\ }\textbf {\bibinfo {volume} {41}},\ \bibinfo {pages} {1509} (\bibinfo {year} {1978})}\BibitemShut {NoStop}%
\bibitem [{\citenamefont {Kaplan}\ \emph {et~al.}(1976)\citenamefont {Kaplan}, \citenamefont {Chang}, \citenamefont {Jani}, \citenamefont {Chang}, \citenamefont {Langenberg},\ and\ \citenamefont {Scalapino}}]{Kaplan1976Quasiparticle}%
  \BibitemOpen
  \bibfield  {author} {\bibinfo {author} {\bibfnamefont {S.~B.}\ \bibnamefont {Kaplan}}, \bibinfo {author} {\bibfnamefont {C.~C.}\ \bibnamefont {Chang}}, \bibinfo {author} {\bibfnamefont {N.~F.}\ \bibnamefont {Jani}}, \bibinfo {author} {\bibfnamefont {W.~Y.}\ \bibnamefont {Chang}}, \bibinfo {author} {\bibfnamefont {D.~N.}\ \bibnamefont {Langenberg}}, \ and\ \bibinfo {author} {\bibfnamefont {D.~J.}\ \bibnamefont {Scalapino}},\ }\href {\doibase 10.1103/PhysRevB.14.4854} {\bibfield  {journal} {\bibinfo  {journal} {Physical Review B}\ }\textbf {\bibinfo {volume} {14}},\ \bibinfo {pages} {4854} (\bibinfo {year} {1976})}\BibitemShut {NoStop}%
\bibitem [{\citenamefont {Faoro}\ and\ \citenamefont {Ioffe}(2015)}]{Faoro2015}%
  \BibitemOpen
  \bibfield  {author} {\bibinfo {author} {\bibfnamefont {L.}~\bibnamefont {Faoro}}\ and\ \bibinfo {author} {\bibfnamefont {L.~B.}\ \bibnamefont {Ioffe}},\ }\href {\doibase 10.1103/PhysRevB.91.014201} {\bibfield  {journal} {\bibinfo  {journal} {Phys. Rev. B}\ }\textbf {\bibinfo {volume} {91}},\ \bibinfo {pages} {014201} (\bibinfo {year} {2015})}\BibitemShut {NoStop}%
\end{thebibliography}%

\clearpage
\pagebreak
\onecolumngrid
\appendix

\section{Measurement Setup and Data Analysis}\label{appendix:setup}
In this section we describe: (1) the physical operating environment of the KIPM detector, (2) the photon-pulse measurement parameters and data processing pipeline, and (3) the method used to convert from MEMS-controller voltage settings to beam-spot location on the detector surface.

\subsection{Measurement setup}
\begin{figure}[b]
    \centering
    \includegraphics[width=0.9\linewidth]{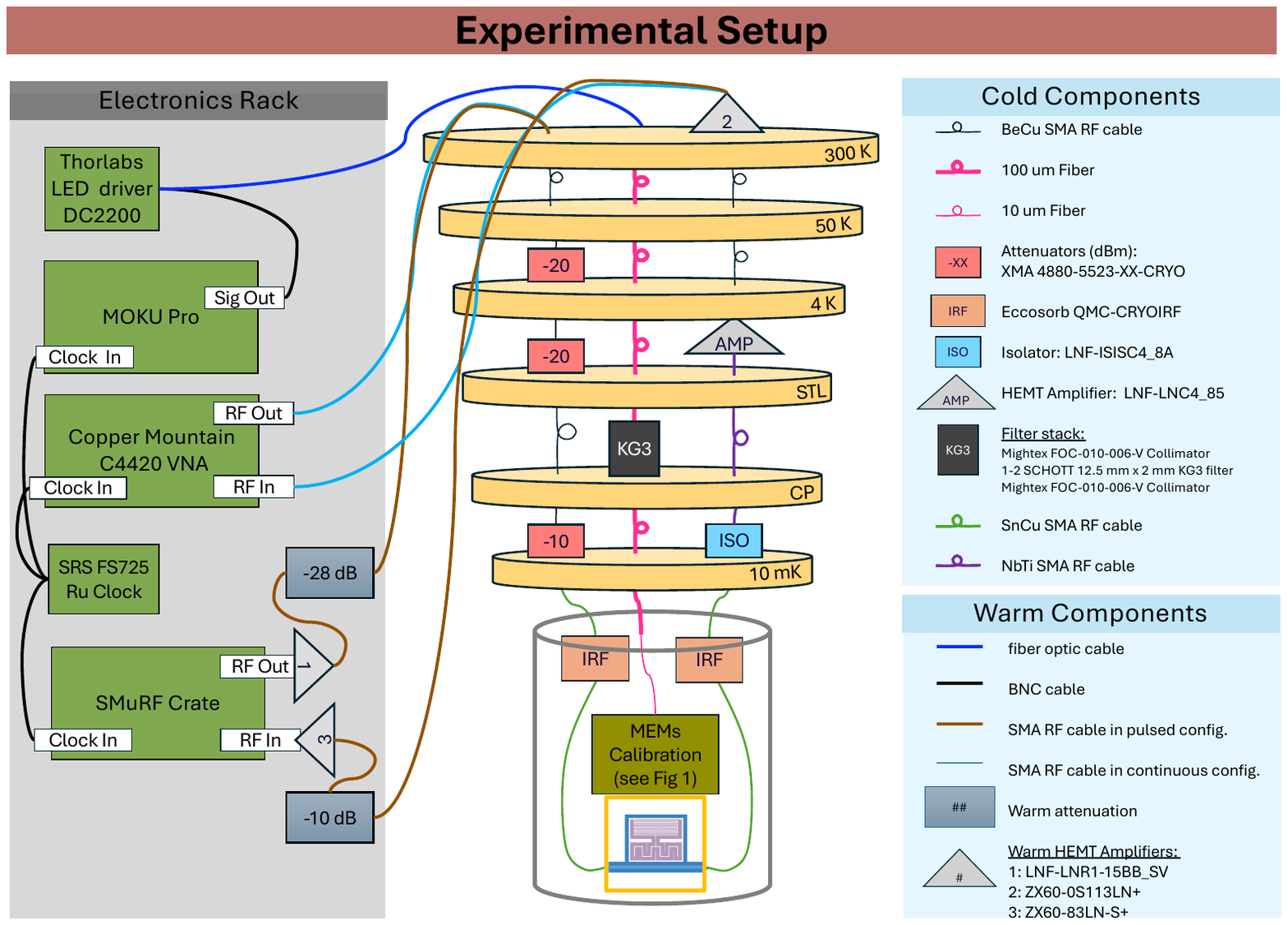}
    \caption{The cryostat configuration for the KIPM measurements from Runs 1 and 2. Readout configurations for continuous optical illumination and pulsed measurements are depicted simultaneously. Attenuation values, amplifier models, and filter placements are indicated at each temperature stage. In the optical filter stack, Run 1 had two 2\,mm thick KG3 filters in place, while Run 2 had only one in order to improve optical transmission to the KIPM.  }
    \label{fig:fridge_configuration}
\end{figure}

The experiment was run at the mixing chamber stage of a Proteox MX500, with a 4\,K cryoperm can and room temperature Metglas blanket installed for magnetic shielding. Additionally, the combined optical calibration unit and KIPM detector were installed in a IR-black coated cryoperm can, heatsunk to the mixing chamber stage of the cryostat, with the KIPM detector oriented so its patterned surface faced the bottom of the can. Commercially available QMC-CRYOIRF eccosorb filters were inserted into both the input and output RF readout lines of the KIPM at their penetrations into the can. The RF configuration is shown schematically in Fig.~\ref{fig:fridge_configuration}. 

Optical light at 625\,nm was routed from a room temperature fiber-coupled LED source to the mixing chamber stage via 100\,$\mathrm{\mu}$m diameter optical fibers that were thermalized at each stage. At the 100\,mK stage, free space 2\,mm thick KG3 filters were installed in a Mightex collimator filter stack. The 100\,$\mathrm{\mu}$m fiber was coupled to a 10\,$\mathrm{\mu}$m fiber at the 10 mK cryoperm can lid to improve the beam profile within the MEMs system. More information on the optical setup of the MEMs system can be found in Ref~\cite{tabassum2025broadbandopticalmodulationcontrol}.



Pulsed measurements presented in sections \ref{res:pulsed} and \ref{res:tempsweeps} were taken using crate-based SMuRF electronics, which was run in the ``debugging" mode described in Ref. \cite{Yu2023}. This mode records up to $2^{25}$ orthogonal IQ samples at a fixed rated of 2.4\,MHz, but was modified to decimate the data and set to stream at 200\,kHz for the base temperature measurements presented in this work. With the maximum DAC output power of -21\,dBm, warm attenuation and amplification were applied to achieve the desired readout power of -70\,dBm at the KIPM detector. To achieve nominal ADC power levels, two follow-on amplifiers (ZX60-0S113LN+ \& ZX60-83LN-S+) were installed on the signal chain. Variable attenuators on the transmit and receive ports of SMuRF were tuned to ensure the system's noise was limited by the 4\,K HEMT. The final warm electronics setup is shown in Fig.~\ref{fig:fridge_configuration}.

For the L-, V-, and temperature-scan optical pulse measurements, the following sequence was performed at each position and temperature. First, a single VNA-style sweep was recorded to determine the device's resonant frequency. Then, short off-resonant tones were recorded to correct for any phase drift between the two readout modes. The LED was then switched on at fixed pulse width, amplitude, and repetition rate and allowed to settle for 1 second. Next, a resonant tone was switched on and on-resonance IQ data was recorded. Finally, the LED and readout tones were turned off before moving to the next scan state. Scan-dependent parameters are summarized in Table \ref{tab:pulse_params}.
For the temperature scans, the cryostat was PID controlled to the desired temperature and allowed to equilibrate for 40 minutes before resuming data collection. The temperature scans were collected in two batches, the first from a base of 10\,mK up to 270\,mK, and the second from 200\,mK up until 350\,mK. 

\begin{table}[htbp]
  
  \centering
  \begin{tabular}{|c|c|c|c|}
    \hline
    \textbf{Parameter} & \textbf{L-scan} & \textbf{V-scan} & \textbf{Temperature scan}\\
    \hline
    Probe tone offsets ($\times$ linewidth) & $-1, 0, 1$ & $-0.2, -0.1, 0, 0.1, 0.2$ & $-0.2, -0.1, 0, 0.1, 0.2$\\
    \hline
    LED pulse width ($\mathrm{\mu}$s) & 10 & 12 & 12\\
    \hline
    LED excitation amplitude (V) & 2.5 & 2.5 & 2.5\\
    \hline
    LED repetition rate (Hz) & 20 & 0.5 & 2\\
    \hline
    data acquisition length (samples) & $2^{22}$ & $2^{22}$ & $2^{25}$\\
    \hline
    Sample rate & 200\,kHz & 200\,kHz & 2.4\,MHz\\
    \hline
    Repeats per position/temp & 1 & 1 & 5\\
    \hline
    Readout power from SMuRF (dBm) & -21 & -21 & -24\\
    \hline
  \end{tabular}
  \caption{Scan-dependent parameters for optical pulse measurements. }\label{tab:pulse_params}
\end{table}

\subsection{Pulsed Data Reduction}
To process the raw timestream data, the VNA-style scan taken at each position and temperature was fit to
Eq. \ref{eqn:S21}, and the transmission parameters were extracted to convert the data into the ideal IQ basis. The SMuRF electronics operating mode used was not time-synced to the LED trigger, so pulse start times were aligned on a per-dataset basis by chunking, smoothing and averaging all traces given the optical pulse repetition rate, and identifying the rising edge threshold crossing of the averaged trace. The raw trace was then shifted so that the trigger consistently fell at a designated trigger time of 5\,ms for the V-scan, and temperature scans, and 0.5\,ms for the L-scan. 

For each trace, the ``pre-trigger region" was defined to be the first 75\% of the trigger time. All cuts were applied using the linearly approximated $\frac{df}{f}$ basis. 
For each scan, traces in which  the pre-trigger RMS and mean fell outside the $99^\text{th}$ percentile were removed to mitigate extreme outliers in noise and coincidental events. 
Pile-up events were cut by estimating an expected peak window spanning 1000 samples past the trigger time, and estimating the pulse height as the maximum within it. Traces were removed if more than 20 samples outside the pre-trigger window exceeded 1.5x that estimated pulse height. For the L-scan and the temperature scans, which had higher optical pulse repetition rates, the trace immediately following an elevated pulse-height signal was removed as well. In the base temperature L- and V- scans, traces were removed if any samples reached 15x the RMS of the pre-trigger region. In order to avoid bias against the changing signal level present in the temperature scans, this 15x RMS cut was applied to just the pre-trigger region and last half of the trace.

These cuts resulted in a median cut fraction of 2.6\% across 3077 positions for the L-scan, consistent with the high number of optical pulses per dataset. The temperature scan had a median cut fraction of 7.7\% across both batches, while the V-scan had a median cut fraction of 25.0\%. This was due to significantly lower number of pulses collected per dataset at the 0.5\,Hz optical repetition rate. Pulse amplitude was found to be uncorrelated with cut fraction, for all scans. After quality cuts were performed, the pulse amplitude at each position was extracted as the maximum of the averaged, low-pass filtered pulse trace.

\subsection{Position Calibration Validation}
\begin{figure}[ht]
    \centering
    \includegraphics[width=0.45\linewidth]{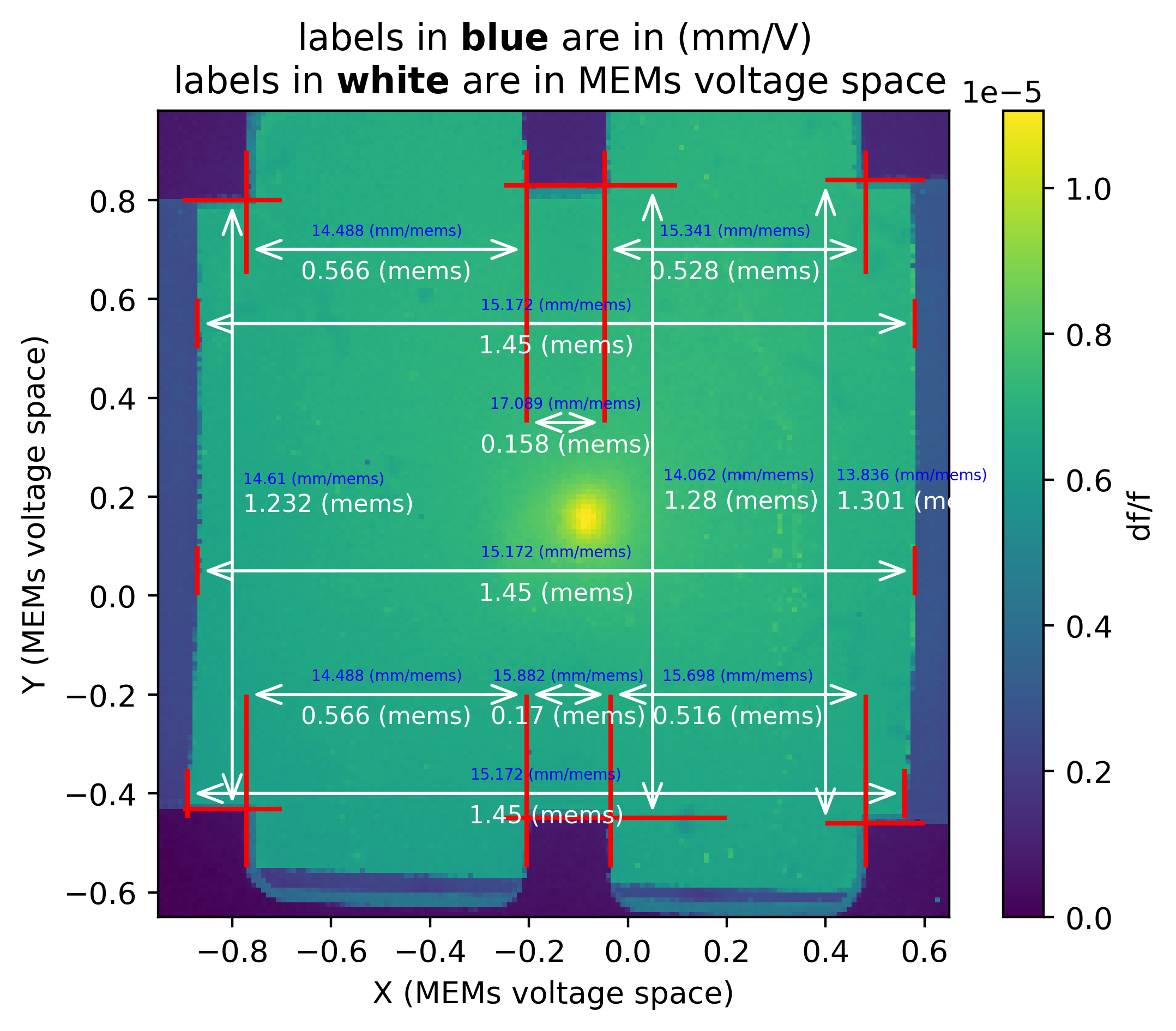}
    \caption{Data are taken with a known voltage applied to the MEMS mirror, but it is well established that the beam spot to position calibration is somewhat nonlinear (see Ref~\cite{tabassum2025broadbandopticalmodulationcontrol}). Physical position for the L scan discussed in the main text is thus established by using the known dimensions of the copper enclosure to establish vertical and horizontal conversions from fractional MEMS voltage units (relative to maximum drive voltage) to physical units in millimeters.}
    \label{fig:positionCalibration}
\end{figure}

The conversion from fractional MEMS voltage units to optical spot deflection in millimeters was done by comparing the machined copper housing dimensions to the voltage positions of those features in fractional MEMS voltage units from the continuous illumination scan, the results are shown in Fig.~\ref{fig:positionCalibration}. Conversion ratios were then averaged in the vertical and horizontal dimensions to achieve the conversion ration of 15.5\,mm/$\Delta$\,V and 14.2\,mm/$\Delta$\,V, respectively. To determine the KIPM location, a 2D gaussian was  fit to the continuouly illuminated position scan and the KIPM center position was set at the mean of (-0.08, 0.159) in fractional MEMS voltage units.

\section{Phonon Collection Efficiency}\label{appendix:analytic-response-modeling}
In order to better understand the response measured, before appealing to simulation, we can first apply some geometric and statistical approaches to understand the general features we observe in the position dependent scans. These will then help us contextualize some of the behavior observed in simulation, discussed at length in the following appendix.

In this model and the simulations, we have a few key parameters that help set the physical limits of our system. These are the probability that the phonon is:
\begin{enumerate}
    \item lost at a surface with probability $p_\mathrm{loss}$,
    \item absorbed by a sensor with probability $p_\mathrm{abs}$, or
    \item reflected back into the substrate with probability $1-p_\mathrm{loss}-p_\mathrm{abs}$.
\end{enumerate}
As discussed at length in e.g. Refs~\cite{Ren_2021,KELSEY2023168473,Yelton2024}, the sensor absorption probability is a complex function of phonon energy, incident angle, and boundary material impedance match, making a simple analytic estimate complex. Here we take simpler approach to gain some intuition for detector behavior before appealing to simulation for a more accurate modeling result. In the following, we use the approximation
\begin{equation}
    p_\mathrm{abs} \approx p_\mathrm{trans}p_\mathrm{break}f_\mathrm{abs}
\end{equation}
where $p_\mathrm{trans}$ is the probability of the phonon entering the film, $p_\mathrm{break}$ is the probability that at least some of the energy is deposited in the quasiparticle channel of the film, and $f_\mathrm{abs}$ is the fraction of energy absorbed during the encounter. The first two parameters we have analytic estimates of, so the third acts as an additional tuning parameter, where $f_\mathrm{abs}\rightarrow 1$ is the best case scenario, and we realistically expect $0 < f_\mathrm{abs} < 1$ to account for some energy lost back to the substrate due to emission of secondary phonons with pair-breaking energy.

\begin{table}[b]
    \centering
    \begin{tabular}{|c|c|c|c|}
    \hline
     & \multicolumn{3}{c|}{Material}\\\hline
     Parameter & Al & Nb & Cu \\
     \hline
     $p_\mathrm{trans}$ & 0.795 & 0.745 & 0.736 \\
     $d$ [nm] & 30 & 70 & $\infty$ \\
     $\Delta$ [$\mathrm{\mu}$eV] & 180 & 1400 & 0 \\
     $\tau_0^\mathrm{ph}$ [ns] & 0.242 & 0.00417 & 1.2 \\
     $v_s$ [$\mathrm{\mu}$m ns$^{-1}$] & 3.58 & 2.44 & 2.61 \\
     $\lambda_\mathrm{pb}$ [nm] & 866 & 10 & 3130 \\
     $p_\mathrm{escape}$& 0.871 & 9$\cdot 10^{-4}$ & 0 \\
     $p_\mathrm{break}$ & 0.129 & $\sim$1 & 1 \\
     $p_\mathrm{abs}$ & 0.103 & 0.745 & 0.736 \\
     $A_\mathrm{film}$ [mm$^2$] & 0.385 & 7.39 & 22.8 \\
     $f_{A}$ & $4.0 \cdot 10^{-4}$ & $7.6 \cdot 10^{-3}$ & $2.4 \cdot 10^{-2}$ \\
     \hline
\end{tabular}
    \caption{Reflection probabilities and film parameters utilized for both analytic estimates and as inputs for the G4CMP simulation described in this paper, largely the same as Ref~\cite{Yelton2024}. Note that the film areas are the contact area either of superconducting films or copper clamps, and the area fraction $f_{A}$ is based on a total surface area of 968 square mm, including both sides of a 22mm square substrate.} 
    \label{tab:baseline-sim-values}
\end{table}

We can estimate $p_\mathrm{trans}$ and $p_\mathrm{break}$ from standard references as in Ref~\cite{Yelton2024} (see also Table~\ref{tab:baseline-sim-values}). We find a pair-breaking mean free path in Al of roughly $\lambda_\mathrm{pb}\approx v_s \tau_{0}^\mathrm{ph} \approx 866$~nm. For a 30~nm thick film, we thus estimate that for each time the phonon enters and transits the film, it crosses a distance $a d$, for $d$ thickness of the film, giving
\begin{equation}
    p_\mathrm{break} = 1-\exp(-a(30)/866) = 1 - (0.965)^a
\end{equation}
where $a \geq 2$ is the mean number of lengths traversed by the phonon. Taking $a\sim 4$ as in Ref~\cite{Yelton2024}, we find $p_\mathrm{break} \sim 0.129$. Combining this with $p_\mathrm{trans}\approx 0.795$, we estimate
\begin{equation}
    p_\mathrm{abs} \approx 0.795(0.129)f_\mathrm{abs}= 0.103f_\mathrm{abs}
\end{equation}
Thus for each encounter with our film we expect at most 10\% of the energy incident to be absorbed, less if some pair-breaking energy is lost back to the substrate. We can see that this issue is likely mitigated substantially just by moving to a thicker film, where both $f_\mathrm{abs}$ and $p_\mathrm{break}$ will approach 1.

Below, we consider two options. First, we consider the case where phonons can only be absorbed on their first encounter with an interface - this is the lower bound on phonon efficiency in the base case scenario that the event occurs below the sensor (but at the far side of the chip). Then, we consider the alternative limit, where phonon transport is diffusive and we can neglect geometric effects, but treat loss with more nuance. We'll contrast these analytic estimates with simulations in the next section, and we use these results in the main text to motivate our conclusions.

\subsection{First-pass absorption}



Assuming a substrate length ($l)$ and width ($w$) much wider than the thickness ($t$), we can compute the solid angle subtended by the KIPM  from a given event position -- this is a proxy for the energy absorbed by the KID on first pass of the prompt phonon cloud. The differential solid angle d$\Omega$ is given by the expression
\begin{equation}
    d\Omega = \frac{\cos(\theta)}{r^2}dA = \frac{t}{((x-x_0)^2+(y-y_0^2)+t^2)^{3/2}}dxdy
\end{equation}
where $x$ and $y$ are the in-plane axes of the KIPM, and $\theta$ is the angle between the radial vector from the point to a given $x,y$ position on the KIPM and the normal vector for the surface. To compute the fraction of energy absorbed, we want to estimate the fraction of the solid angle of the istropic cloud $f_{\Omega}$ incident on the sensor. This is given by the expression
\begin{equation}
    f_{\Omega} = \frac{1}{4\pi}\int_{-w/2}^{w/2}\int_{-l/2}^{l/2}\frac{t}{((x-x_0)^2+(y-y_0)^2+t^2)^{3/2}}dxdy
\end{equation}
This equation is evaluated numerically and over-plotted on data in the main text to illustrate the fact that, when close to the KIPM, the shape of the response vs position curve roughly matches the shape of this fraction.

In the special case of $x_0=y_0=0$, and for small angle, this will just reduce to the limiting case of 
\begin{equation}
    f_{\Omega} \approx \frac{\Omega}{4\pi} \approx \frac{A_\mathrm{sensor}}{4\pi t^2} = \frac{lw}{4t^2}
\end{equation}
which for the KID in this paper is roughly 3\% (assuming a substrates thickness of 1~mm). This is much higher than the area fraction covered by the KID on average, and thus we should expect to see some enhancement in phonon collection for events below the KID.

The total phonon efficiency for this first wave of phonons $\eta_{0}$ is the fraction incident on the sensor times the efficiency of absorbing this fraction, or
\begin{equation}\label{eq:AngleEff}
    \eta_{0} \approx f_{\Omega}p_\mathrm{abs} \sim 0.03(0.106)f_\mathrm{abs} = 3.2\cdot 10^{-3}f_\mathrm{abs}
\end{equation}
which tells us that the relative fraction of phonons absorbed without any reflections is small even without loss back to the substrate. While not surprising, this implies losses per crystal transit need to be small for phonons to have enough attempts at absorption for efficiency to approach even 1\%.

As a side note, in the opposite regime, where $x_0,y_0 >> w,l,t$, we find
\begin{equation}
    f_{\Omega} \approx \frac{1}{4\pi}\frac{lwt}{r^3}
\end{equation}
which is just the relative solid angle of a patch of area $lw$ at distance $r$ multiplied by the cosine from the original integrand. For events very far from the KID, this becomes trivially small; even for thick films, we will still want the distance to the primary phonon sensor to be comparable to the size of the sensor to have a reasonable chance of the first-pass phonons being absorbed. This is the opposite limit we often find ourselves in when trying to optimize for small volume, high-resolution sensing elements, and provides a guide for how sparse we're likely to be able to make our sensing elements when combined with the following sections.

\subsection{Absolute Efficiency}

We can put a lower bound on efficiency in constrast to the previous calculation by assuming the phonons are completely randomized in position - this is the opposite, `many bounce' extreme. In this limit, we treat phonon collection akin to a one-dimensional state machine, in which the three state transitions are determined by the probabilities defined earlier. Of the two free parameters, the absorption is easiest to estimate. Drawing from the geometric example, we can expect an absorption probability of
\begin{equation}
    p_\mathrm{abs,\infty} \approx p_\mathrm{abs}\frac{A_\mathrm{sensor}}{A_\mathrm{substrate}} = p_\mathrm{abs}f_\mathrm{A}
\end{equation}
where $f_\mathrm{A}$ is the fraction of surface area covered by a given film, listed in Table~\ref{tab:baseline-sim-values}. Putting in numbers from above, we thus have
\begin{equation}\label{eq:pabsInf}
    p_\mathrm{abs,\infty} \approx 0.103f_\mathrm{abs}\frac{0.385}{800} = 5.0\cdot 10^{-5}f_\mathrm{abs}
\end{equation}
We can see why a small area, thin film is not going to be an efficient absorber. We also note that this absorption rate is lower than the angular rate in Eq.~\ref{eq:AngleEff}, suggesting that there will in fact be some enhanced collection for events close to the sensor.

To set an overall expectation for efficiency, we have to balance this absorption rate against a loss rate. The expected energy imparted to the sensor is just given by
\begin{equation}\label{eq:epsInf}
    \eta_\mathrm{ph} \approx \frac{p_\mathrm{abs,\infty}}{p_\mathrm{abs,\infty}+p_\mathrm{loss}} = \left[1+\frac{p_\mathrm{loss}}{p_\mathrm{abs,\infty}}\right]^{-1}
\end{equation}
This gives the expected result that equivalent loss and absorption probabilities result in a 50\% efficiency. Inverting this, we can set an expectation for loss rate by comparing to the range of efficiencies discussed in Ref~\cite{Temples_2024}, and we find that
\begin{equation}
    p_\mathrm{loss} = p_\mathrm{abs,\infty}\left[\eta_\mathrm{ph}^{-1}-1\right] \approx 5.0\cdot 10^{-5}f_\mathrm{abs}\left[\eta_\mathrm{ph}^{-1}-1\right]
\end{equation}
For a benchmark efficiency of 1\%, we thus find a predicted loss of around 0.5\% per pass for the best-case $f_\mathrm{abs}=1$ - this is thus an upper bound on the expected loss per scatter. We'll revisit this in the next section when more detailed simulations are done to model overall efficiency in more detail. As the expected loss from the film increases for events that break pairs, this loss rate bound will go down.

\subsection{$\epsilon$ Modeling}

Finally, we can use the above to produce a rough model for the ratio of on-KID to off-KID efficiency, to compare to the simulated results. We can model the `true' efficiency of events initially below the KID as a prompt event fraction $f_\mathrm{prompt}$. If these events all see the constant absorption probability described by $\eta_0$, then we have
\begin{equation}
    \eta_\mathrm{prompt} \approx \left[1+\frac{p_\mathrm{loss}}{\eta_0}\right]^{-1}
\end{equation}
We then approximate the on- and off-KID efficiency as
\begin{equation}
    \eta_\mathrm{on-KID} \approx f_\mathrm{prompt}\eta_\mathrm{prompt} + (1-f_\mathrm{prompt})\eta_\mathrm{ph},\;\;\eta_\mathrm{off-KID} = \eta_\mathrm{ph}
\end{equation}
which gives us the efficiency ratio
\begin{equation}
    \epsilon = \frac{\eta_\mathrm{off-KID}}{\eta_\mathrm{on-KID}} = \frac{\eta_\mathrm{ph}}{ f_\mathrm{prompt}\eta_\mathrm{prompt} + (1-f_\mathrm{prompt})\eta_\mathrm{ph}} = \frac{1}{1+f_\mathrm{prompt}(\eta_\mathrm{prompt}/\eta_\mathrm{ph}-1)}
\end{equation}
Here this is properly a two parameter model for fixed geometry, which is a function both of $p_\mathrm{loss}$ and $f_\mathrm{prompt}$. It has the correct extrema - in the limit $p_\mathrm{loss}\rightarrow 1$ and $p_\mathrm{abs,\infty} \rightarrow 0$, we get the expected limit of $\epsilon \rightarrow 0$. In the limit $p_\mathrm{loss}\rightarrow 0$, we trivially get $\epsilon\rightarrow 1$. In contrast to the simulation, however, $f_\mathrm{prompt}$ is an input, where in the simulation that is a derived parameter, with the resonator area largely driving $f_\mathrm{prompt}$.

If we look at the extremes, we find that $1 > \epsilon > \frac{\eta_\mathrm{ph}}{\eta_\mathrm{prompt}}$ for the full range of $0 < f_\mathrm{prompt} < 1$. To replicate the measurement in this paper, where we find $\epsilon\approx 0.34$, we find that $f_\mathrm{prompt}\sim 3\%$ in conjunction with the 0.2\% loss from above. This too can be contrasted to the simulation in the next section, and implies that the majority of the response is due to largely diffusive phonons with a small position-dependent enhancement. 

%




\clearpage
\pagebreak

\section{G4CMP Simulation}\label{appendix:simulation-details}

In the previous appendix, we described some limiting cases where phonon efficiency can be understood analytically, but it is clear that this is a problem with enough spatial dependence and non-linearity that it will benefit from a Monte-Carlo treatment, to adequately account for different loss mechanisms, energy conversions, and boundary conditions. In this section, we describe in more detail the phonon simulation implemented in \Geant Condensed Matter Physics (G4CMP)\cite{KELSEY2023168473}. This package is capable of simulating three-dimensional phonon propagation and phonon absorption in superconducting thin films, and has been benchmarked for use with Al phonon absorbers. See Ref~\cite{KELSEY2023168473} for more specific details of those validations.

\subsection{Modeling Inputs}\label{subsec:modeling-inputs}

\begin{figure}
    \centering
    \includegraphics[width=0.3\linewidth]{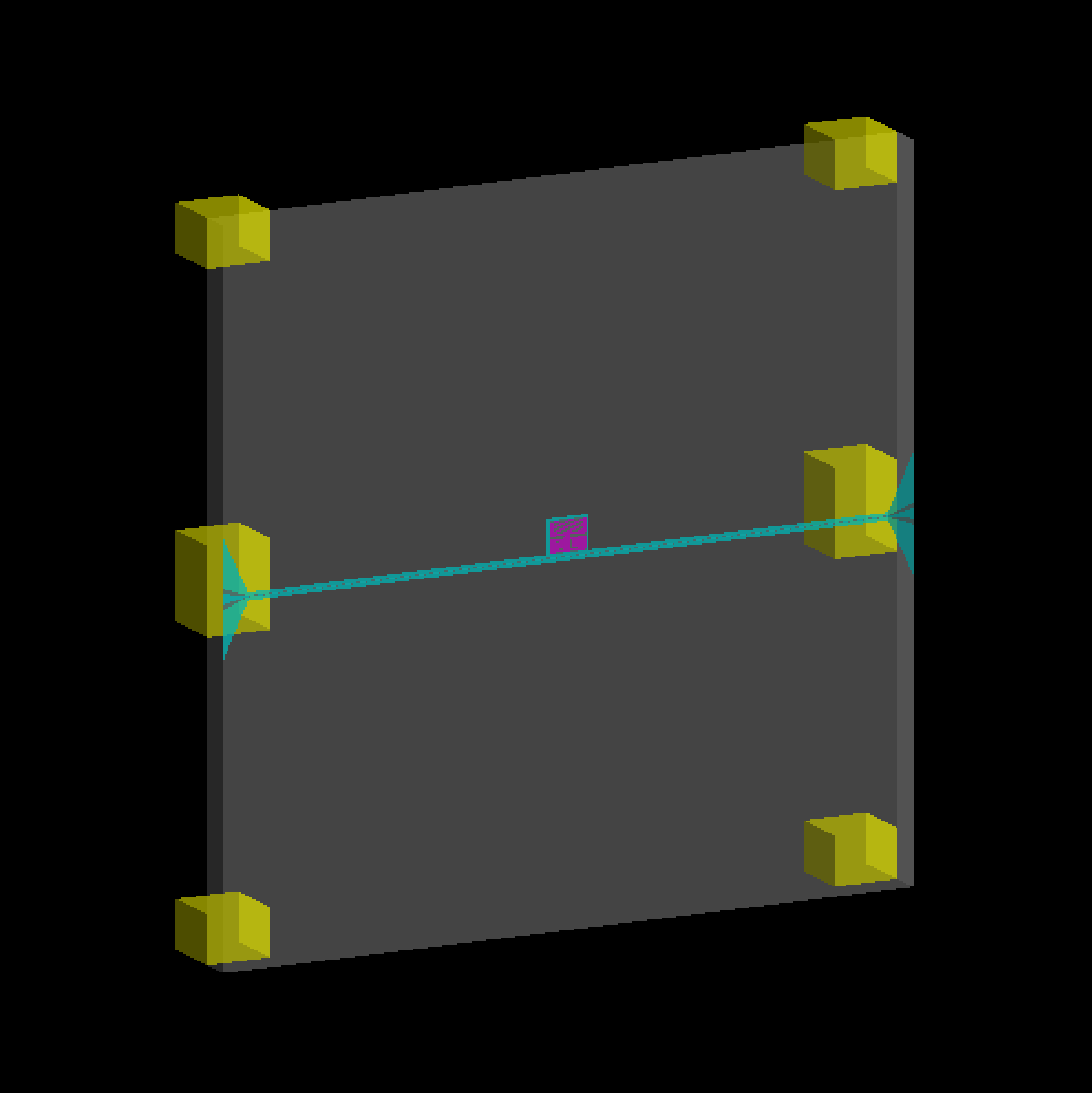}
    \hfill
    \includegraphics[width=0.65\linewidth]{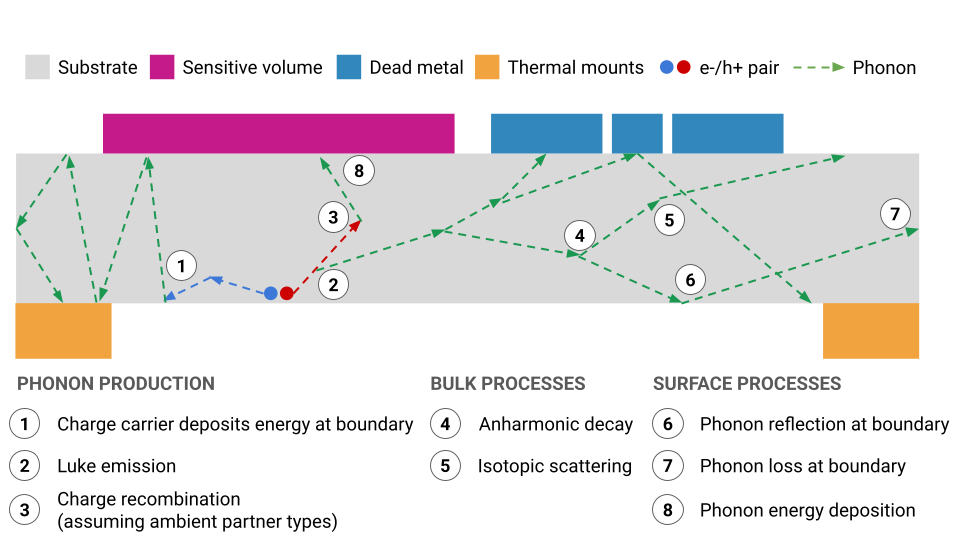}
    \caption{\textit{Left:} Visualization of the chip geometry in G4CMP. \textit{Right:} Schematic of physics processes in our simulation model. }
    \label{fig:g4cmp-vis}
\end{figure}

The detector geometry we simulate in G4CMP consists of five components: the Si substrate, the Al inductor, the Al capacitor, the Nb feedline, and the copper housing mounts, in line with the design provided in Section \ref{sec:kipm-design}. The superconductor features are consistent with the mask provided in \cite{temples2025}, and the detector housing mounts are modeled as copper rectangular prisms on the underside of the substrate. A visualization of the simulated geometry is provided in Fig.~\ref{fig:g4cmp-vis}.

For the model discussed in the main text (Section \ref{sec:comparisonToSimluation}), we have made a number of simplifying assumptions to allow us to tune the model to best match our data - these are the same as those made in Ref~\cite{Yelton2024} for a similar study. We simulate a 625-nm photon impact by producing an $e^-/h^+$ pair with a total kinetic energy of 0.814 eV (corresponding to the photon energy minus the Si band gap at 0 K), partitioned in an ``all-to-one'' scheme according to Ref~\cite{Ramanathan_2020}. These $e^-/h^+$ pairs are generated in a uniform $(x,y)$ distribution across a custom region of interest and an exponential $z$-distribution consistent with the attenuation of a 625-nm optical photon in Si, which has a characteristic length of 2.84 $\mathrm{\mu}$m \cite{Green2008}. We set the angular momentum distribution of the primary charge carriers to be isotropic by giveing the electron-hole pair a random initial direction. 

G4CMP then handles subsequent emission of acoustic phonons as these charges relax to the bandgap energy, producing a large number of acoustic phonons per charge carrier. All phonon reflections are assumed to be diffuse to account for roughness at the surfaces, in keeping with Ref~\cite{Yelton2024}. We set the maximum number of phonon reflections at a high enough value for each simulation that the phonon tracks are always terminated by absorption at some surface or reaching a low energy cutoff at the superconducting gap of aluminum (defined in the simulation to be $2\Delta_\textbf{Al}=360$ $\mathrm{\mu}$eV). Charges are discarded once they have finished emitting phonons, and are subsequently ignored in the remaining simulation. A schematic of the main physics processes involved in our model is provided in Fig.~\ref{fig:g4cmp-vis}. 


Film modeling for this study was handled in the same simplified manner as described in the previous section, with the caveat that the geometric aspect is handled implicitly by G4CMP. For each film type, the probability that the phonon is absorbed (and thus removed from the simulation) is described by the probability $p_\mathrm{abs}$, with the default values given in Table~\ref{tab:baseline-sim-values}. For the simulations where Nb or Cu loss is included, these are the default values used. For this study, we opt for the simplified model instead of built-in methods for including details of phonon absorption and down-conversion in G4CMP. Future work will utilize these methods, but we do not anticipate them changing our conclusions, as the underlying calculations are the same in this regime.

\begin{table}[h]
    \centering
    \begin{tabular}{|c||c|c|c|c|c|c|}
    \hline
    Model & $p_\mathrm{abs, inductor}$ & $p_\mathrm{abs, capacitor}$ & $p_\mathrm{abs, Nb}$ & $p_\mathrm{abs, Cu}$ & $p_\mathrm{abs, polished faces}$ & $p_\mathrm{abs, side walls}$ \\
    \hline
    \hline
    A & 0.103 & 0.103 & 0.745 & 0.736 & 0 & 0 \\ 
    B & 0.103 & 0 & 0.745 & 0 & 0 & 0 \\ 
    C & 0.103 & 0 & 0 & 0.736 & 0 & 0 \\ 
    D & 0.103 & 0 & 0 & 0 & 0.0025 & 0 \\ 
    E & 0.103 & 0 & 0 & 0 & 0 & 0.02 \\ 
    F & 0.103 & 0 & 0.1 & 0 & 0.002 & 0 \\
    \hline
\end{tabular}
    \caption{Modeling parameters for absorption at surfaces and interfaces, corresponding to results in Fig.~\ref{fig:lossMechanisms}.}
    \label{tab:sim-models}
\end{table}

\begin{figure}
    \centering
    \includegraphics[width=0.45\linewidth]{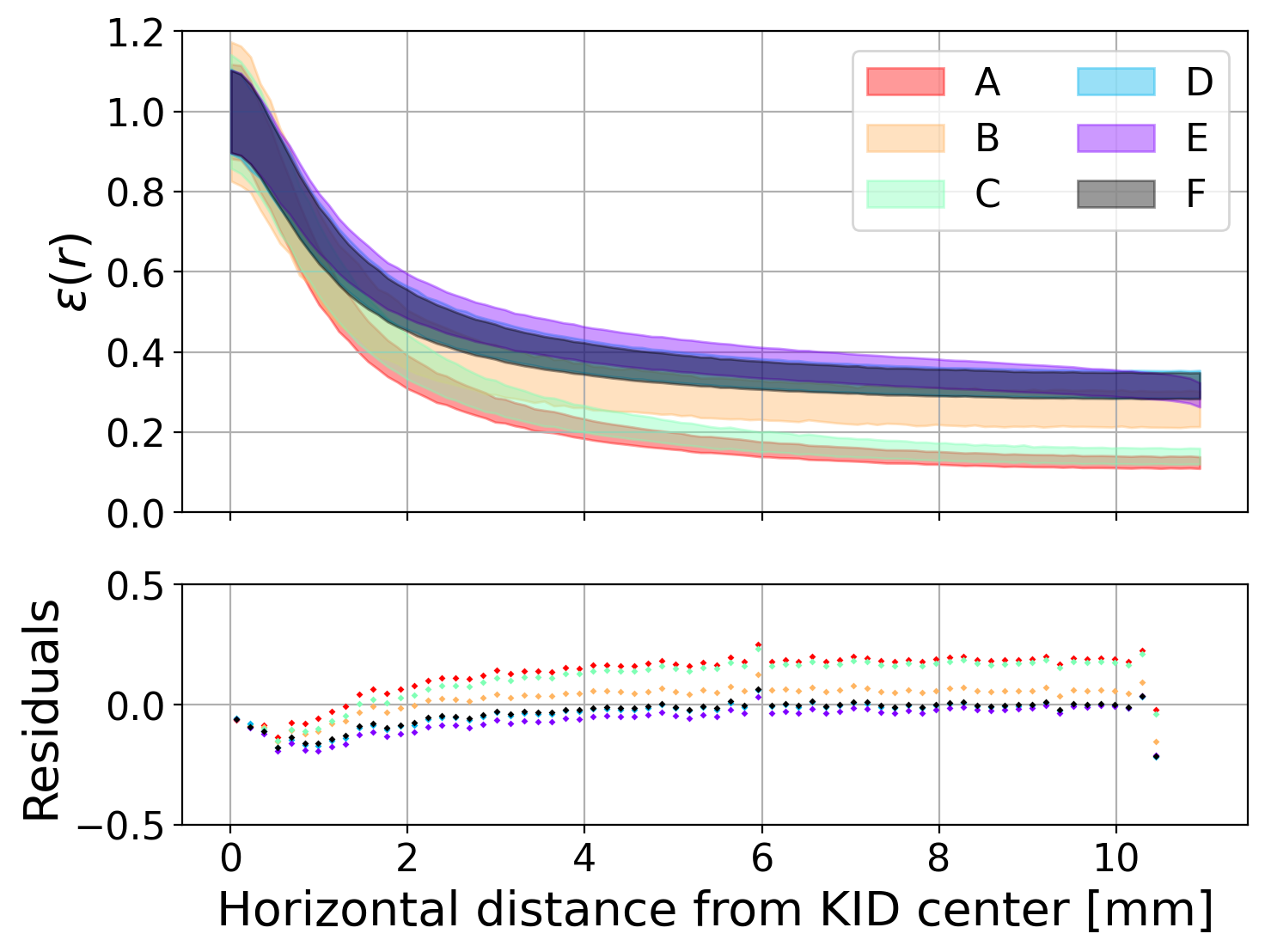}
    \includegraphics[width=0.45\linewidth]{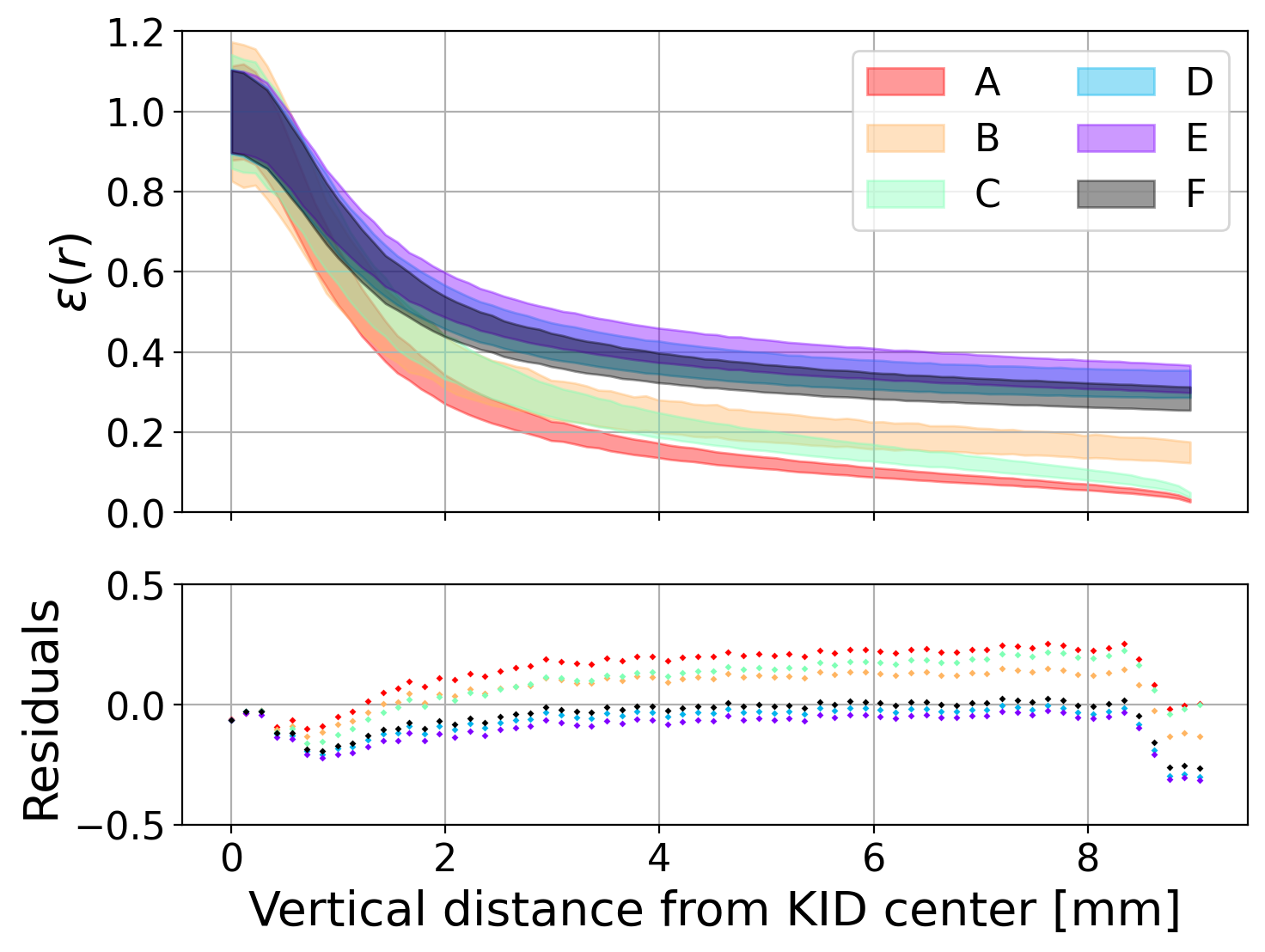}
    \caption{Simulation results of peak-normalized phonon collection efficiency $\epsilon$ as a function of $e^-/h^+$ pair distance from the inductor center across the horizontal (left) and vertical (right) branches of the L scan for differnt phonon loss models provided in Table \ref{tab:sim-models}. Model F is shown in Fig.~\ref{fig:pulsedMeasurements} in comparison with the pulsed measurement. }
    \label{fig:lossMechanisms}
\end{figure}

\subsection{Simulation Results}
\subsubsection{Position-Dependent Phonon Collection for Loss Models}
Within our region of interest, we can create a 2D histogram of total $e^-/h^+$ pair energy generated within $(x,y)$ bins. Similarly, we can create a 2D histogram of total phonon energy absorbed by the sensor originating from charge carrier production events within these position bins. We define the phonon collection efficiency $\eta_\mathrm{ph}(x,y)$ as the ratio of the absorbed phonon energy to the generated $e^-/h^+$ pair energy for each $(x,y)$ bin. Analogously, we can create 1D histograms of total primary energy and total absorbed energy binned by the distance of $e^-/h^+$ pairs from the center of the inductor, and from their ratio we can obtain phonon collection efficiency as a function of radial distance $\eta_\mathrm{ph}(r)$. We obtain the peak-normalized phonon collection efficiency $\epsilon_\mathrm{ph}(r)$, shown in Fig.~\ref{fig:pulsedMeasurements}, by dividing $\eta_\mathrm{ph}(r)$ by the value in the bin closest to the inductor.

Two sets of simulations were completed for this paper to illustrate what the MEMS measurements can say about relative phonon losses in a range of scenarios. First, we used the default paramteters in Table~\ref{tab:baseline-sim-values} to simulate a range of potential loss scenarios, including: (A) Loss to Al capacitor, Nb feedline, and copper mounts with default strength; (B) loss only to Nb feedline with default strength; (C) loss only to copper mounts with default strength; (D) loss to either top/bottom polished faces; (E) loss to side walls; (F) weak loss to Nb feedline and polished faces. 
The results of these simulations are shown in Fig.~\ref{fig:lossMechanisms}. The simulations that best replicate the data are those with an isotropic surface loss; loss tied too strongly to distinct features tends to have higher residuals close to those features. In particular, our data does not show any evidence that we can resolve excess loss from the Nb feedline, while simulations that include a nominal probability of feedline loss tend to show noticeable evidence of the feedline in the 2D scans.

\subsubsection{Projections for Efficiency from Parameterized Studies}

The second simulation simplified the treatment to just consider surface loss, which allowed us to explore a wide range of models parameterized by the combination of $p_\mathrm{abs}$ and $p_\mathrm{loss}$, where $p_\mathrm{loss}$ was treated as the loss probability at the top and bottom faces of the detector. We select $p_\mathrm{abs}\in\{0.01, 0.03, 0.1, 0.3, 1\}$ and $p_\mathrm{loss}\in\{1\times10^{-1}, 2.5\times10^{-2}, 1\times10^{-2}, 2.5\times10^{-3}, 1\times10^{-3}, 2.5\times10^{-4}, 1\times10^{-4}, 0\}$, and we run a simulation for each pair of ($p_\mathrm{abs}$, $p_\mathrm{loss}$) on this grid. In each simulation, $e^-/h^+$ pairs are generated at two $(x,y)$ positions in the substrate corresponding to the on-KID and off-KID locations in Fig.~\ref{fig:VNAscan}a with a $z$ position sampled from the exponential distribution consistent with Appendix \ref{subsec:modeling-inputs}. We then calculate the phonon collection efficiency at the two positions, where 
\begin{align}
    \eta_\mathrm{ on-KID} &=\frac{\text{total energy absorbed from on-KID source}}{\text{total energy generated in on-KID source}},\\
    \eta_\mathrm{ off-KID} &=\frac{\text{total energy absorbed from off-KID source}}{\text{total energy generated in off-KID source}},\\
    \epsilon_\mathrm{ off-KID} &=\frac{\eta_\mathrm{ off-KID}}{\eta_\mathrm{ on-KID}}.
\end{align}
We present simulated results of $\eta_\mathrm{ on-KID}$ and $\epsilon_\mathrm{ off-KID}$ from all combinations of ($p_\mathrm{abs}$, $p_\mathrm{loss}$) in Fig.~\ref{fig:positionDependenceVsCollectionEfficiency}. 

To sketch an intuitional picture for the role of sensor area coverage in these kinds of efficiency studies, we introduce the metric $p_\mathrm{abs, \infty}$ as defined in Eq. \eqref{eq:pabsInf}. Our previous metric of $p_\mathrm{abs}$ then becomes a weak proxy of the sensor film thickness. Within this picture, a hypothetical phonon-mediated detector with $p_\mathrm{abs, \infty}=1$ would require the substrate to be fully covered by the sensor and have all phonons that reach the substrate-sensor interface deposit energy in the sensor. For our device with 0.08\% sensor area coverage, we have $p_\mathrm{abs,\infty} \lesssim 0.001$ even for $p_\mathrm{abs}=1,$ which is much lower than other existing architectures such as the HVeV detector in Ref~\cite{Ren_2021} with a 25\% sensor area coverage and a projected $p_{\text{abs, }\infty} \sim 0.18$. This motivates systematically exploring the scaling of phonon collection efficiency with sensor volume for future work on modeling phonon-mediated detectors.

\begin{figure}
    \includegraphics[width=0.5\linewidth]{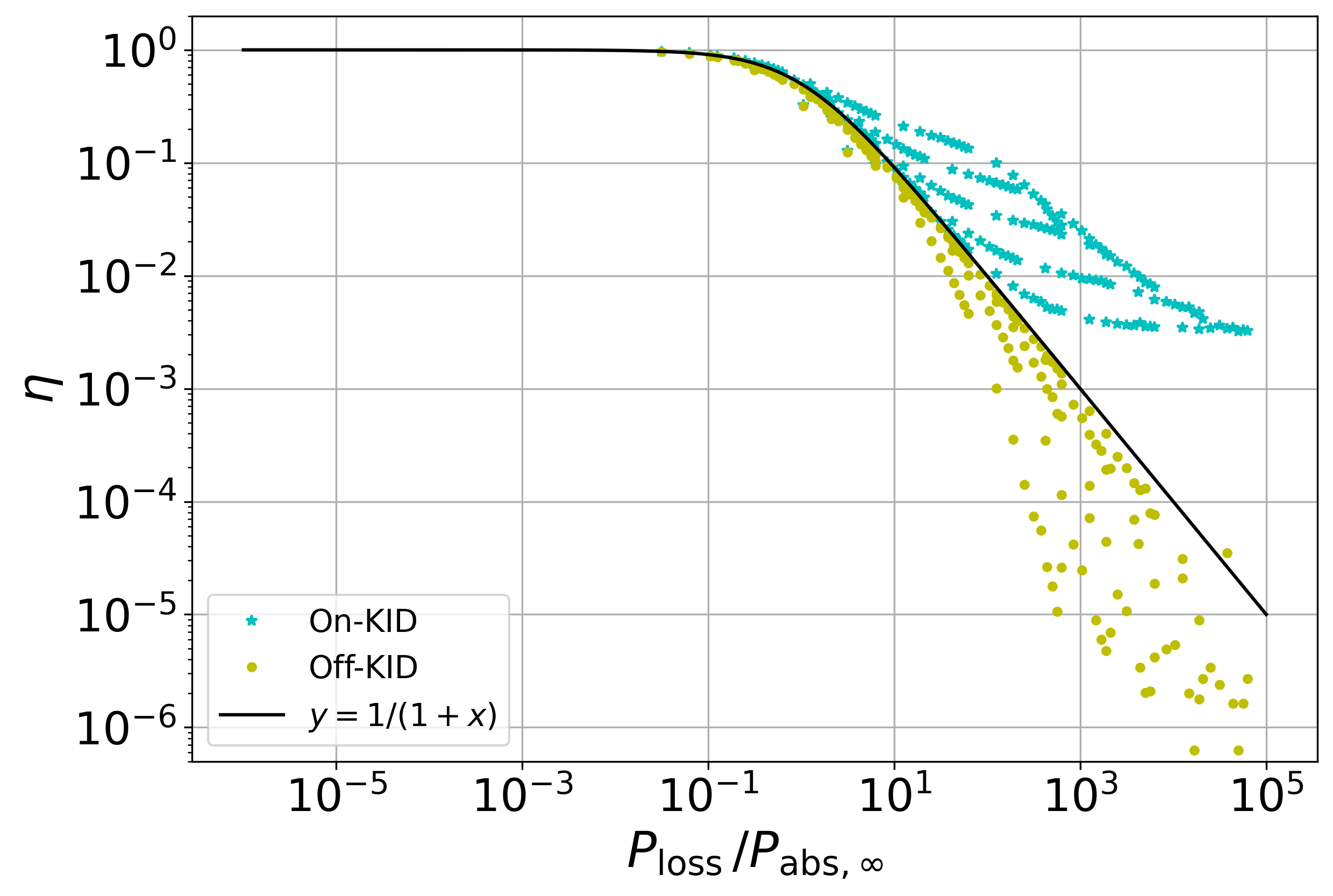}
    \caption{Efficiency computed from the set of simulated geometries described in this appendix for the on-KID (cyan) and off-KID (yellow) points compared to the expected trend from Eq~\ref{eq:epsInf} in the limit of position-independent energy response. In the high-efficiency limit, where either the loss is very small or absorption probability high, we expect both position indepdenent response ($\epsilon\rightarrow 1$) and high overall collection efficiency. As the loss probability begins to exceed absorption, we find more position dependence, with expected 'average' efficiency representing an idealized upper bound for events far from the KID, and a lower bound for events close to the KID. 
    }
    \label{fig:loss-capture-ratio}
\end{figure}

As an illustration of the importance of achieving high $p_\mathrm{abs,\infty}$ through high area coverage, we present in Fig. \ref{fig:loss-capture-ratio} an additional view of the data discussed in Fig.~\ref{fig:positionDependenceVsCollectionEfficiency}, showing $\eta$ as a function of $x = p_\mathrm{loss}/p_\mathrm{abs,\infty}$. This simulated grid was run for two discrete positions, but illustrates two main trends. First, in the limit $x \ll 1$, we find that energy collection is high and $\epsilon$ approaches 1. We also find that the predictability of Eq.~\ref{eq:epsInf} of imputed efficiency is good for on- and off-KID devices. 

In the $x\gg1$ limit, low $\epsilon$ implies  large variation in collection efficiency. This means that Eq.~\ref{eq:epsInf} becomes an upper bound of off-KID efficiency and lower bound for on-KID efficiency. It can thus still be used as an effective guide for setting average energy collection scales in low-coverage devices as a way of estimating the order of magnitude of average device response. We leave further exploration of these efforts to more modeling-focused work, where we can draw on other measurements of $\eta$ to make statements with further benchmarking.

\section{Gap Smearing Modeling of KID Response}\label{app:gs}




\begin{figure}[h!]
    \centering
    \includegraphics[width=1\linewidth]{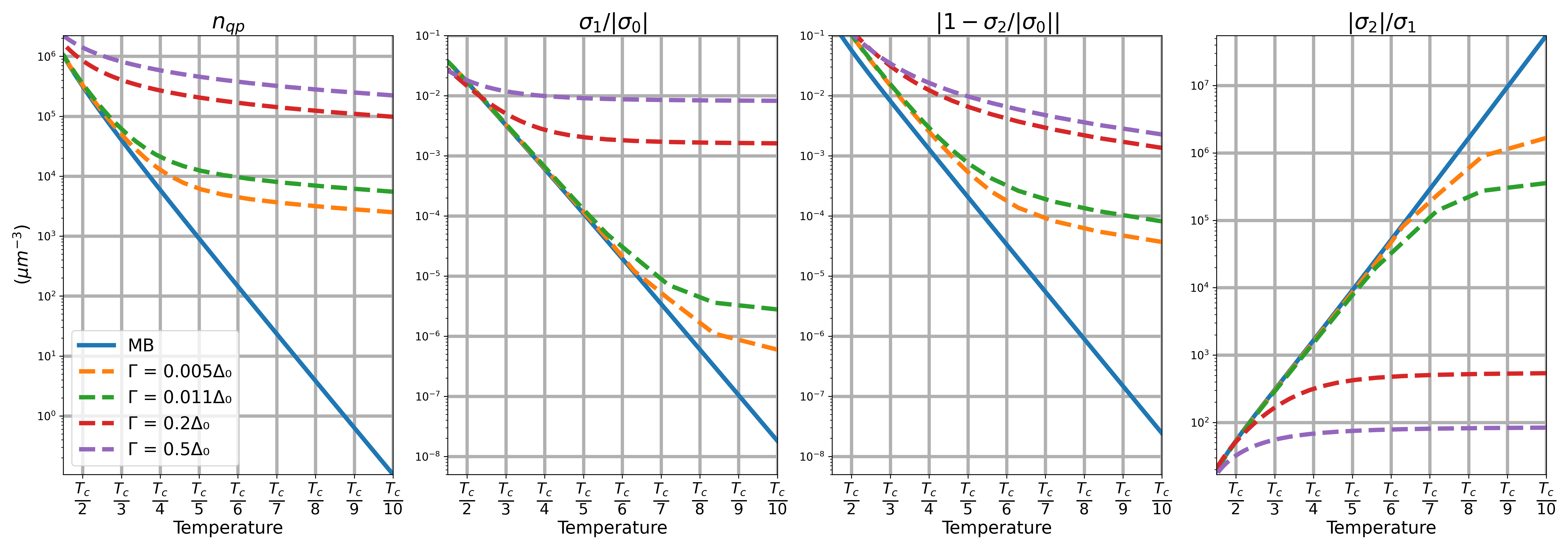}
    \caption{Quasiparticle density and complex conductivity versus a reciprocal temperature axis. Fractional deviations from $|\sigma_0| = \frac{\pi \Delta}{\hbar \omega}$ are computed with and without gap smearing. The $\Gamma=0.011\Delta_0$ traces correspond to the best fit value presented in the main text.}
    \label{fig:nqp_and_sigmas}
\end{figure}

The superconducting gap $\Delta$ and the kinetic inductance fraction $\alpha$ of the resonator are extracted by fitting the bath-temperature dependence of the resonant frequency and the internal quality factor to Mattis-Bardeen (MB) theory. In its idealized form, MB theory \cite{MattisBardeen1958} assumes the BCS quasiparticle density of states (DOS), given by:

\begin{equation}
n(E,\Delta) = \frac{E}{\sqrt{E^{2}-\Delta^{2}}},
\label{eq:bcs_dos}
\end{equation}
which is strictly zero for $|E|<\Delta$ and diverges at $E=\Delta$, exhibiting a hard gap. In the local, thin-film limit, expressions for the normalized complex conductivities are given by \cite{MattisBardeen1958}

\begin{equation}
        \frac{\sigma_1}{\sigma_n} = \frac{2}{\hbar\omega} \int_{\Delta}^\infty \left(f(E) - f(E+\hbar\omega)\right) \frac{E}{\sqrt{E^2-\Delta^2}}\frac{(E+\hbar\omega)}{\sqrt{(E+\hbar\omega)^2-\Delta^2}}\left(1+\frac{\Delta^2}{E(E+\hbar\omega)} \right) dE
\end{equation}
\begin{equation}
\frac{\sigma_2}{\sigma_n} = \frac{1}{\hbar\omega} \int_{\Delta-\hbar\omega}^\Delta\left(1 - 2f(E+\hbar\omega)\right) \frac{E}{\sqrt{\Delta^2-E^2}}\frac{(E+\hbar\omega)}{\sqrt{(E+\hbar\omega)^2-\Delta^2}}\left(1+\frac{\Delta^2}{E(E+\hbar\omega)} \right) dE,
\end{equation}

Where $\sigma_1$ and $\sigma_2$ are respectively the real and imaginary components of the superconducting complex conductivity and $\sigma_n$ is the normal-state conductivity. In the limits $\hbar\omega \ll\Delta$ and $k_BT \ll\Delta$, the conductivity response to a small excess quasiparticle density can be reduced to a closed form in terms of modified Bessel functions, the derivation of which are detailed in Ref~\cite{GaoThesis}, and are denoted as $\kappa_1$ and $\kappa_2$, respectively. Fig.~\ref{fig:VNAscan} shows that this analytical MB model matches the resonator data well below 100 mK but is unable describe the observed frequency shifts at higher temperatures. This behavior has been observed in similar films fabricated with various materials, where it has been demonstrated that the addition of a gap-broadening model yielded a better description of the data \cite{Temples_2024,deVisser2014,Zemlicka2015,li2025, Gavidia2026}. 

Dynes \textit{et al.} \cite{Dynes1978} phenomenologically described the broadening of the gap edge by adding a small imaginary parameter $\Gamma$ to the quasiparticle energy, $E \rightarrow E + i\Gamma$.
We reproduce the inclusion of the broadened density of states to Nam's generalization of MB theory \cite{Zemlicka2015, Nam1967}, which defines the complex functions

\begin{equation}
n(E, \Delta, \Gamma)=\text{sgn}(E)\frac{E+i\Gamma}{\sqrt{(E+i\Gamma)^2-\Delta^2}}
\label{eq:DOS_smeared}
\end{equation}
\begin{equation}
p(E, \Delta, \Gamma)={\text{sgn}(E)\frac{\Delta}{\sqrt{(E+i\Gamma)^2-\Delta^2}}},  
\end{equation}
where $\operatorname{sgn}(E)$ is $-1$, $0$, $1$ for $E<0$, $E=0$, $E>0$, respectively.
The modified expressions for the normalized complex conductivities are
\begin{equation}
\frac{\sigma_1}{\sigma_n} = \frac{1}{\hbar\omega} \int_{-\infty}^{\infty}
\left[f(E,T) - f(E+\hbar\omega,T)\right]
\left[\Re(n(E))\,\Re(n(E+\hbar\omega)) + \Re(p(E))\,\Re(p(E+\hbar\omega))\right] dE
\end{equation}
\begin{equation}
    \frac{\sigma_2}{\sigma_n} = \frac{1}{\hbar\omega} \int_{-\infty}^{\infty}
\left[1 - 2f(E+\hbar\omega,T)\right]
\left[\Im(n(E))\,\Re(n(E+\hbar\omega)) + \Im(p(E))\,\Re(p(E+\hbar\omega))\right] dE.
\end{equation}
In the limit $\Gamma \rightarrow 0$, Eq. \eqref{eq:DOS_smeared} recovers Eq. \eqref{eq:bcs_dos}, the clean BCS form. 
In both cases the quasiparticle density follows from integrating the
occupied DOS,
\begin{equation}
    n_\mathrm{qp}=4N_0\int_0^\infty\Re(n(E,\Delta,\Gamma))f(E,T) dE
\end{equation}
where $N_0$ is the single-spin normal-state density of states at the Fermi
level and $f(E,T) = 1/(1+e^{E/k_BT})$ is the Fermi distribution function. 

The measured quantities follow from the standard relations, 
 \begin{equation}
 f(T) = f_0 \left(1 + \frac{\alpha}{2} \frac{\sigma_2(T) - \sigma_2(0)}{\sigma_2(0)}\right),  \label{eq:f_T} \end{equation}

\begin{equation}
    Q_i(T)^{-1} = \alpha\frac{ \left(\sigma_1(T) - \sigma_1(0)\right)}{|\sigma_1(0) - i\sigma_2(0)|} + Q_\mathrm{i,0}^{-1},
    \label{eq:Qi_T}
\end{equation}
where  $Q_{i,0}$ collects loss mechanisms that exhibit no temperature dependence. We work in the thin film limit, $\gamma = 1$, throughout.

We apply a treatment similar to that described in Ref. \cite{li2025} to numerically integrate the complex conductivity responses and extract resonator observables and allow all remaining parameters to float.
Fitting Eqs. \eqref{eq:f_T} and \eqref{eq:Qi_T} jointly to the Run 2
data yields $\Delta_0 = 196.1 \pm 0.7~\mathrm{\mu}$eV and
$\Gamma = 2.17 \pm 0.06~\mathrm{\mu}$eV (Fig.~\ref{fig:VNAscan}), corresponding to
\begin{equation}
\Gamma/\Delta_0 = (1.11 \pm 0.03)\times 10^{-2} ,
\end{equation}
and reduces $\chi^2_\nu$ from $615.11$ to $0.82$. The best-fit value is approximately represented by the $\Gamma = 0.001\Delta_0$ curve of Fig.~\ref{fig:nqp_and_sigmas}, the most weakly broadened case shown; the $0.012\Delta_0$ and $0.2\Delta_0$
curves are also included to illustrate the qualitative trend of increasing $\Gamma$. 

As a final note, we can comment on $Q_i$ trends seen for this device prior to the data presented here. The quality factor of this device is the highest of all measurements made in this cryostat without the MEMS system. We observe significant run to run variation in $Q_i$ that we attribute to magnetic flux trapping in these devices - more detailed studies of these effects are the subject of future work. The high quality factor in this case is likely due to the KID being located close to the bottom of our magnetic shielding can, where field attenuation is the largest. The additional optical filter added at the output of the MEMS system likely contributes to more substantial IR isolation than in other runs in which fibers were directly coupled to the device box lid. We find it likely that the residual flux trapping, the result of a low critical field, is related to our need for gap-smearing fits, and may also have a connection to the other anomalous behavior observed in our amplitude modeling.

\newpage

\begin{figure}
\includegraphics{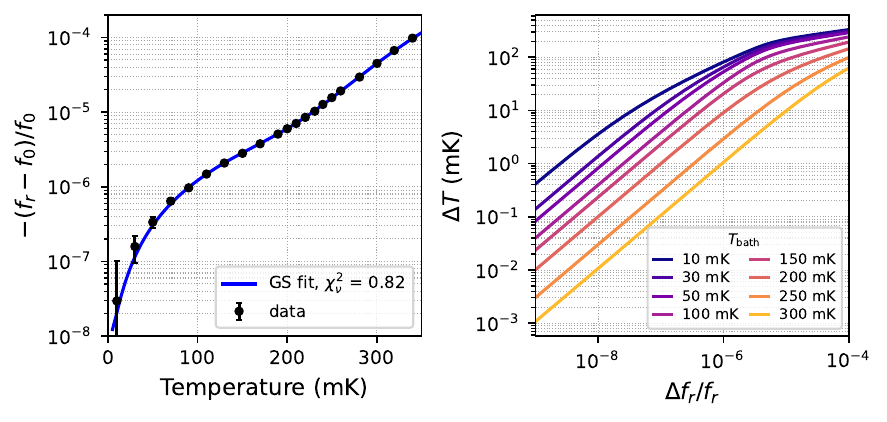}
\caption{\textit{Left}: same as Fig.~\ref{fig:VNAscan} Left with a logarithmic $y$-axis.
\textit{Right}: conversion of $\Delta f_r/f_r$ into $\Delta T$ for several bath temperatures,
using the gap-smearing lookup. Lower bath temperatures give a larger $\Delta T$ for a fixed
$\Delta f_r/f_r$, because $f_r(T)$ is nearly flat at low temperature.}
\label{fig:lookup_semilogy}
\end{figure}
\section{$\Delta T$ conversion}\label{app:dT}


We would like to calculate $\Delta T \equiv T_\mathrm{L} - T_\mathrm{bath}$ as an estimator for absorbed energy, where the subscript $L$ refers to the resonators under some phonon loading from an LED pulse that induces a resonant frequency shift $\Delta f_r/f_r$ and $T_\mathrm{bath}$ refers to the MXC stage temperature. 
We use the following formula to extract $T_L$:
\begin{equation}
\Delta T = g\left(g^{-1}(T_\mathrm{bath})-\frac{\Delta f_r }{f_r}\right) - T_\mathrm{bath},
\end{equation}
where $T= g((f_r-f_0)/f_0)$ is a linear interpolation of the gap smearing fit with a step size of $0.5$\,mK.  $f_0= f(T=0K)$ is the resonant frequency calculated from the gap smearing fit presented in Fig.~\ref{fig:VNAscan}. The expression inside the large parentheses can be understood in the following way:
\begin{equation}
g^{-1}(T_\mathrm{bath})-\frac{\Delta f_r }{f_r} = \frac{f_g-f_0}{f_0}- \frac{f_g-f_r(t)}{f_r(t)}\approx \frac{f_r(t)-f_0}{f_0},
\end{equation}
where $f_g$ is the frequency of the generator tone chosen at temperature $T_\mathrm{bath}$, which is just the un-loaded resonant frequency at that temperature. In other words, since the on-resonance timestreams are explicitly measured relative to the readout tone set at the KID's unilluminated resonance state, we convert the data in terms of a frequency shift relative to $f_0$ in order to use the gap smearing fit $g$ for any pulse at any temperature. This procedure is robust against shifts in $f_0$, so long as we assume that the shape of the $T= g((f_r-f_0)/f_0)$ is constant between and during cooldowns.

\begin{figure*}[ht]
    \centering
    \includegraphics[width=\linewidth]{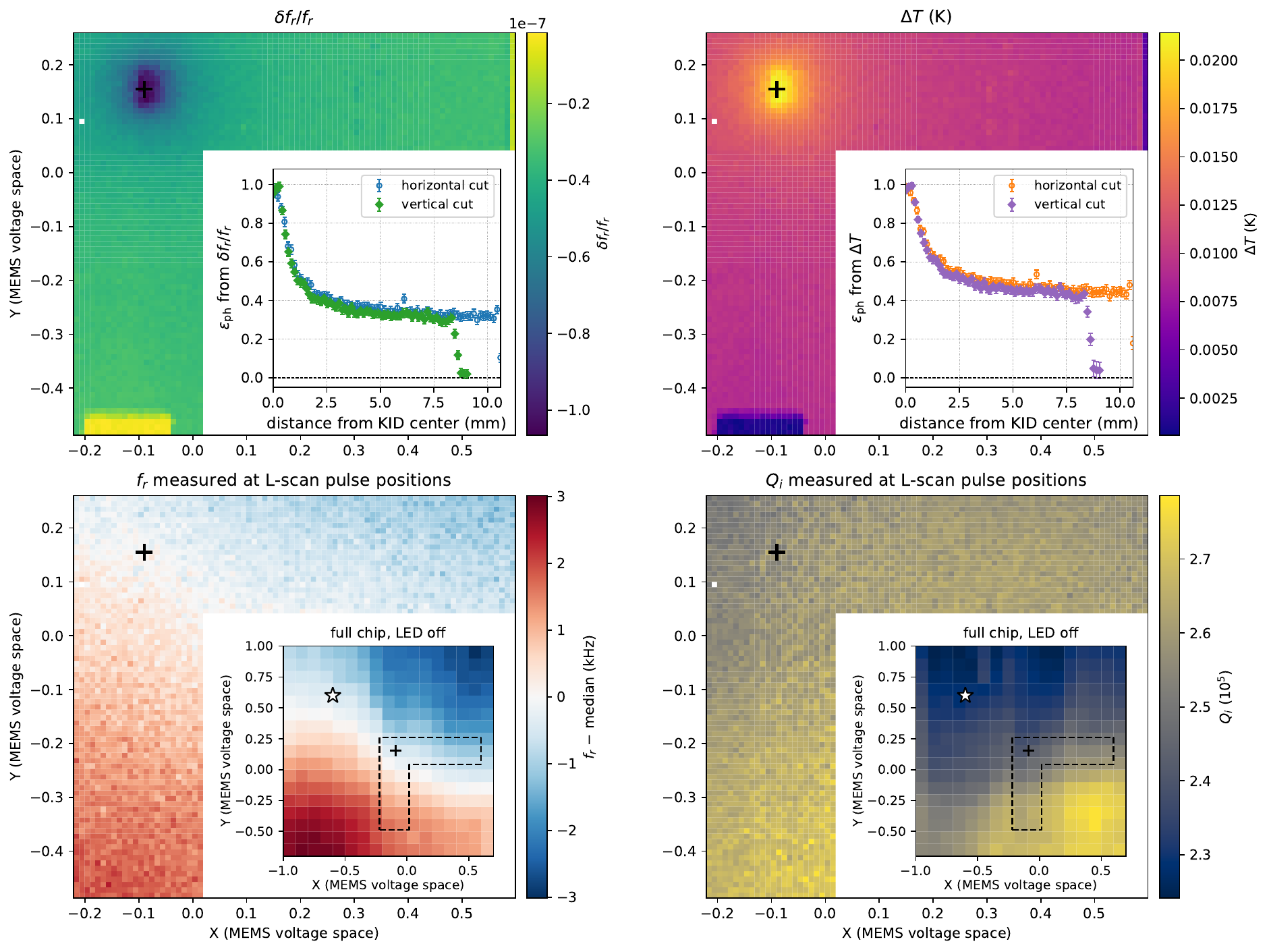}
    \caption{Heatmaps of the L-scan data. \textit{Top Left} Pulse heights in linearized fractional frequency shift units. \textit{Top Right}: $\Delta T$ pulse heights, using the conversion described in this Appendix. The relative phonon collection efficiency, for both energy estimators and in both the horizontal and vertical cuts are shown in their respective insets. \textit{Bottom}: $f_r$ and $Q_i$ versus MEMS position, with no LED on, both over the L scan and over the entire chip. The $f_r$ heatmap is centered around its median whereas the $Q_i$ heatmap is absolute.  The star is the off-KID position used in the temperature analysis. The plus is the on-KID position.}
    \label{fig:dT_lscan}
\end{figure*}

We show the L-scan data converted to $\Delta T$ in Fig.~\ref{fig:dT_lscan}. The data was taken at 10.5\,mK, near the base temperature of the fridge, so the conversion assumes a constant $T_\mathrm{bath}$ for the entire dataset. At far distances from the KID, $\Delta T$ shows a higher $\epsilon_\mathrm{ph}$ by about 40\% as compared to the $\delta f_r/f_r$ curve.

We also highlight that the measured $f_r$ and $Q_i$ change with the MEMS mirror position, at the level of several kHz for $f_r$ over the entire chip; see Fig.~\ref{fig:dT_lscan} Bottom. At base temperature, a 6\,kHz shift corresponds to an 80\,mK discrepancy in the inferred bath temperature of the KID. However, we do not believe that this is a thermal effect because the $f_r$ and $Q_i$ maps are not correlated in a way that would indicate thermal behavior. 



In the temperature-scan data, we measure $\Delta f_r/f_r$ using the non-linear and ring-up corrected conversion described in Appendix \ref{app:pulse_fitting}. The resonant frequency is only measured once, before the on-KID data taking, so the discrepancy between the on-KID and off-KID LED-off $f_r$'s is not fit for. Fig.~\ref{fig:dT_lscan} indicates that the resonant frequency between the star and the plus shape is less than a kHz. However, we do not know if this gradation could have shifted between when the Fig.~\ref{fig:dT_lscan} inset data was taken and when the temperature data was taken. Although the $f_r$ was not measured between LED positions, we note that the pre-pulse baselines at the on- and off-KID positions agree in $S_{21}$ to within $2 \times 10^{-5}$ at every temperature, which corresponds to resonance frequencies equal within 4\,Hz and $Q_r$ equal within $1.3 \times 10^{-4}$.

\clearpage

\newpage

\begin{figure}
\includegraphics{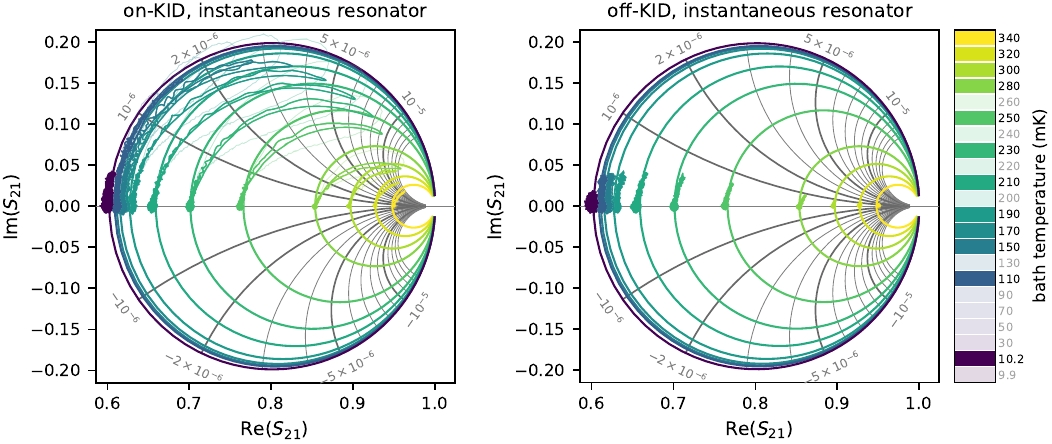}
\caption{Pulse trajectories in $S_{21}$, with an overlain Smith chart based on a M\"{o}bius transformation that converts $\Delta f_r/f_r$ into $S_{21}$. Caption is the same as Fig. \ref{fig:pulseShapeModeling} Left, with more temperatures shown. }
\label{fig:pulses}
\end{figure}
\section{Run 2 Pulse Fitting}\label{app:pulse_fitting}
On- and off-KID $S_{21}$ pulse trajectories are shown over temperature in Fig. \ref{fig:pulses} for data taken during Run 2. As mentioned in the main text, since the pulses are comparable to the size of the circle, we use a nonlinear transformation to extract $\delta f_r/f_r$. The excursions of the pulse beyond a particular $S_{21}(f)$ circle is due to the resonator ring-up time, which we discuss below. The data are converted into timestreams and fit to a combined on-KID and off-KID model. The results of these fits are used in the main text. Here we describe the details of the pulse timestream fit analysis.

We remind the reader that this is not required for the low temperature data, where the pulses are much slower than the resonator ring-up and the temperature analysis leads to identical conclusions as the $\delta{f_r}/{f_r}$ analysis, as described in Appendix~\ref{app:dT}. The analysis in this section shows that the time constants found from fitting the lowest temperature data are consistent with this picture.

\subsection{Binning and error estimation}

As can be deduced from Table \ref{tab:pulse_params}, an individual timestream of data is approximately 14\,s at 2.4\,MHz. The 2\,Hz LED repetition rate was chosen because of the observed $\sim100\,$ms fall times that were observed in the lowest temperature pulses. There are thus about 28 pulses that populate an individual timestream, and there are five such timestreams taken at each temperature and at both on- and off-KID positions. 2.4\,MHz was chosen for this study so that the rise time of the pulses, which can be no faster than the resonator ring-up time (about 1\,$\mu$s at the highest temperature), could be well resolved.

The timestreams are divided into 0.5\,s chunks with one pulse each and averaged over a particular 14\,s timestream. We also implement averaging across the five identical timestreams, which we refer to as trials, but the alignment between different timestreams is not known because a trigger was not used to align the LED pulses to a known time sample of the timestream during data taking. The pulses occur about 5\,ms into each chunk. We perform a shift in $S_{21}$ to align the pre-pulse baseline to $1-Q_r/Q_c$ before averaging different trials. The averaged, 0.5\,s pulse timestream is then split into two regimes: early pulse and late pulse. The early pulse is defined to be $-30\,\mu$s to 200\,$\mu$s relative to the 5\,ms alignment point. The late pulse is any time after that. 

\textit{Early pulse:} The readout noise is correlated between neighboring samples, so the uncertainty in the early pulse points is treated with generalized least squares. The covariance matrix is computed from the 4.2\,ms pre-pulse baseline and is used to calculate the appropriate weighted residual for a particular fit candidate. The autocorrelation function that is built into the covariance matrix quantifies the level of correlation between different points in time and so calculates the residual at a particular point in time to be appropriately weighted by the residuals at neighboring times. We find that this increases the uncertainties of the fitted parameters by factors of 1.0 to 2.2 compared with assuming independent samples. 

\textit{Late pulse:} when the correlation between different time points cannot be used because there is not enough pre-pulse baseline to estimate the correlation at longer timescales, we use the trial-to-trial scatter. Since there are only five trials, this metric can be noisy for a single time point. Instead, we put the data into 255 logarithmically-sized bins, compute the standard deviation $\sigma_\text{bin}$ of the five trials' bin means, and assign $\sigma_\text{bin}\times\sqrt{\text{\# of points in bin}}$ to each point within that bin. 255 bins from 200\,us to 0.5\,s means that each bin covers approximately 3\%. Since the pulses are not decaying rapidly beyond 200\,$\mu$s, 3\% bin width is sufficient for error estimation. We note now that we perform this entire analysis and fitting pipeline with different bin widths and find no substantive change to the results. Furthermore, we comment that, throughout the model-building process, we fit to the average values of the logarithmic bins themselves. In other words, instead of a half second at 2.4\,MHz, we used logarithmic bins to dramatically reduce the number of points to fit to. The final results that are presented in this section are all fits to unbinned data, but the errors at late times are binned.

Throughout this appendix, we show timestreams of the fitted pulses. For longer timescale data, we apply additional binning for plotting purposes.

\subsection{Resonator ring-up time and LED pulse width corrections in the on-KID rise times}

As we describe in Subsection \ref{sec:onKID}, the on-KID data is fit to a sum of a prompt component and the off-KID contribution. The prompt component has a few corrections that have to do with proper peak estimation of the pulse. In general, if there is a rise time that is comparable or not much faster than the fall time of the pulse, then the peak of the data will underestimate the energy delivered to the sensor. We start with a discussion of rise times in the pulses, which is relevant for fitting all of the data, but is most visibly impactful in the on-KID data, particularly at middling and high temperatures.

The resonator ring-up time $\tau_r = Q_r/\pi f_r$ is visible in the excursions of the pulse trajectories away from the steady-state $S_{21}(f)$ resonance circles: the resonator cannot follow changes in its resonant frequency and quality factor that are faster than its ring-up time. The subsequent effect in the IQ plane is for the pulse to form a loop rather than a line in IQ space. For a resonator driven at a fixed readout tone, the field strength in the resonator is proportional to $y(t) = 1 - S_{21}(t)$ and obeys the relationship
\begin{equation}
    \frac{dy}{dt} = \frac{\omega_r}{2}\left[\frac{1}{Q_c} - z(t)\,y(t)\right], \qquad
    z(t) = \frac{1}{Q_i(t)} + \frac{1}{Q_c} + 2i\,\frac{f - f_r(t)}{f_r(t)},
\end{equation}
so that $y$ relaxes toward its steady-state value $y_\text{ss} = (1/Q_c)/z$ (the resonance circle) on the ring-up time. Here, $y$ is our measured field strength and $z$ is the field strength we want to estimate, which contains the frequency and dissipation information with the ring-up measurement effect removed. Solving this equation for $z$ at each instant gives the instantaneous response,
\begin{equation}
    \left(\Delta\frac{1}{Q_i}-2i\frac{\Delta f_r}{f_r}\right)_\text{inst}
    = \left(\Delta\frac{1}{Q_i}-2i\frac{\Delta f_r}{f_r}\right)_\text{ss} - \frac{2}{\omega_r}\frac{d\ln y}{dt},\label{eq:res_ringup}
\end{equation}
where the steady-state term is the usual conversion of $S_{21}$ through the resonance circle,
$z_\text{ss} = (1/Q_c)/y$, and the second term corrects for the lag of the resonator behind its changing parameters. In steady state $dy/dt = 0$ and the two coincide. 

\begin{figure}[p]
\includegraphics[width=0.9\textwidth]{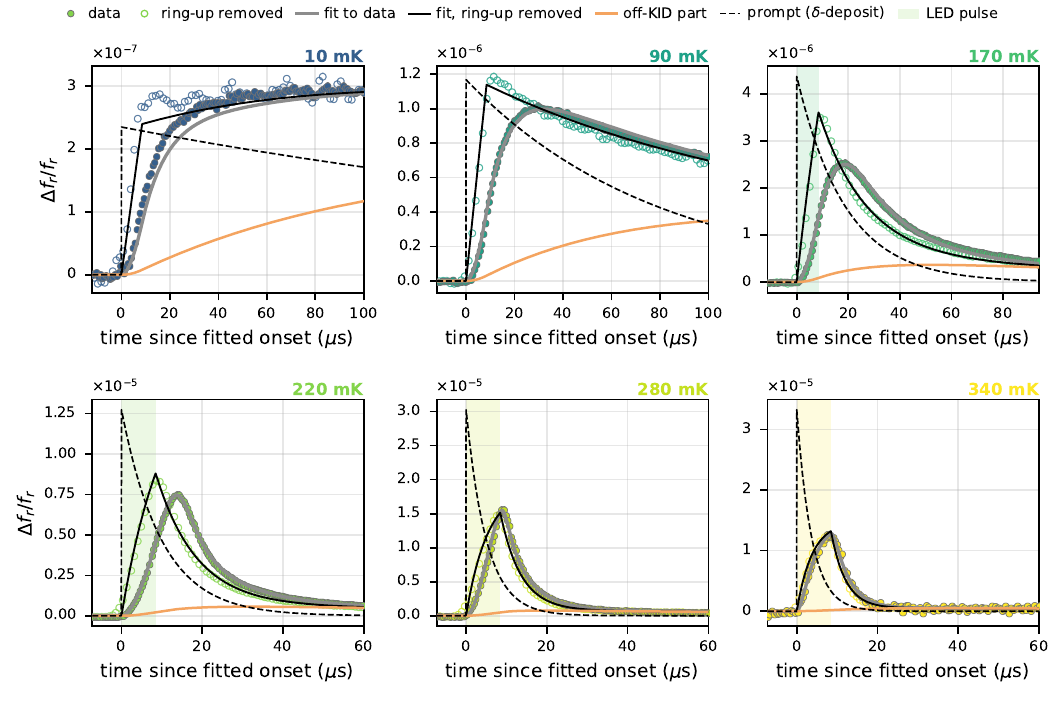}
\includegraphics[width=0.9\textwidth]{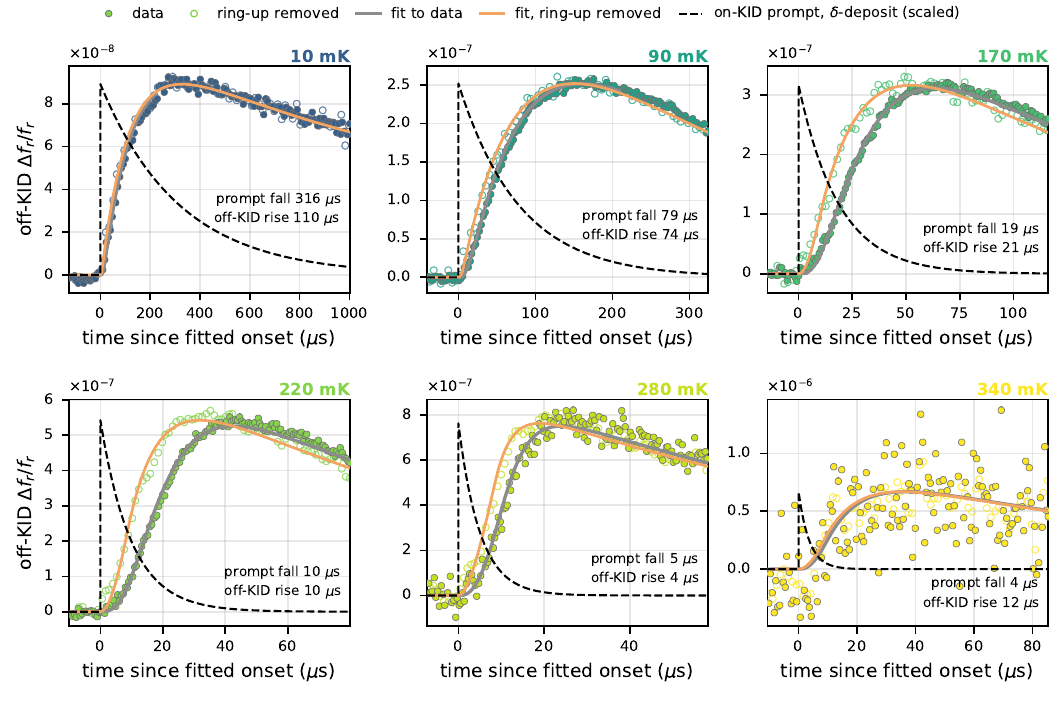}
\caption{On-KID (top) and off-KID (bottom) rise time fits at different temperatures. We compare the rise time with the decay of the prompt component from the on-KID data and note their similarity.}\label{fig:riseFits}
\end{figure}

The left-hand side of  Eq.~\ref{eq:res_ringup} is just $\Delta z$, which is our ring-up corrected pulse with the baseline subtracted. Evaluating this numerically involves a time derivative of the measured trace, which amplifies the high-frequency readout noise and correlates neighboring samples. We therefore fit the measured data and forward-model the resonator instead. For each candidate set of pulse parameters, the model $\Delta f_r/f_r(t)$ is converted to $z(t)$, with the dissipation term $\Delta(1/Q_i)=\beta\,2\Delta f_r/f_r$ using the ratio $\beta$ measured from the data. The equation of motion is then integrated to obtain the $y(t)$ the resonator would produce, and $y(t)$ is converted to $\Delta f_r/f_r$ with the same steady-state relation, $z_\text{ss}=(1/Q_c)/y$, that is applied to the data. The fit as the resonator records it is shown in gray in Fig.~\ref{fig:riseFits}; the fitted pulse itself, with the ring-up removed, is shown in black and used in the rest of the analysis. The ring-up-corrected data are shown only for visual comparison.

We also fit the LED pulse width $W$ with the following model:
\begin{equation}
\mathrm{prompt}(t) =
\begin{cases}
0, & t \le t_0,\\[2pt]
\dfrac{\tau_p}{W}\left(1 - e^{-(t-t_0)/\tau_p}\right), & t_0 < t < t_0 + W,\\[8pt]
\dfrac{\tau_p}{W}\left(e^{W/\tau_p} - 1\right)e^{-(t-t_0)/\tau_p}, & t \ge t_0 + W,
\end{cases}
\end{equation}
where $\tau_p$ is the response time of the prompt component and $t_0$ is the onset of the pulse. This is a convolution assuming a flat-top injection of energy with width $W$ and a temperature dependent KID response time. $W$ is fit globally while $t_0$, $\tau_p$, and $A_0$ are fit per temperature. The basic idea is that energy is leaking out of the resonator faster than the LED pulse, so the same time constant is seen both during the rise and fall of the pulse. We show the fit of this model to the data in Fig. \ref{fig:riseFits} with the solid black line. We also fit for the amplitude.

Once the single time constant, pulse width, and amplitude are fit for, the energy deposition from a $\delta$-function deposit ($W\rightarrow0$) can be estimated directly from the amplitude. We show the hypothetical pulse shape from such a deposition with the dashed black curves in Fig. \ref{fig:riseFits}. The discrepancy between the $\delta$-function deposit and the fitted curves is largest at the highest temperatures, where $\tau_p$ is shortest, so most of the prompt signal decays while the LED is still on. This also shows that the amplitude at the lowest temperatures is an accurate energy estimator without needing these corrections, allowing us to retain the simpler processing for our spatial mapped base temperature data. In Fig. \ref{fig:riseFits}, the off-KID component of the fit is also visible, but we save that discussion for Subsection \ref{sec:onKID}. We introduced the rise time components first because they are relevant for the off-KID fitting but are less visibly impactful.

\subsection{Off-KID pulse shape model}
The first step of the fitting process is to fit to the off-KID data. The ring-up is forward-modeled as we mentioned in the previous section, and should thus already be considered in the fit. We define a two-pole exponential in the following way:
\begin{equation}
p(t;\tau_r,\tau_f,t_0) = \Theta(t')\left(e^{-t'/\tau_f}-e^{-t'/\tau_r}\right)\quad t'=t-t_0
\end{equation}
We identify $\tau_f$ as a fall time and $\tau_r$ as a rise time; $t_0$ is the onset time of the pulse.

\begin{figure}[p]
\includegraphics[width=0.9\textwidth]{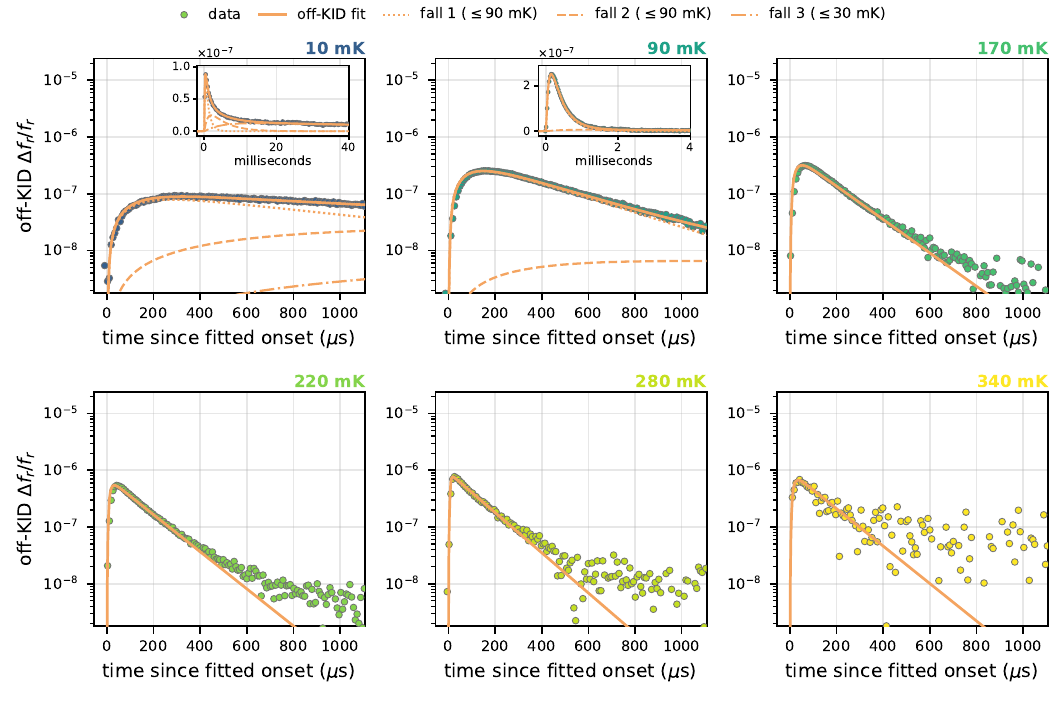}
\includegraphics[width=0.9\textwidth]{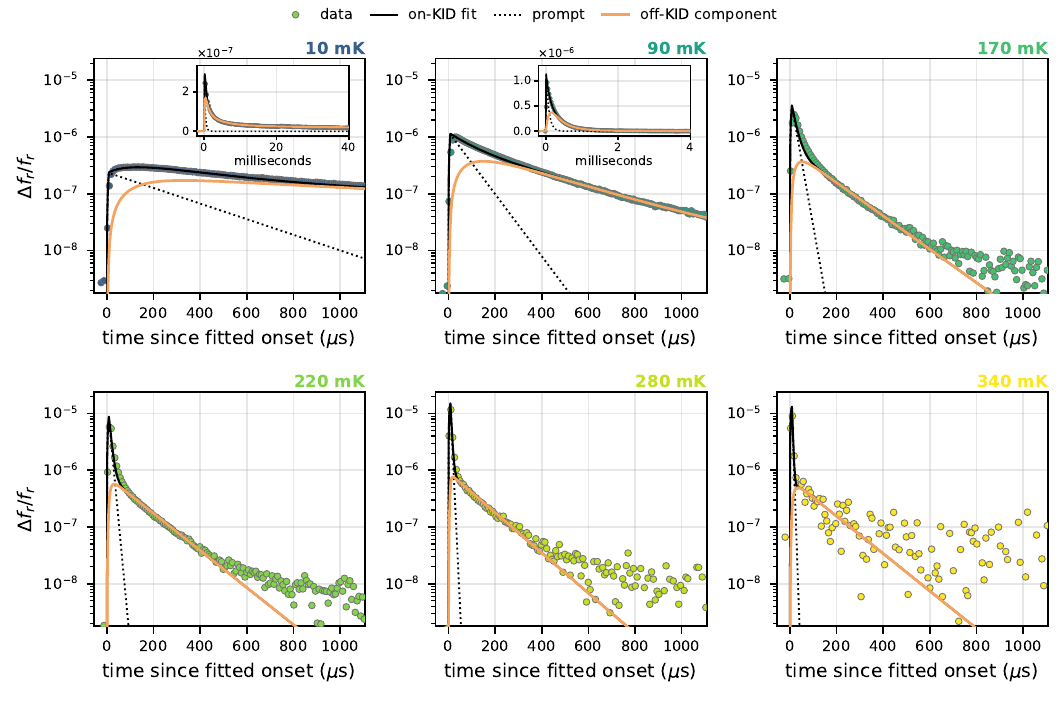}
\caption{Off-KID (top) and on-KID (bottom) fall time fits. Insets show the performance of the fit for lower temperatures where the long-timescale features are most notable.}\label{fig:fallFits}
\end{figure}


The off-KID data is fit for one to three fall times, depending on the temperature:
\begin{equation}
x(t) = 
\begin{cases}
A_1 p(t;\tau_r,\tau_{f1},t_0), & T > 90\,\mathrm{mK},\\
A_1p(t;\tau_r,\tau_{f1},t_0)
 + A_2p(t;\tau_r =\tau_{f1},\tau_{f2},t_0), & 30\,\mathrm{mK} < T \le 90\,\mathrm{mK},\\
 A_1p(t;\tau_r,\tau_{f1},t_0)
 + A_2p(t;\tau_r =\tau_{f1},\tau_{f2},t_0) + A_3p(t;\tau_r=\tau_{f2},\tau_{f3},t_0) & T \le 30\,\mathrm{mK},\\
\end{cases}
\end{equation}
We set $\tau_{f3}>3\tau_{f2}$ and $\tau_{f2}>3\tau_{f1}$. The rise times of the longer fall time pulses are given by the fall times of the preceding fall time. $A_1$, $A_2$, and $A_3$ are amplitudes. $\tau_{f2}$ is included only for the six lowest temperatures. The longest fall time that is around 100\,ms is only applied at the three lowest temperatures. These boundary choices are arbitrary, but we note that the amplitude of these slow-time components are negligible to the peak of the pulse at early times.

The amplitude of the off-KID pulse is defined to be the response to a $\delta$-deposit of energy, though as with the on-KID data, the finite LED width is included in the modeling. The effect of the square pulse on the off-KID response is much less than the on-KID fits because the pulse fall times are all around 100\,$\mu$s or more, so the pulse has sufficient time to rise to its peak without energy leaking out over the 8.5\,$\mu$s LED pulse width. We show the results of the rise time fits in the off-KID data in Fig. \ref{fig:riseFits}. We also show the prompt component decay in the same plots. The fact that they are similar at the higher temperatures indicates that this unified time constant is potentially a sensor response time, which could be a quasiparticle lifetime. 


The quality of the fall time fits is highlighted in Fig. \ref{fig:fallFits}. When applicable, multiple fall times are shown. We find that a simple rise and fall model for the off-KID pulses is sufficient to model most of the higher temperature data, and the longest timescales require a time constant on order of 100\,ms, but the pulses may be better explained by other models. The increasing number of time constants at lower temperature is evidence that one potentially better model would be to have a stretched exponential of the form $\exp(-(t/\tau)^\gamma)$, where $\gamma<1$ causes the later times to decay less quickly than at earlier times. We leave more careful investigation of the long time constants for future study because this present study is focused on peak amplitudes, which is important for the energy resolution of the detector. At most, the second fall time contributes 10\% to the peak of the off-KID fit, and only at the lowest temperatures.

\subsection{On-KID pulse shape model}\label{sec:onKID} 
The second step is to fit the on-KID data to a sum of the off-KID component and the prompt component:
\begin{equation}
\text{on-KID}(t) = B_1\;\text{prompt}(t) + B_2\;\text{off-KID}(t)\label{eq:onKIDpulse}
\end{equation}
$B_1$ and $B_2$ are amplitudes. We follow \cite{Florang2026} with this model, where they observe a constant off-KID component over the entire chip. Further heritage for this two-component model can be found in \cite{Moore2012} and \cite{Temples_2024} where the data is fit to ``prompt'' and ``delayed'' components. We synthesize these models into Eq. \ref{eq:onKIDpulse}.

In an earlier subsection, we examined the performance of $\text{prompt}(t)$ in its capacity to extract the peaks of the pulses. In Fig.\;\ref{fig:fallFits}, the fall times are brought into view, as well as the overall ability of Eq. \ref{eq:onKIDpulse} to describe the data. The peak of the pulse is dominated by the prompt component, and the off-KID component dominates the long timescale behavior. At the lowest temperatures, the peak of the pulse is actually at some intermediate time between the peak of the $\delta$-deposit response and the peak of the off-KID component. The off-KID component subtraction is thus important for proper estimation of the zero-bounce phonon flux into the KID.

\begin{figure} 
\includegraphics{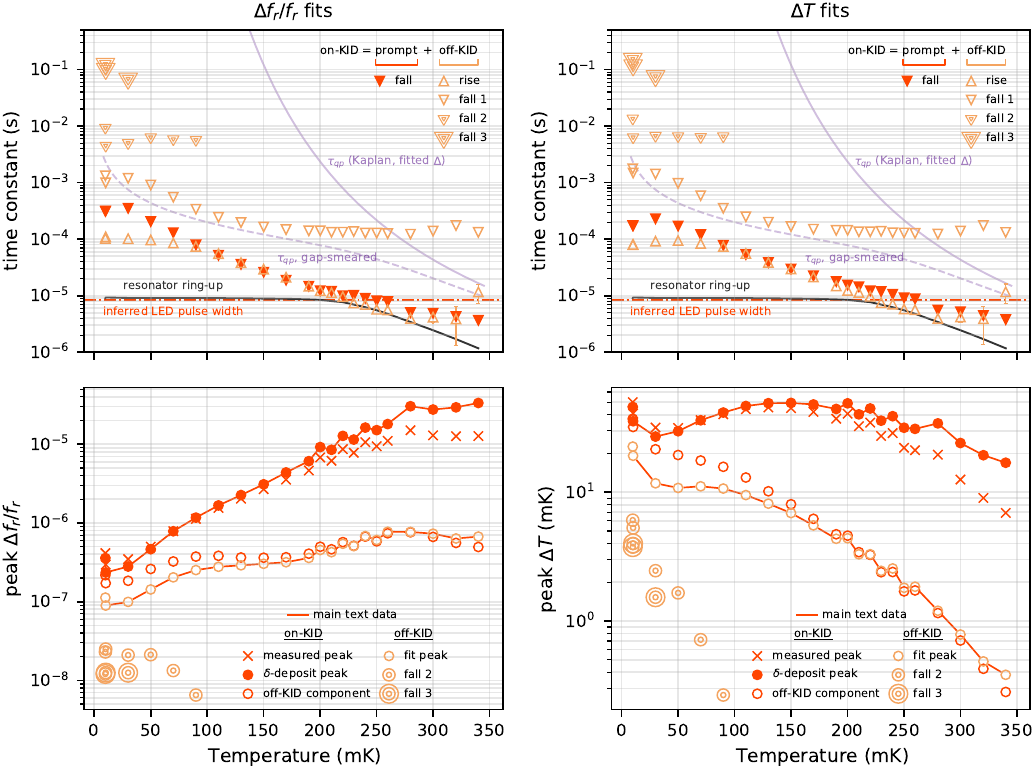}
\caption{Time constants and amplitudes for $\Delta f_r/f_r$ and $\Delta$T. off-KID parameters are shown in faded color and hollow markers, and on-KID parameters are shown as solid markers. The orange-red line in the bottom row of plots indicate the peak $\Delta f_r/f_r$ and $\Delta T$ that are used in the main text.}\label{fig:fit_params}
\end{figure}

\subsection{Understanding Temperature Trends}

We plot the global time constant and amplitude fits versus temperature in Fig. \ref{fig:fit_params}. The orange lines indicate what is used in the main text. Given the varying number of fall times required at low temperatures, and the discussion of the simpler high-temperature data in the main text, we can comment on some meaningful trends. So far, the fitting has not assumed anything physical about the time constants, aside from decomposing the on-KID response into prompt and off-KID components. Here we make some assertions about the nature of the fastest time constant in the high-temperature data, and show that the conclusions that follow suggest this is a reasonable assumption. In particular, we show that phonon collection efficiency is not temperature dependent, and the apparent reduction from the $\Delta$ T ratios has more to do with better time resolution of the prompt and delayed phonon components.

First, we note that the prompt component is only characterized by a single time constant. In the main text, we treat this as our estimate for linearized KID response, and our main results presented in Sections IV-V suggest it is likely dominated by thermal dynamics in the KID. At high temperatures, we show that the response is consistent with a fixed pair-breaking density, and thus this time constant could become limited by quasiparticle recombination. However, it is much shorter than the quasiparticle lifetime that can be computed from \cite{Kaplan1976Quasiparticle}. We also show what the gap-smeared $\tau_\text{qp}$ would be for comparison. In the main text, we posit that intrinsic quasiparticle dynamics begin to limit the KID response only at the highest temperatures studied.

We can explore this further by deriving total energy from the off-KID model to show how the fall times couple to the meaning of the amplitude. For a thermal sensor, we know that the sensor response to a time-dependent power injection $\Delta P(t)$ is given by \cite{Kurinsky}
\begin{equation}
    \Delta T(t) = \int_{-\infty}^{\infty} \frac{1}{C}e^{-\tau/\tau_r} \Delta P(t-\tau) d\tau
\end{equation}
where we have labeled the faster fall time from the off-KID response as $\tau_r$. The impulse response is given by setting $\Delta P(t) = E\delta(t)$, such that
\begin{equation}
    \Delta T(t) = \frac{E}{C}e^{-t/\tau_r}
\end{equation}
If we integrate this response, we get
\begin{equation}
    \int_0^\infty \Delta T(t) dt = \frac{\tau_r E}{C} \approx \frac{E}{G}
\end{equation}
where $G$ is the thermal conductance of the sensor to the bath, showing us that the amplitude, not the integral, is the optimal energy estimator for a fixed heat capacity model. 

The off-KID model also has a longer fall time component, which we can think of as a non-trivial time-dependent power absorbed by the KID. This has the form
\begin{equation}
    \Delta P = \frac{E}{\tau_f} e^{-t/\tau_{f}}\theta(t)
\end{equation}
which integrates to: $\int_0^\infty \Delta P(T) dt = \tau_{f} \frac{E}{\tau_{f}} = E$. If we convert these both to units of effective power, we find that the phonon flux has the same form, but the sensor impulse response is converted to power as $\Delta P \approx G\Delta T$ and thus the G in the denominator cancels.

Let's combine these by convolving sensor response with phonon flux. In that case, we have
\begin{equation}
    \Delta T(t) = \int_{-\infty}^{\infty}d\tau \frac{E}{C\tau_f}e^{-\tau/\tau_r}e^{-(t-\tau)/\tau_{f}}\theta(t-\tau) = \frac{E}{C} \frac{r}{r-1}\left[e^{-t/\tau_r}-e^{-t/\tau_f}\right]
\end{equation}
where $r=\tau_r/\tau_f$. We thus end up with the model for the off-KID component; here though, if we interpret the faster time constant as the sensor response, as suggested by our fits, we find that there's a correction containing the time-constants we need to apply. In the limit $r << 1$ that the response time is much shorter than the phonon flux, we find that the amplitude is modified as
\begin{equation}
    \Delta T(t) \approx \frac{r}{C} e^{-t/\tau_f}
\end{equation}
and thus a finite phonon flux, even for a flat heat capacity, can appear as a reduced temperature change for an increasingly faster response despite no actual change in energy incident on the device.

\begin{figure}
    \centering
    \includegraphics[width=\linewidth]{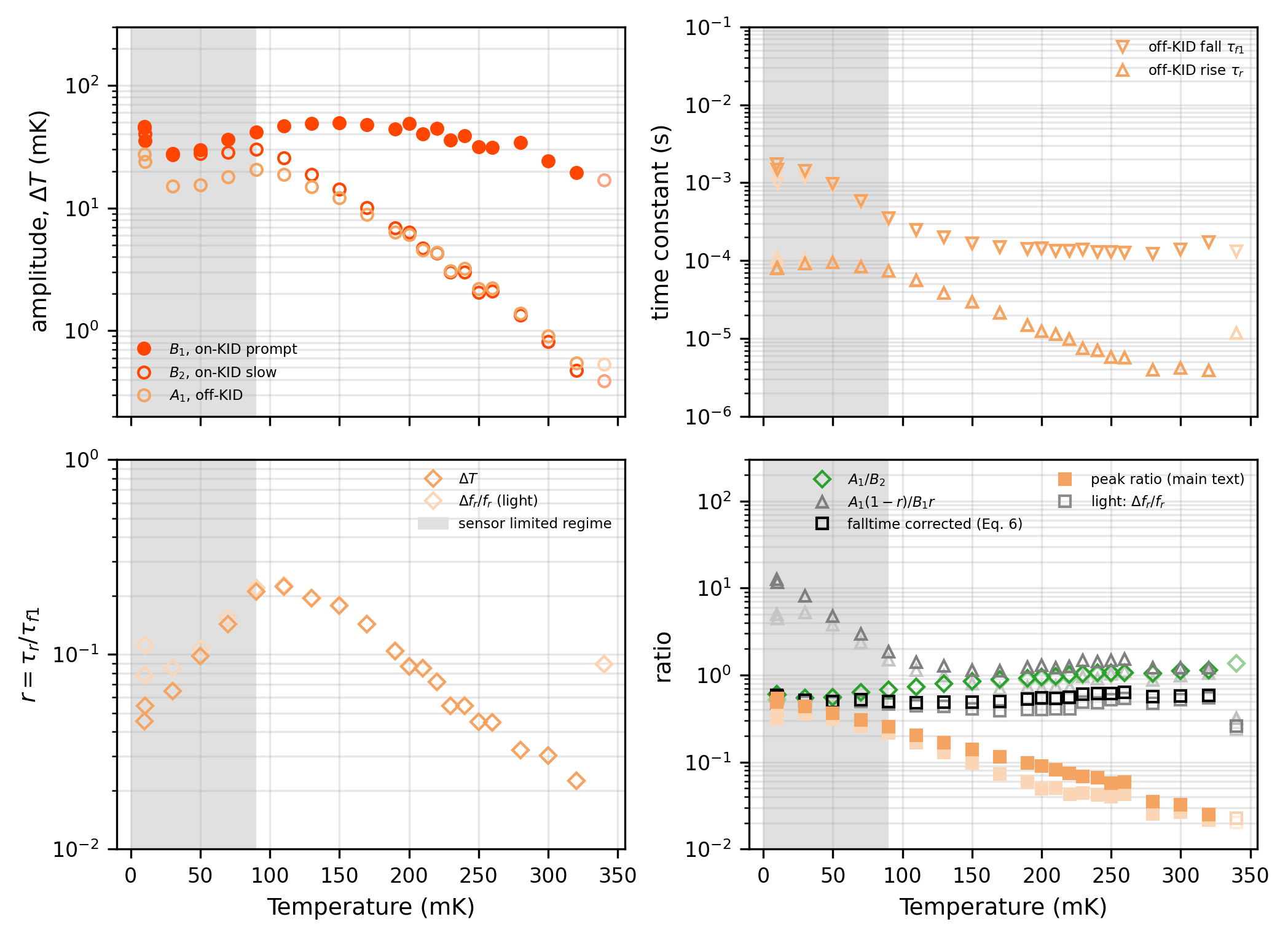}
    \caption{Process of converting raw $dT$ amplitudes into an efficiency ratio, showing that the interpretation of sensor response suggested in this section is consistent with a temperature-independent phonon efficiency. The top row is the same data from Fig.~\ref{fig:fit_params} but excludes the additional fall times in the slow response fit for simplicity, to highlight the main amplitudes and trends studied. We highlight in grey the region where multiple fall times are used in the fits, coinciding with the regime where the response is sensor limited rather than phonon limited. The bottom row shows the imputed fall time ratio $r$, and the comparison of different ratios exploring alternate interpretations that show different aspects of inferred behavior. The result from the main text is shown as black squares. Also shown are the slow amplitude ratio ($A_1/B_2$) which implies that the slow pulse amplitude is identical at high temperature, and begins to drop where pole mixing occurs. We also see the peak ratio dropping exponentially as the fast and slow components become more distinct. Computing the correction without the slow amplitude suggests a non-physical rise at low temperature as the denominator becomes a worse estimator of total energy.}
    \label{fig:efficiencyRatio}
\end{figure}

We can connect this to $\epsilon_{\mathrm{off-KID}}$ as follows. If we take only the results from the two-pole model for the off-KID data, the energy estimator takes the form
\begin{equation}
    E_{\mathrm{off-KID}} = A_1 C \frac{1-r}{r}
\end{equation}
The on-KID energy is the sum of an off-KID pulse shape with amplitude $B_2$ and on-KID simple energy estimator $B_1$, so we find
\begin{equation}
    E_{\mathrm{on-KID}} = B_1 C + B_2 C \frac{1-r}{r}
\end{equation}
Finally, this gives
\begin{equation}
    \epsilon_{off-KID} = \frac{E_{off-KID}}{E_{on-KID}} = \frac{A_1(1-r)}{B_1 r + B_2(1-r)}
\end{equation}
We see that Figs.~\ref{fig:fit_params}, \ref{fig:efficiencyRatio} show both $A_1$ and $B_2$ falling nearly identically above 100~mK, and the time constants are the same by construction, so these factors are identical. $B_1$ is nominally flat, but $r$ drops  at a similar rate as $B_2$, and thus we can see how a nominally changing amplitude ratio can be expected to become roughly constant when the dynamics of the sensor are taken into account.

We can thus use the main result-- that our phonon efficiency, which we expect to be temperature independent in accordance with the model of Ref~\cite{Tamura}-- to come to some conclusions about the likely source of the different time constants in these models. We have shown that the high-temperature, fast time constant is the thermal sensor response, and inferred that the flat time constant is the phonon collection time. This would suggest that a more physics informed model should use a global fit to phonon collection time and allow only sensor responsivity to change with temperature. In this limit, the time constants in the top right of Fig.~\ref{fig:efficiencyRatio} imply that sensor response time dominates the base temperature data. 

We have also shown in the main text that the response is thermal, and thus this is just the film thermalization time, and not a quasiparticle lifetime. As shown in Fig.~\ref{fig:fit_params}, the sensor response time is always faster than the idealized quasiparticle lifetime, but is approaching that limit for the gap-smeared solution at higher temperature. Future data taking to even higher temperatures will confirm whether the transition to the qp-limited regime is complete; those studies would require a much lower coupling Q to remove the required ring-up correction discussed here.

Finally, we note that, as with other similar low-temperature phonon detectors, the very long time constants seen in Fig.~\ref{fig:fit_params} are most likely additional thermal degrees of freedom in either the substrate or KID that become more apparent as thermal conductances are reduced with temperature. Given that the $df$ response and bandwidth both increase with temperature, it seems likely that operating these at elevated substrate temperature would provide resolution enhancements. This requires substantial further study to see how baseline resolution would correlate with bath temperature and if GR noise would become a relevant limitation. The net result is a drastic change from our existing assumptions but is a promising development for understanding a KID response that, until now, has made little sense.



Future KID designs should therefore aim to absorb phonons in a time comparable to or shorter than expected KID response time, which motivates work to further understand the intrinsic KID response trends seen in these data. The hope would be that, with less disordered films, we see behavior more in line with that expected from MB theory. Alternatively, this may just exacerbate the problem, with quasiparticle lifetimes following the expected Kaplan trend more closely. It's possible that KID-based phonon sensing will only then work in the regime where either disorder allows for rapid thermalization, or the KID is operated at a finite temperature, such that the intrinsic response is fast enough to be utilized as an efficient calorimeter.


\section{Heat capacity of a kinetic inductance detector}
\label{app:heat_capacity}

\subsection{Pair breaking and temperature change}
\label{app:heat_capacity_pb}

In the case of pair-breaking by a non-thermal energy deposition $E_\text{abs}$ (e.g., photons or phonons) in a KID, the resulting perturbations in the resonant frequency and quality factor are approximately linear with perturbations to the quasiparticle density \cite{Gao2008}:
\begin{equation}
\delta\frac{1}{Q_i}-2i\frac{\delta f_r}{f_r} \approx\alpha\gamma\left(\kappa_1+i\kappa_2\right)\delta n_\mathrm{qp},\label{eq:dnqp}
\end{equation}
where $\gamma$ relates fluctuations in complex conductivity $\sigma=\sigma_1+i\sigma_2$ to those in surface impedance ($dZ_s/Z_s=\gamma d\sigma/\sigma$) and $\kappa=\kappa_1+i\kappa_2=\left(d\sigma_1/dn_\mathrm{qp}-id\sigma_2/dn_\mathrm{qp}\right)|\sigma_0|^{-1}$ is the fractional responsivity of complex conductivity to changes in quasiparticle density. The main result of Ref~\cite{Gao2008} is that there is an approximate equivalence between $\partial\sigma/\partial n_\mathrm{qp}$ and $\partial\sigma/\partial T \div \partial n_\mathrm{qp}/\partial T$. In other words, a perturbation $\Delta f_r/f_r$ due to excess quasiparticles is approximately equivalent to  a $\Delta f_r/f_r$ perturbation that arises from a temperature increase $\Delta T$ that produces an equal excess of quasiparticles. The equivalent $\Delta T$ for a fixed $\Delta f_r/f_r$ is exponentially suppressed in $1/T$. We show this behavior with the intersection of the blue dashed lines with the green dotted lines in Fig \ref{fig:pulseShapeModeling}; it is also evident in Fig \ref{fig:VNAscan} Left.


An equivalent statement of $ E_\text{abs} \propto \delta n_\text{qp}$ is that the heat capacity in a superconductor is exponentially suppressed in $1/T$:
\begin{align}
E_\text{abs} &\propto \delta n_\text{qp}\nonumber\\
\lim_{\delta T\rightarrow 0}\frac{E_\text{abs}}{\delta T}&\propto\lim_{\delta T\rightarrow 0}\frac{\delta n_\text{qp}}{\delta T}\nonumber\\
C_\text{pb} = \frac{dU_\text{qp}}{dT} & \propto \frac{dn_\text{qp}}{dT}\propto e^{-\Delta/k_\text{B}T},\label{eq:heat_capacity}
\end{align}
where we introduce $C_\text{pb}$ as the heat capacity due to pair-breaking and $U_\text{qp}$ as the quasiparticle internal energy. $U_\text{qp}$ can be computed from a MB integral:
\begin{align}
U_\mathrm{qp} &= \sum_k E_k f_k\nonumber\\
&\rightarrow  4N_0\int_{0}^{\infty} \operatorname{Re}(n(E,\Delta,T)) E f(E,T)dE,
\label{eq:intE_qp}
\end{align}
where the $f_k$ is the occupation probability of a quasiparticle state at energy $E_k$ and $n(E,\Delta,T)$ is the density of states. We calculate $C_\text{pb}$ from Eqs \ref{eq:heat_capacity},  \ref{eq:intE_qp}, and \ref{eq:DOS_smeared} via numerical integration and differentiation. The main feature of the ensuing $C_\mathrm{pb}(T)$ curve is that it is exponentially suppressed in $1/T$: $C_\text{pb}\propto e^{-\Delta/k_\text{B}T}$, a hallmark of a quantity proportional to $n_\text{qp}$.

\subsection{A TLS-based interpretation for the heat capacity at low temperature }
\label{app:heat_capacity_tls}

Here we present an interpretation to the heat capacity behavior in Fig. \ref{fig:C_comparison} at temperatures below 150\,mK. We hypothesize that it may be dominated by a narrow-mode population of TLSs, where the heat capacity would follow a Schottky anomalous heat capacity:
\begin{equation}
C_\mathrm{Sch}(T) = N k_B \left(\frac{E_\mathrm{TLS}}{k_B T}\right)^{2}
\frac{e^{-E_\mathrm{TLS}/k_B T}}{\left(1 + e^{-E_\mathrm{TLS}/k_B T}\right)^{2}}
= N k_B \left(\frac{E_\mathrm{TLS}}{2 k_B T}\right)^{2}
\mathrm{sech}^{2}\!\left(\frac{E_\mathrm{TLS}}{2 k_B T}\right),
\end{equation}
where $E_\text{TLS}=\sqrt{\delta^2+\epsilon^2}$ and $N$ is the number of TLSs. $\delta$ and $\epsilon$ are the tunneling amplitude and energy-splitting of the TLS, respectively. 

More generally, we would expect that TLSs are more broadly distributed in $E_\text{TLS}$ \cite{Faoro2015}. If $\delta \gg \epsilon$, then the expectation is usually a log-uniform distribution. If $\delta \ll \epsilon$, then a uniform distribution would be expected. A uniform distribution in $E_\text{TLS}$ produces a heat capacity that goes as $T$, whereas a log-uniform distribution produces a flat heat capacity.

\begin{figure}
\includegraphics{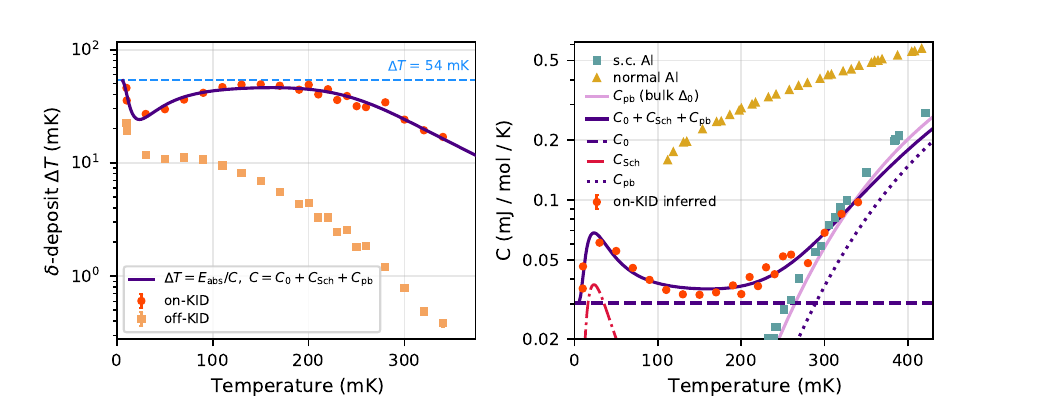}
\caption{Fit of the entire range of $\Delta T$ data to a three-component heat capacity, including a Schottky term. }\label{fig:Schottky}
\end{figure}

We model the $\Delta T $ data with a Schottky component, a pair-breaking component, and a flat component:
\begin{equation}
\Delta T(T) = \frac{E_\mathrm{abs}}{C_0 + C_\mathrm{Sch}(T) + C_\mathrm{pb}(T)}.
\end{equation}
The fit is shown in Fig. \ref{fig:Schottky}. The parameters of the fit are $E_\text{TLS}/h=1.2\,\text{GHz}$, which peaks at about $T=E_\text{TLS}/2.4k_B=23\,\text{mK}$. The fit value for N is 7.4$\times10^9$.

\clearpage
\pagebreak

\end{document}